\documentclass[twocolumn]{aastex63}
\usepackage{natbib}
\usepackage{graphicx}
\usepackage[space]{grffile}
\usepackage{multirow} %make multiple rows in latex table
\usepackage{changepage} %use for adjusting width of tables in the column
\usepackage{chngcntr} %restart counter for appendix figures
\usepackage{cleveref} %reference multiple figures and tables
\usepackage{afterpage} %keep sections from starting before figures
\usepackage{hhline} %make nice double bottom borders in tables
\usepackage{subfigure} %create subfigures within a single figure
\usepackage{rotating} %used to rotate tables by 90 
\usepackage{placeins} %use \FloatBarrier to keep floats from passing it
\shorttitle{Disentangling Stellar and Airglow Emission Lines from \textit{HST}-COS Spectra}
\shortauthors{Cruz Aguirre et al.}

\begin{document}
\setlength{\skip\footins}{1.2pc plus 5pt minus 2pt}

\title{Disentangling Stellar and Airglow Emission Lines from \textit{HST}-COS Spectra}
\author{Fernando Cruz Aguirre}
\affiliation{Laboratory for Atmospheric and Space Physics\\CU Boulder, Boulder, CO}
\author{Allison Youngblood}
\affiliation{NASA Goddard Space Flight Center\\Greenbelt, MD}
\author{Kevin France}
\affiliation{Laboratory for Atmospheric and Space Physics\\CU Boulder, Boulder, CO}
\author{Vincent Bourrier}
\affiliation{Geneva Observatory\\Versoix, Switzerland}
\date{\today}

\begin{abstract}

H I Ly$\alpha$ ($1215.67$ \AA) and the O I triplet ($1302.17$, $1304.86,$ and $1306.03$ \AA) are bright far-ultraviolet (FUV) emission lines that trace the stellar chromosphere. Observations of stellar Ly$\alpha$ and O I using the \textit{Hubble Space Telescope}'s (\textit{HST}) most sensitive FUV spectrograph, the Cosmic Origins Spectrograph (COS), are contaminated with geocoronal emission, or airglow. This study demonstrates that airglow emission profiles as observed by COS are sufficiently stable to create airglow templates which can be reliably subtracted from the data, recovering the underlying stellar flux. We developed a graphical user interface to implement the airglow subtraction on a sample of 171 main sequence F, G, K, and M-type dwarfs from the COS data archive. Correlations between recovered stellar emission and measures of stellar activity were investigated. Several power law relationships are presented for predicting the stellar Ly$\alpha$ and O I emission. The apparent brightness of the stellar emission relative to the airglow is a critical factor in the success or failure of an airglow subtraction. We developed a predictor for the success of an airglow subtraction using the signal-to-noise ratio (SNR) of the nearby chromospheric emission line Si III ($1206.51$ \AA). The minimum attenuated Ly$\alpha$ flux which was successfully recovered is $1.39\times10^{-14}$ erg cm$^{-2}$ s$^{-1}$, and we recommend this as a minimum flux for COS Ly$\alpha$ recoveries.

\vspace{1cm}
\end{abstract}

\section{Introduction}\label{s:intro} %1

The hydrogen Lyman alpha (Ly$\alpha$, $\lambda = 1215.67$ \AA) emission line is the dominant feature of the far ultraviolet (FUV; 912 - 1700 \AA) spectrum of F, G, K, and M dwarf stars. For M dwarfs, Ly$\alpha$ flux alone can account for up to 75\% of the total ultraviolet (UV) stellar flux \citep{2013ApJ...763..149F}. Ly$\alpha$ and O I triplet ($\lambda = 1302.17, 1304.86,$ and $1306.03$ \AA) emission act as tracers for stellar chromospheric activity  \citep{2004ApJ...604L..69V, 2013AnA...553A..52B, 2010ApJ...717.1291L,2015ApJ...804..116B,2018AnA...615A.117B}. The line cores of Ly$\alpha$ and the O I triplet are formed in the transition region and chromosphere, both with a formation temperature of $\sim$30,000K \citep{1997AnAS..125..149D,2012ApJ...744...99L}. The wings of the line profiles originate from a range of cooler temperatures within the chromosphere \citep{2008ApJS..175..229A,2013ApJ...766...69L}. 
\par
Ly$\alpha$ photons, and O I triplet photons to a lesser but non-negligible extent, act as drivers for photochemistry in exoplanet atmospheres. This is primarily due to the large photoabsorption cross sections across the FUV for many molecules typically present in exoplanet atmospheres \citep{2015MNRAS.446..345M,2013AnA...551A.131V}. FUV photochemical effects include the UV sensitive formation of haze layers, which can obscure key indicators of atmospheric habitability in exoplanet transmission spectra \citep{2015ApJ...815..110M,2019ApJ...877..109K}, and the alteration of the abundances (and therefore the mixing ratios) of biosignature molecules (e.g., CH$_4$, O$_2$, O$_3$) \citep{2015MNRAS.446..345M,2015ApJ...812..137H,2018ApJ...854...19R,2014E&PSL.385...22T,2022ApJ...930..131R}. 
\par
Understanding the UV radiation environment, and in particular the dominating Ly$\alpha$ emission, around exoplanet host stars is crucial for properly interpreting biosignatures and habitability indicators \citep{2019ApJ...886...77P,2022ApJ...926..129Y,2022ApJ...927...90T}. \cite{2015ApJ...815..110M} explores how the photodissociation of CH$_4$ ultimately results in the formation of haze layers in super earth atmospheres. This study found that for low incident UV flux, the rate of CH$_4$ dissociation was also low. For high incident UV flux, the carbon photochemistry is altered and produces a CO dominated atmosphere, blocking critical chemical pathways between CH$_4$ dissociation and haze formation.
\par
The spectral type of the host star is a key factor in interpreting O$_2$ features as a biosignature. \cite{2015ApJ...812..137H} found that for Earth-like exoplanets orbiting around F and G type stars, the presence of both strong FUV and near ultraviolet (NUV; 1700 - 4000 \AA) flux from the host star resulted in little accumulation of abiotic O$_2$ from CO$_2$ photolysis. Due to the decreased NUV continuum flux from K and M type stars, the O$_2$ dissociation rate decreases and an abiotic buildup of O$_2$ on Earth-like exoplanets is possible, a false positive for life. A similar effect is also seen in the O$_3$ abundance of Earth-like planets \citep{2018ApJ...854...19R,2014E&PSL.385...22T}. 
\par
In addition to driving photochemistry, Ly$\alpha$ and O I are also probes of exoplanet atmospheric escape \citep{2003Natur.422..143V,2019MNRAS.490.3760A,2010AnA...514A..72L,2012AnA...543L...4L,2013AnA...553A..52B,2014ApJ...786..132K,2015Natur.522..459E,2017AnA...605L...7L,2018AnA...620A.147B} and predictors of extreme ultraviolet (EUV; 100 - 912 \AA) emission \citep{2014ApJ...780...61L}. \cite{2004ApJ...604L..69V} details the detection of Ly$\alpha$, O I triplet, and C II doublet ($\lambda = 1334.53$ and $1335.71$ \AA) transits of the exoplanet HD 209458b. The presence of O I and C II transits at high altitudes constrains the escape method to hydrodynamic escape \citep{2013Icar..226.1678K,2013Icar..226.1695K}. The key driver of escape in exoplanet atmospheres is stellar EUV flux; characterizing the host star's EUV emission is necessary for proper modeling of atmospheric escape. There are few stars for which an EUV spectrum has been observed, with strong attenuation from the interstellar medium (ISM) in the range $\sim$400 - 912 \AA. As shown in \cite{2014ApJ...780...61L}, Ly$\alpha$ can be used to predict the EUV flux of dwarf stars.
\par
While Ly$\alpha$ can be used to estimate the unobserved EUV flux of a star, it too suffers from ISM absorption. For nearby stars, the line core of Ly$\alpha$ is completely attenuated when the radial velocity difference between the stellar and ISM features are within 80 km s$^{-1}$ from each other, leaving only the wings present in FUV spectra \citep{2022ApJ...926..129Y}. The severity of the attenuation varies depending on the direction and distance to the target star due to the nonuniform structure of the ISM. The O I triplet lines can also be affected by ISM attenuation. Due to smaller ISM abundances, a weaker oscillator strength, and requiring that the stellar and ISM velocities be similar to align the absorption with the more narrow emission feature, it is normally not as pronounced in FUV spectra.
\par
The observed spectra from FUV instruments also display bright Ly$\alpha$ and O I emission features originating from Earth's exosphere and upper thermosphere; this is known as geocoronal emission or airglow. Hydrogen and oxygen atoms within these regions are excited by either solar photons or from collisions. These atoms will then radiatively de-excite, releasing Ly$\alpha$ and O I triplet photons in the process. Geocoronal Ly$\alpha$ is observed on both the day and night sides of the planet, while geocoronal O I emission is only strong on the day side \citep{1991SSRv...58....1M}. This different behavior is driven by the large difference in the scale heights of both species, which is inversely proportional to atomic weight. Both species are excited by solar UV photons on the day side, but the density of oxygen atoms is too low in the exosphere for collisional excitation to occur on the night side. The hydrogen density is still large enough in the exosphere for geocoronal Ly$\alpha$ to be generated through collisions. Typical airglow intensities of during orbital day are 20,000 Rayleighs (R) for Ly$\alpha$ and 2,000 R for the individual O I triplet lines, whereas during orbital night typical values decrease to 2,000 R for Ly$\alpha$ and effectively 0 R for the O I triplet lines \citep{COS_IHB_2022}.
\par
The orbit of the \textit{Hubble Space Telescope} (\textit{HST}; $\sim$540 km) lies within the Earth's exosphere, which extends beyond 38 Earth radii \citep{2017GeoRL..4411706K}. The lower boundary of the exosphere is defined by the altitude in which atmospheric density is low enough such that particles become collisionless, which varies with stellar activity. Airglow fully illuminates the slits of the Cosmic Origins Spectrograph (COS) and the Space Telescope Imaging Spectrograph (STIS). All spectra observed by these spectrographs become contaminated with airglow, obscuring stellar Ly$\alpha$ and O I emission. 

\begin{figure*}[!ht]
\centering
\subfigure{\epsscale{0.5} \plotone{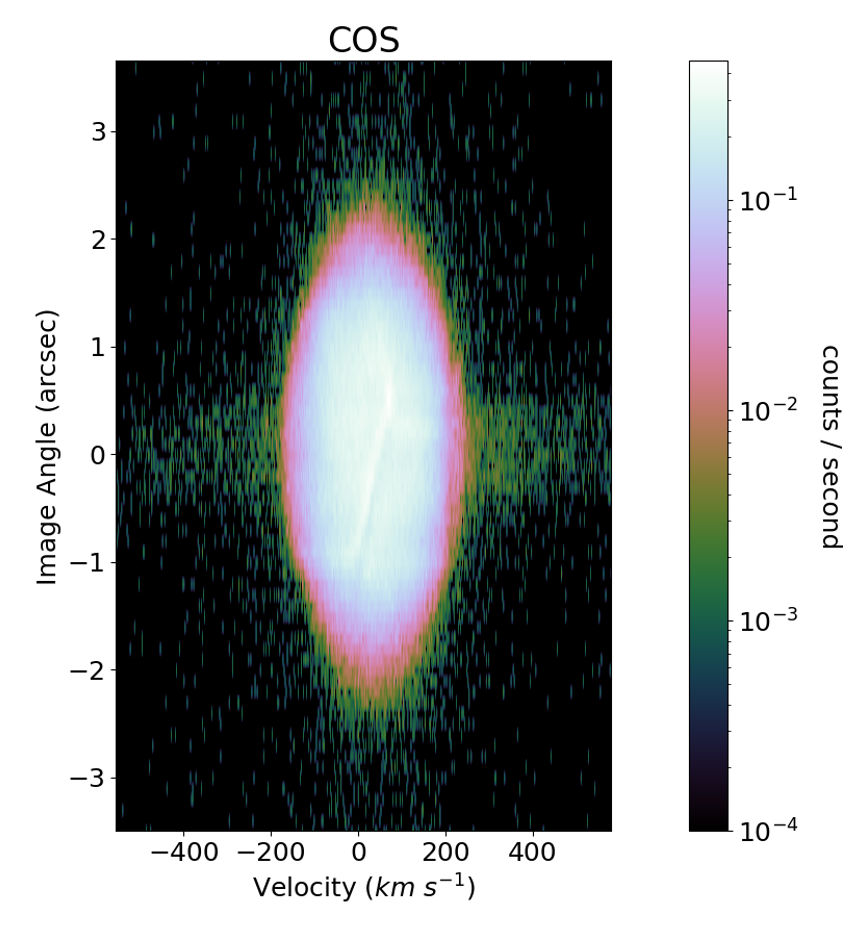} \label{sfig:cosD}} %1a
\subfigure{\epsscale{0.5} \plotone{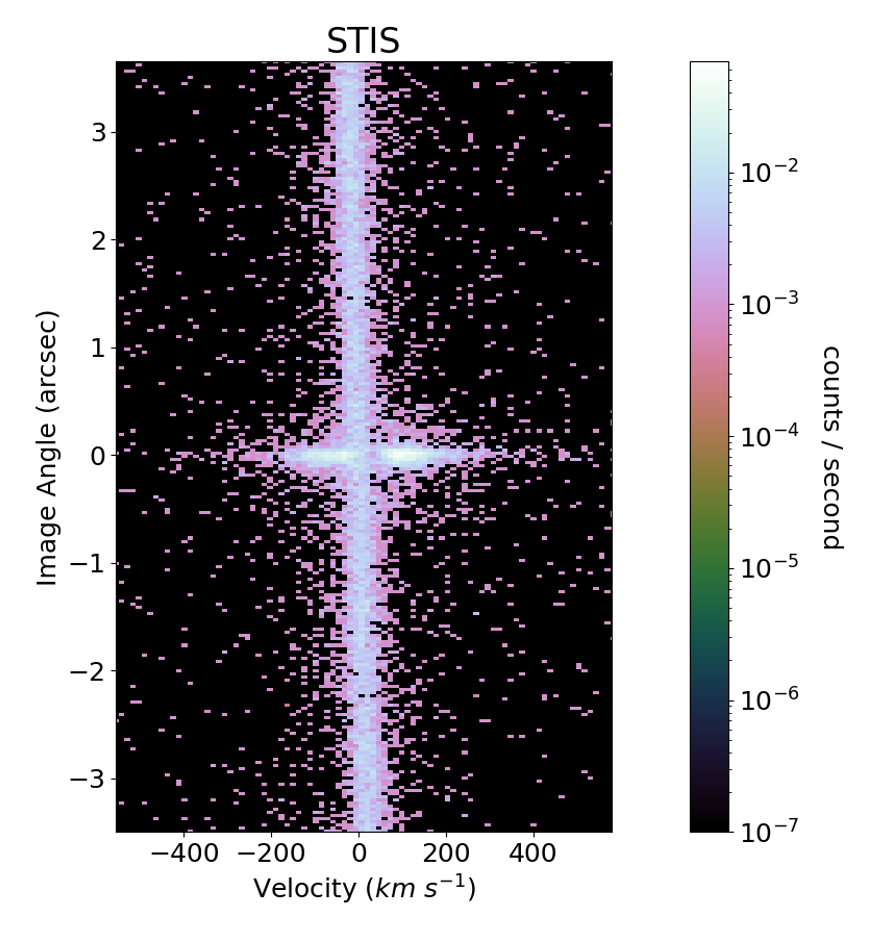} \label{sfig:stisD}} %1b
\caption{2D detector images of COS (left) and STIS (right) observations of GJ 876, centered on the Ly$\alpha$ stellar and airglow emission features. The stellar spectrum runs along the X axis in both images. The circular COS and rectangular STIS apertures are fully illuminated by airglow, contaminating the stellar spectrum. Aspect ratios of both detector images were altered for a better visual comparison. \label{fig:detector}} %1
\end{figure*}

Due to the different designs of these spectrographs, they experience airglow contamination differently. STIS is an imaging spectrograph, with a spatial resolution capable of distinguishing the target spectrum from the off-target background spectrum. Airglow can then be subtracted during standard background removal procedures, recovering the stellar emission. The optical design of COS prioritized the spectral resolution of faint point sources through the use of a single optic which corrects for multiple aberrations \citep{2012ApJ...744...60G}, but suffers from strong vignetting  and uncorrected astigmatism in the spatial axis, so an off-target background spectrum is impossible \citep{COS_IHB_2022}. The large, circular COS aperture (2.5'' diameter) also results in more airglow captured by the instrument, compared to the narrow, rectangular slits (52'' length and 0.05 - 2'' width for first-order spectral imaging, 0.2 - 1'' length and 0.025 - 0.5'' width for echelle spectroscopy) employed by STIS  \citep{2012ApJ...744...60G,1998PASP..110.1183W}. The instrument design of COS results in severe airglow contamination, without the ability to subtract the airglow from a single exposure. Examples of COS and STIS 2D detector images of the M3.5 V dwarf GJ 876 are shown in Figure \ref{fig:detector}, illustrating the differences in how the two spectrographs experience airglow.
\par
Few stars with COS spectra have had an airglow subtraction performed \citep{2013AnA...553A..52B,2017AnA...599A..75W,2018AnA...615A.117B,2019AnA...629A..47D,2021AnA...650A..73B}, however there are more than 170 dwarf stars in the COS data archive. This paper describes the airglow subtraction of archival F, G, K, and M dwarf star COS spectra and the subsequent analysis of the recovered profiles. Section \ref{s:observations} describes the data collection and coaddition of the stellar sample, and the creation of airglow templates from COS observations. Section \ref{s:removal} details the development of a graphical user interface (GUI) which performs the airglow subtraction for contaminated Ly$\alpha$ and O I triplet line profiles. The integrated fluxes obtained from recovered stellar profiles are compared to other stellar chromospheric fluxes and activity indicators in Section \ref{s:correlations}.

\section{COS and STIS Observations}\label{s:observations} %2

We identified 171 stars of F, G, K, or M type from available COS G130M spectra in the Mikulski Archive for Space Telescopes (MAST) data archive. All COS observations analyzed in this work can be accessed through this\dataset[DOI]{http://dx.doi.org/10.17909/0sc9-b207}. There are 78 confirmed exoplanet host stars in this sample \citep{NEA}. Characterizing the Ly$\alpha$ and O I emission of exoplanet host stars is critical for interpreting star-planet interactions, requiring a proper airglow subtraction of COS stellar data. 
\par
Stars with both COS and STIS observations allow for a verification of proper airglow subtraction of the star. We identified 35 stars that have STIS G140M or E140M spectra at comparable wavelengths which can be used for comparison with our airglow subtracted COS spectra. These STIS observations were obtained from MAST, and can be accessed through a separate\dataset[DOI]{http://dx.doi.org/10.17909/nsag-sq06}. There were 20 STIS spectra observed as part of the same programs as the COS data; this assists in minimizing the effect of stellar variability, as the COS and STIS observations were acquired at similar times. 
\par
Settings for the COS G130M grating and detector are the Lifetime Position (LP), the focal plane position (FP-POS, FP), and the Cenwave (in \AA). The LP refers to the location where the spectrum falls along the cross-dispersion direction of the detector. The different LPs were created to address regions on the detector that began to lose  efficiency after prolonged photon exposure, a phenomenon known as gain sag. As gain sag increases over time, the ability to distinguish a real photon from a background event is no longer possible. This manifests as a depression in the spectrum due to real photon events being mistakenly discarded by the COS data calibration pipeline (CALCOS) \citep{COS_IHB_2022}. These heavily gain sagged regions are marked with a gain sag data quality (DQ) flag in the CALCOS-generated data files, indicating that the data in the region may not be reliable \citep{COS_DHB_2021}. The first region to experience gain sag is where bright geocoronal Ly$\alpha$ emission lands on the detector. Over time, gain sag is seen across the detector face, and a new LP is chosen \citep{COS_IHB_2022}, shifted above or below the previous LP (Table \ref{tab:LP} lists each LP start date/cycle). The LPs are numbered 1 through 6, however all stellar data used in this work was taken using LPs 1-4.
\par
The Cenwave refers to the central wavelength of the spectrum that falls on the detector. The gratings on COS can be tilted such that different wavelength ranges fall onto the detector. The COS detector consists of two separate plates, referred to as Segment A (longer wavelengths) and Segment B (shorter wavelengths), and the Cenwave number is approximately the shortest wavelength observed on Segment A \citep{COS_IHB_2022}. Because the detector consists of two separate plates, there is a gap in the spectrum. Observations of the same target with different Cenwaves make up for these gaps. The five Cenwaves considered in this work are 1291, 1300, 1309, 1318, and 1327. While all five Cenwaves are available for use at LP4 (with Segment restrictions), the only available LP4 stellar spectra in the COS data archive that were relevant to this work were taken using Cenwave 1291 \citep{COS_IHB_2022}.
\par
In addition to the Cenwave, the different FPs allow for finer adjustments of the spectral location on the detector. There are four FP positions, numbered 1 to 4. For a given Cenwave, the four FPs offset the spectrum in the dispersion direction by a small amount (250 px in the FUV channel, or $\sim$2.5 \AA\ for G130M). The nominal position is FP3. Taking multiple observations of the same target with various FPs can increase the overall signal-to-noise ratio (SNR) by smoothing over the fixed pattern noise of the FUV detector \citep{COS_IHB_2022}. 

\subsection{Stellar Observations}\label{ss:stellar obs} %2.1

The COS stellar observations addressed in this work were obtained using the G130M grating. The STIS stellar observations used in this work were obtained using the G140M or E140M gratings. Tables \ref{tab:literature} and \ref{tab:literature_STIS} contain the observing programs of the stellar COS and STIS data, respectively.

For each star, we created a coadded COS spectrum consisting of all available observations taken within a $\pm$2 week time frame. The spectral location of the airglow is known to change over time with the spacecraft's orbital motion, however it is not expected to change significantly within this time frame. In the case where a star has multiple possible $\pm$2 week windows, we use the set of observations which results in the maximum usable exposure time. We exclude certain observations within these time frames due to the effects of gain sag (see discussion below) or because a Cenwave other than the ones considered in this work were used. Exposures of exoplanets mid-transit were not excluded. In the few cases where we would expect a detectable transit, the exclusion of transit exposures from the coadded spectrum did not significantly change the results of this work.
\par
The coaddition process used a modified version of the \texttt{COADD\_X1D} IDL code developed for the coaddition of individual COS G130M \texttt{x1d} files.\footnote{\url{https://casa.colorado.edu/$\sim$danforth/science/cos/coadd\_x1d.pro}} The airglow contaminated stellar Ly$\alpha$ and O I triplet profiles were coaligned through a cross correlation process. Changes in the airglow profile are minimized within the $\pm$2 week window, allowing for stable coadditions. A weighted average was taken within each wave bin to produce the final coadded spectrum of each star. Nearby spectral emission features were used to determine that the cross correlation had properly aligned individual spectra. 
\par
We do not perform the cross correlation on these nearby emission features to minimize smearing of the Ly$\alpha$ and O I profiles. We later found that within these $\pm$2 week windows, a cross correlation on either the nearby emission features or the airglow contaminated airglow features produce similar spectra. Applying the procedures detailed in Section \ref{s:removal} on both cross correlated spectra produced results consistent within 1$\sigma$, and we continue with Ly$\alpha$ and O I cross correlations throughout this work. 
\par
We use the coadded G140M and E140M spectra of MUSCLES (v2.2) and Mega-MUSCLES (v2.3) targets within our sample \citep{2016ApJ...820...89F,2016ApJ...824..101Y,2016ApJ...824..102L,2019ApJ...871L..26F,2021ApJ...911...18W}. For the remaining STIS targets, a weighted average coaddition is performed with all available G140M or E140M exposures for a given target. As with COS data, the exclusion of transit exposures did not result in a significant change. 
\par
For faint G140M spectra, the STIS data calibration pipeline (CALSTIS) fails to discern the dim stellar spectral trace from the surrounding background \citep{2016ApJ...824..102L}. This is a known issue in the CALSTIS pipeline which affected 12 targets observed with the G140M grating. In these instances, spectra were manually extracted from the 2D detector image of the exposure through the use of the \texttt{stistools} python package distributed by the Space Telescope Science Institute (STScI).\footnote{\url{https://stistools.readthedocs.io/en/latest/}} For the M dwarf L 980-5, the stellar spectra were dim enough that a manual extraction was impossible, as the stellar trace was not discernible from the background by eye.
\par
As mentioned above, gain sag can heavily affect the quality and reliability of COS stellar spectra. Within our sample, gain sag was observed in LP3 and LP4 data. A visual inspection of the shapes of Ly$\alpha$ profiles for these stars was conducted to determine if the datasets were of acceptable quality. All LP4 exposures observed with FP3 were discarded due to heavy gain sag. 
\par
We note that for the SNAP 14633 program data, we ignored the DQ flag relating to detector gridwire shadows. This program only took two exposures per star, at FP3 and FP4. The FP3 exposures had the detector gridwire shadow DQ flag appear within the Ly$\alpha$ profile, and the \texttt{COADD\_X1D} code will deweight the flux in this region. With only two exposures, this directly affect the weighted average and causes depressions in the coadded spectrum. At worst, we expect a flux decrease up to 20\% due to the detector gridwire shadows in these regions \citep{2011cos..rept....3E}.
\par
Six stars within our sample were too bright for COS to observe with both detector segments, and were observed only using Segment A which does not contain Ly$\alpha$. This reduced the total number of stars for which a Ly$\alpha$ airglow subtraction was attempted to 165, where 28 of these stars had STIS spectra available for comparison. For observations using Cenwave 1309, the O I triplet lies within the detector gap. Stellar O I data was not available for 19 targets observed with Cenwave 1309. The STIS G140M grating cannot cover both Ly$\alpha$ and the O I triplet in a single exposure; all G140M spectra used in this work cover Ly$\alpha$. STIS E140M spectra cover both Ly$\alpha$ and the O I triplet. The total number of stars for which an O I triplet airglow subtraction was attempted was 152, and 12 had STIS spectra available for comparison.

\vspace{-0.2cm}

\subsection{Airglow Observations}\label{ss:airglow obs} %2.2

The airglow observations used in this work were also obtained from the MAST data archive.\footnote{\url{https://www.stsci.edu/hst/instrumentation/cos/calibration/airglow}} We build upon the airglow template creation and subtraction techniques developed in \cite{2013AnA...553A..52B}, \cite{2017AnA...599A..75W}, and \cite{2018AnA...615A.117B}. Both intentional and serendipitous observations were used, for combinations of LPs 1, 2, and 3, with Cenwaves 1291, 1309, and 1327. The serendipitous observations are only available for LP1, and show no significant difference from intentional observations. For a given LP and Cenwave, all available observations were shifted, scaled, and coadded together into a single LP/Cenwave template using another modified version of the \texttt{COADD\_X1D} code. The shifts are obtained through the same cross correlation method used for the stellar profiles, centered on the geocoronal emission. The individual observations were all scaled to the same flux level, and the coadded flux was determined as the median flux in each wave bin. The flux errors for these templates were calculated using the Median Absolute Deviant (MAD) \citep{Leys2013}, defined as the median of all absolute flux deviations from the median flux. The MAD is multiplied by 1.4826 to obtain an equivalent estimator for $\sigma$ \citep{Leys2013}. This median method was chosen over a simple mean due to being less sensitive to flux outliers within each wave bin, producing more stable coadded airglow templates with flux errors similar to that of an unweighted average. 
\par
We first compared templates for Cenwaves within a given LP to each other in order to determine if there were any differences in the airglow profiles of different Cenwaves. Afterwards, we compared representative airglow templates of each LP to representatives of the other LPs. For LP 1-3, airglow was not observed for every possible LP/Cenwave combination, and there is no airglow data available for LP4. The missing combinations are inferred from available templates.

\subsubsection{Cenwave Comparison}\label{sss:cen} %2.2.1

\begin{figure}[!ht]
\epsscale{1.2}
\plotone{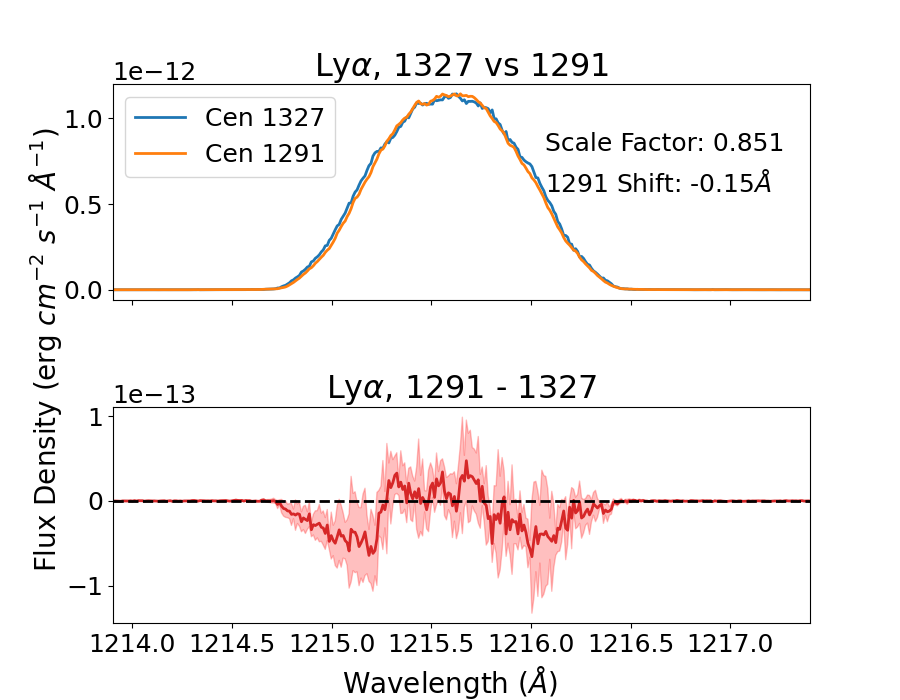}
\caption{A comparison of LP3 1291 and LP3 1327 Ly$\alpha$ airglow templates on the upper plot, and the difference between the two templates on the lower plot. The templates are shifted such that the templates are centered on one another, then scaled to each other. The 1327 template is broader in the line wings. This discrepancy is seen in the difference plot, where the variations tend to be larger than the flux subtraction error (shaded region). \label{fig:Cen_comp}} %2
\end{figure}

Within each LP, airglow templates were created for each available Cenwave. These templates were then shifted in wavelength and scaled on top of one another to examine differences in the shapes of the airglow profiles. An example is shown in Figure \ref{fig:Cen_comp} (see full collection in Figures \labelcref{fig:All_LP1,fig:All_LP2,fig:All_LP3}).

For LPs 1 and 2, there were no significant differences in the different Cenwave templates for all four airglow lines. The LP3 1327 Ly$\alpha$ template displays enhancements in the wings of the profile when compared to that of LP3 1291, but no differences were observed between the Cenwaves for O I. Therefore, for each O I line, one template can be used within each LP. One general Ly$\alpha$ template can be used for LP 1 and 2, but individual templates are needed for LP3 (See Figure \ref{fig:Cen_comp}). No airglow observations are available for LP4 and beyond, so no comparison could be made.

\subsubsection{Lifetime Position Comparison}\label{sss:LP} %2.2.2

\begin{figure}[!ht]
\epsscale{1.2}
\plotone{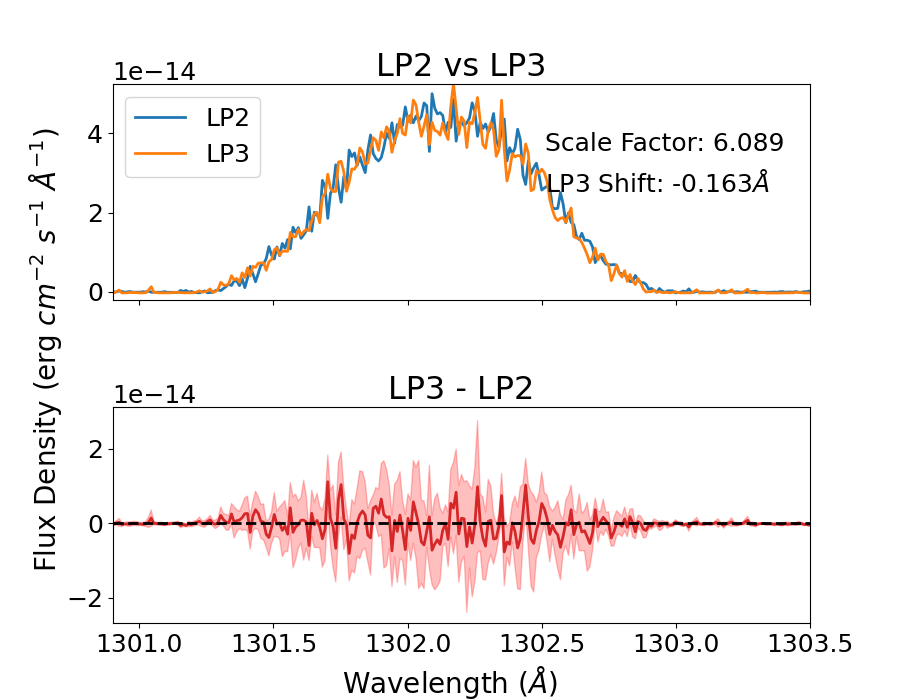}
\caption{A comparison of LP2 and LP3 airglow templates of O I 1302 on the upper plot, and the difference between the two templates on the lower plot. The flux variation in the difference plot is consistent with zero. \label{fig:LP_comp}} %3
\end{figure}

After determining the representative templates for each LP, these were compared to each other. An example is shown in Figure \ref{fig:LP_comp} (see full collection in Figures \labelcref{fig:All_LyA,fig:All_OI2,fig:All_OI5,fig:All_OI6}). The shift and scale factors are determined in the same way as in the previous section. These comparisons show that each O I triplet line only needs a single airglow template across LPs 1-3 and Cenwaves 1291-1327, with the exception of Cenwave 1309 which places the triplet in the gap between detector plates. For Ly$\alpha$, one general template can be used for most configurations, except for LP3 1327 which requires a separate template. We examine the location of the Sun, the Moon, and the orientation of \textit{HST} during all airglow observations, however these were unable to explain the observed difference in the shape of the LP3 1327 profiles.  

\begin{figure*}[!ht]
\epsscale{1.15}
\plotone{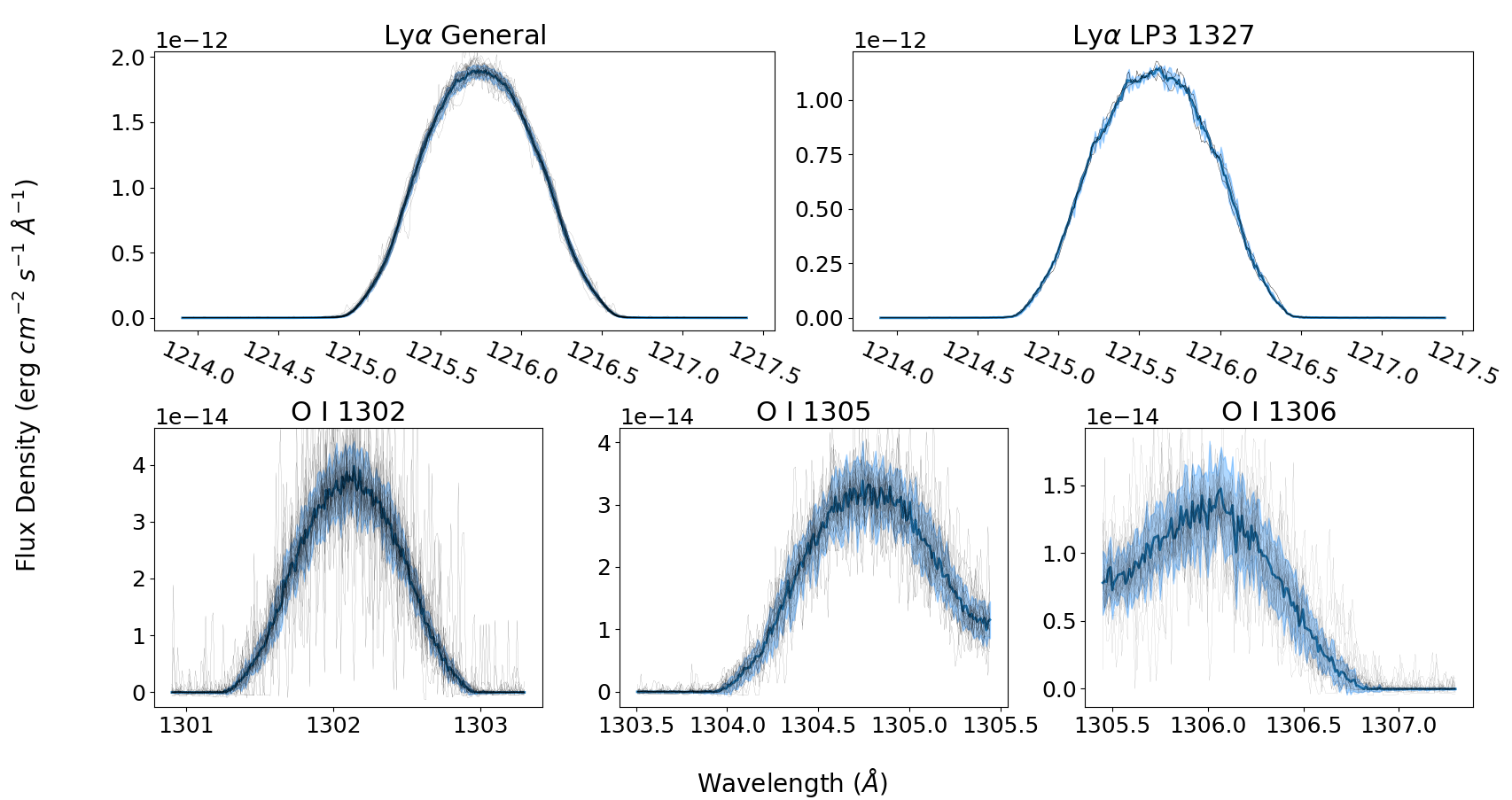}
\caption{The airglow templates developed and used throughout this work. Shaded regions show the 1$\sigma$ uncertainties. The general Ly$\alpha$ template can be used for all LP and Cenwave combinations, except LP3 1327 which requires its own template. The O I triplet templates can be used for all LP and Cenwave combinations for which the triplet is not within the gap between detector plates. Individual airglow exposures are overlain in each subplot as thin black curves. \vspace{1.25cm} \label{fig:templates}} %4
\begin{adjustwidth}{0.3cm}{}
\epsscale{1.18}
\plotone{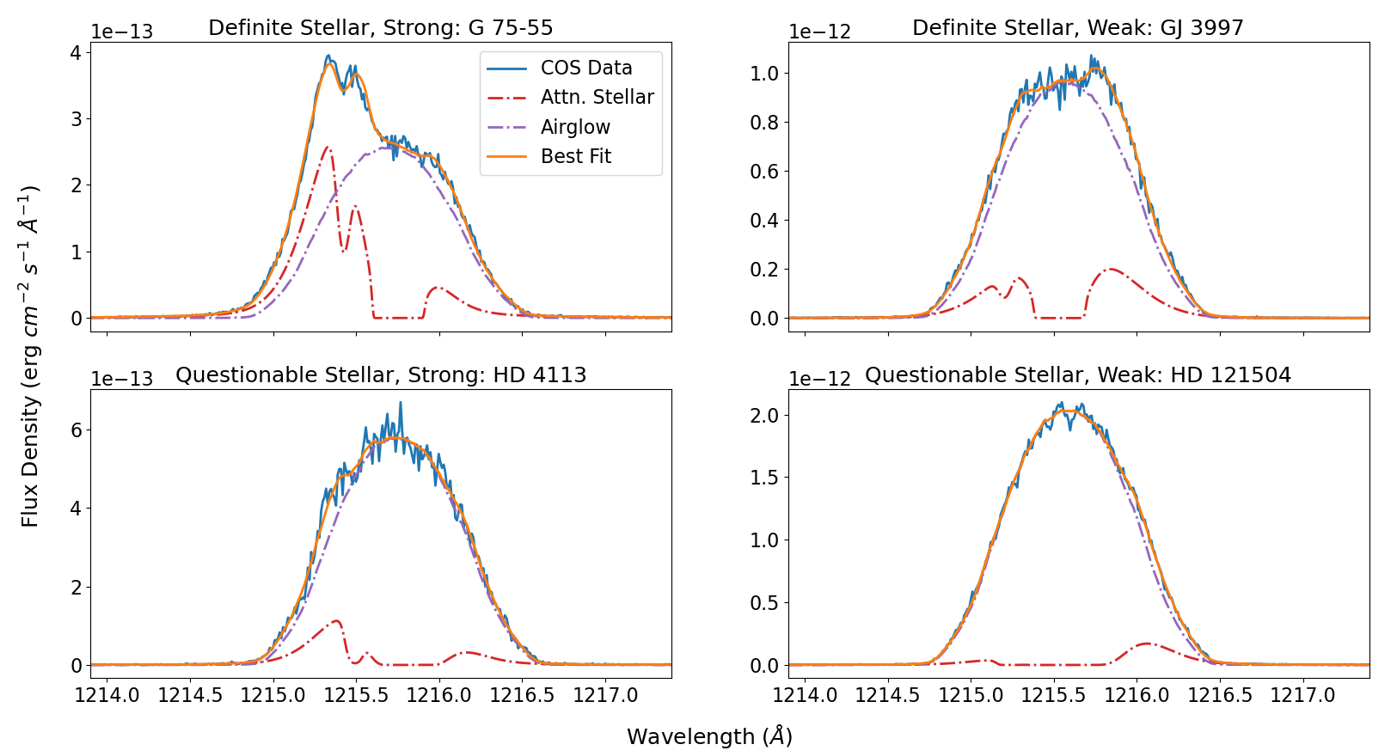}
\caption{The four general cases of the contaminated Ly$\alpha$ profiles we observe from our sample of stars. \label{fig:morphology}} %5
\end{adjustwidth}
\end{figure*}

\begin{table}[!ht]
 \begin{adjustwidth}{-1.5cm}{}
\centering
\resizebox{4in}{!}{
\begin{tabular}{|c|c|c|}
\hline
\textbf{Template Name} & \textbf{Applicable LP} & \textbf{Applicable Cenwave}\\ 
\hline
Ly$\alpha$ - General & 1,2,3,4 & 1291,1300,1309,1318,1327* \\ \hline
Ly$\alpha$ - LP3 1327 & 3 & 1327 \\ \hline
O I 1302 & 1,2,3,4 & 1291,1300,1318,1327 \\ \hline
O I 1305 & 1,2,3,4 & 1291,1300,1318,1327 \\ \hline
O I 1306 & 1,2,3,4 & 1291,1300,1318,1327 \\ \hline
\end{tabular}}
\end{adjustwidth}
\caption{A summary of the five airglow templates created for airglow subtraction, and which detector settings each template is applicable to. \\ **Cenwave 1327 not applicable for LP3.\vspace{-0.6cm}}
\label{tab:templates} %1
\end{table}

No data is available for LP3 1300 - 1318, so it was unclear which Ly$\alpha$ template would be appropriate to use for stars observed using these detector settings. Through the least squares minimization technique which will be described in Section \ref{ss:GUI}, we determined that the general Ly$\alpha$ template produced a lower chi squared ($\chi^2$) than the LP3 1327 template when modeling airglow contaminated LP3 1300 - 1318 spectra. This method also indicated that the general Ly$\alpha$ template and the O I templates are appropriate for LP4 1291 data, without significant gain sag present at Ly$\alpha$. The airglow templates are shown in Figure \ref{fig:templates}, and Table \ref{tab:templates} summarizes the detector settings these templates are applicable for.

\section{Recovering Stellar Emission}\label{s:removal} %3

Using the appropriate airglow template from Section \ref{ss:airglow obs}, we perform airglow subtractions on the stellar spectra described in Section \ref{ss:stellar obs}. The templates must be shifted and scaled onto the stellar spectrum in order to properly subtract the airglow and recover the underlying stellar profiles. Building upon the techniques developed in \cite{2013AnA...553A..52B}, \cite{2017AnA...599A..75W}, and \cite{2018AnA...615A.117B}, we developed airglow subtraction methods which recover the stellar Ly$\alpha$ and O I triplet profiles. These methods were first tested on a subset of stars for which both COS spectra and airglow-subtracted STIS spectra were available. This allows for a way to verify that the subtraction method is functioning as expected and is producing satisfactory results. 

\subsection{General Cases of Airglow Contamination}\label{ss:contamination} %3.1

We have generally categorized contaminated Ly$\alpha$ profiles into four groups, which can be seen in Figure \ref{fig:morphology}. The first category is Definite Stellar, Strong. This refers to cases where the stellar profile is clearly visible above the airglow profile. The second category is Definite Stellar, Weak, where the stellar wings rise above the airglow, however their amplitude is comparable to that of the airglow profile. For these two categories, separation of the stellar and airglow signals is straightforward.
\par
The third category is Questionable Stellar, Strong. This refers to a profile in which the airglow dominates, however the inclusion of an attenuated stellar profile produces a better fit than the airglow template alone. When the observed profile could potentially be well fit by the airglow template alone, stars are placed in the fourth category, Questionable Stellar, Weak. Recovering the comparatively weak stellar signal in these cases is where the difficulty lies.
\par
There are three general cases for O I triplet airglow emission in contaminated COS spectra. The first is Stellar Dominant, where the stellar emission feature dominates. These observations were taken during orbital night, where geocoronal O I emission is either very dim or undetectable. The second case is Mixed Emission, a mixture of a narrow stellar emission feature and a broad geocoronal emission feature. The stellar line noticeably peaks above the airglow line, making airglow template placement straightforward (See Figure \ref{fig:recover2}). The final case is Airglow Dominant, where the airglow dominates, and the stellar oxygen line is difficult to separate from the noise after airglow removal.

\subsection{Airglow Subtraction Model}\label{ss:model} %3.2

In order to determine the spectral location and amplitude of the airglow templates, models of contaminated stellar Ly$\alpha$ and O I triplet emission as observed by COS were fit to the data using the \texttt{LMFIT} package in python \citep{LMFIT}. \texttt{LMFIT} is a non-linear least squares minimization and curve fitting package that uses the Levenberg-Marquardt (LM) algorithm. This approach allows for simultaneous fitting of stellar emission, ISM attenuation, and contaminating airglow.
\par
We verified the accuracy of the model by using a Markov chain Monte Carlo (MCMC) for three stars spanning the range in Ly$\alpha$ data quality; 55 Cnc (Definite Stellar, Strong), HD 121504 (Questionable Stellar, Strong), and K2-3 (Questionable Stellar, Weak). The stellar Ly$\alpha$ profile of K2-3 could not be recovered through our recovery tool. We utilize the MCMC method within \texttt{LMFIT}, which employs the \texttt{emcee} package \citep{LMFIT,2013PASP..125..306F}. 
\par
When comparing our model results with those of the MCMC, we find excellent agreement between best fit parameters, Ly$\alpha$ fluxes, and parameter errors in all three cases. The MCMC was run to convergence for 55 Cnc and HD 121504. While the MCMC shows the potential to extract the dim stellar emission of K2-3 from the dominating airglow contamination, the computational time to run the MCMC to convergence was prohibitively expensive, and its application would be unfeasible for the full sample of failed recoveries. The MCMC results give us confidence that the results of our model can provide similar results to those of a computationally intensive MCMC in significantly less time.

\subsubsection{Model Components}\label{sss:components} %3.2.1

The first of the model components is the underlying stellar emission. For Ly$\alpha$, this is represented as a Voigt profile \citep{2017AnA...599A..75W, 2018AnA...615A.117B}. A Gaussian profile was instead used for the O I triplet's stellar component \citep{2004ApJ...602..776R,2018AnA...615A.117B}. The free parameters for this component are the star's radial velocity, the Gaussian FWHM, the Lorentzian FWHM for the Voigt profile only, and the flux amplitude of the Lorentzian/Gaussian. We bound the Gaussian FWHM for the O I triplet profiles to 41.6 km s$^{-1}$ based on the measured widths of the 55 Cnc O I triplet. This upper bound prevents the model from trying to fit a Gaussian to the airglow signal rather than the stellar signal. 
\par
The stellar profiles were produced using \texttt{Astropy}'s \texttt{Gaussian1D} and \texttt{Voigt1D} functions. We also include a self-reversal subcomponent to the Ly$\alpha$ profile, represented as a Gaussian absorption centered on the Voigt profile, as its exclusion can result in an overestimation of the underlying stellar Ly$\alpha$ flux \citep{2022ApJ...926..129Y}. The FWHM of this absorption component is fixed at 47\% of the Gaussian FWHM of the Voigt profile. For FGK stars, the absorption depth is fixed at 50\%. These fixed values were selected to produce self-absorbed profiles that are qualitatively similar to solar Ly$\alpha$ profiles across a range of typical Gaussian FWHM values that were observed in fits to stellar profiles. For M dwarfs, we determined that a shallower absorption depth of 35\% was more appropriate based on Kapteyn's Star. With a radial velocity of 245 km s$^{-1}$, the Ly$\alpha$ profile of Kapteyn's Star is shifted away from both the contaminating airglow and the ISM absorption, allowing for a rare view at the shape of the stellar Ly$\alpha$ profile \citep{2016ApJ...821...81G,2016ApJ...824..101Y,2022ApJ...926..129Y}.
\par
The second model component is the ISM attenuation for Ly$\alpha$. This component includes attenuation from both hydrogen and deuterium. While ISM attenuation can be present in the O I lines, this feature is only seen in a few stars within our sample and is not generally applicable. The free parameters for the ISM model component are the hydrogen column density and the radial velocity of the ISM. There is a known degeneracy between the column density and the Doppler b parameter \citep{2018MNRAS.474.3649S}. Because accurate ISM parameters are not a priority of our fits, for simplicity we fix the Doppler b parameter to 11.5 km s$^{-1}$ for Ly$\alpha$. This value is based on typical values from local ISM observations, and the bounds for the column densities and ISM radial velocities were also based on the range of typically observed values \citep{2004ApJ...609..838W, 2005ApJS..159..118W, 2004ApJ...602..776R}.
\par
An optical depth profile is created using these ISM parameters, where the absorption cross section is modeled by a Voigt function \citep{1948ApJ...108..112H}. The deuterium column density is determined as the hydrogen column density multiplied by the $D/H$ ratio. For this work, $D/H=1.5\times10^{-5}$ was used \citep{1998SSRv...84..285L,2003ApJ...599..297H,2006ApJ...647.1106L}. The optical depths of hydrogen and deuterium are added together to produce a final Ly$\alpha$ optical depth ($\tau$), and the underlying stellar emission is multiplied by $e^{-\tau}$ to create the ISM attenuated stellar emission. 
\par
Finally, the modeled profile is then convolved with the COS Line Spread Function (LSF) that corresponds to the LP that the stellar data was observed with. Because the coadded spectra can be created from observations of multiple Cenwaves, we utilize the Cenwave 1291 LSF for the given LP. These LSFs were obtained from STScI.\footnote{\url{https://www.stsci.edu/hst/instrumentation/cos/performance/spectral-resolution}}

\begin{table}[!ht]
 \begin{adjustwidth}{-1.6cm}{}
\centering
\resizebox{4in}{!}{
\begin{tabular}{|c|c|c|}
\hline
\textbf{Parameter Name} & \textbf{Lower Bound} & \textbf{Upper Bound}\\ 
\hhline{|=|=|=|}
\multicolumn{3}{|c|}{\textbf{Stellar}} \\ \hline
Gaussian & \multirow{2}*{0.0} & N/A (Ly$\alpha$) \\ 
FWHM (km/s) & & 41.6 (O I) \\ \hline
Lorentzian & 0.0 (Ly$\alpha$) & \multirow{2}*{N/A} \\ 
FWHM (km/s) & N/A (O I) & \\ \hline
\multicolumn{3}{|c|}{\textbf{ISM}} \\ \hline
Column Density & 17.1 (Ly$\alpha$) & 19.1 (Ly$\alpha$) \\ 
($10^x$ cm$^{-2}$) & N/A (O I) & N/A (O I)\\ \hline
ISM Velocity & -50.0 (Ly$\alpha$) & 50.0 (Ly$\alpha$) \\ 
(km/s) & N/A (O I) & N/A (O I) \\ \hline
\multicolumn{3}{|c|}{\textbf{Airglow}} \\ \hline
Airglow & \multirow{2}*{-1.5} & \multirow{2}*{1.5} \\
Shift (\AA) & & \\ \hline
Airglow & \multirow{2}*{0} & \multirow{2}*{10} \\
Scale Factor & & \\ \hline
\end{tabular}}
\end{adjustwidth}
\caption{Upper and lower bounds for the model parameters which are bounded. \vspace{-0.5cm}}
\label{tab:bounds} %2
\end{table}

The final model component is the airglow template, where the appropriate template developed in Section \ref{ss:airglow obs} is added onto the convolved stellar (and ISM attenuated, if appropriate)  emission. The two parameters of this model component are a wavelength shift and a scale factor to the template, accounting for changes in the location and amplitude of the profile. The airglow template is subtracted from stellar data in order to produce a recovered profile. The error of the airglow template and the error of the COS data are added in quadrature, producing the error of the recovered profile. Bounded parameter limits are shown in Table \ref{tab:bounds}. 

\subsection{The Stellar Emission Recovery Tool} \label{ss:GUI} %3.3

A GUI was developed to combine the model components, simultaneously fitting for the underlying stellar emission, the attenuating ISM, and the contaminating airglow \citep{GUI}. We explored an alternate model which performs a second fit after the airglow was removed, simultaneously fitting the stellar and ISM components to the recovered profile which reflects the errors in the airglow subtraction. We found that there were no significant differences in the best fit parameters between the two fitting methods. We also explored manual placement of the airglow template on the subset of stars with both COS and STIS observations, and found that this also did not produce a significantly different fit. All airglow subtractions presented in this work were done by simultaneous fitting of all available model components for each emission line.
\par
The GUI takes in \texttt{x1d} COS G130M files or output ascii files produced from the coaddition process described in Section \ref{ss:stellar obs}, automatically searching the data header for the target name, the LP, and detector Segment(s) used during the observation. The user is prompted for this information if it cannot be obtained from the input file. The GUI optionally takes in \texttt{x1d} STIS G140M or E140M files for the purpose of making visual comparisons with STIS data. The GUI currently does not support stellar reconstructions on STIS data.
\par
After selecting which emission line to remove airglow from, the user must then input the radial velocity of the star as an initial guess for the models. For this work, nearly all radial velocities were obtained from the SIMBAD astronomical database.\footnote{\url{http://simbad.u-strasbg.fr/simbad/}} The user must also input an initial guess for the ISM velocity. We calculated ISM velocities using the Local ISM (LISM) Kinematic Calculator \citep{2008ApJ...673..283R}.\footnote{\url{http://lism.wesleyan.edu/LISMdynamics.html}} We input the Right Ascension (RA) and Declination (Dec) of the target star into the Kinematic Calculator, which results radial velocities of clouds in the LISM. We calculate a weighted average of radial velocities of clouds in the line of sight, and if there are no such clouds then we calculate the weighted average of nearby ($<20^{\circ}$) clouds.
\par
Four sliders are available to make adjustments to the fit. The primary two sliders allow the user to provide initial guesses to the airglow template shift and scale factor. The secondary two sliders allow for an overall wavelength shift to the total fit which shifts all of the individual components, and an overall scale factor to the best fit which scales all of the individual components. This can serve as a guide for determining initial guesses for the airglow parameters. The user also has direct control over the wavelength range in which the fit is performed on, which then corresponds to the integration region when calculating integrated fluxes. While other model parameters cannot be accessed by the user, their initial guesses and parameter ranges have been selected such that the model is generally stable and widely applicable to COS stellar spectra.
\par
Once a model is fit to the data, the user has the ability to display individual model components along with the fit and the COS data, the best fit parameters and 1$\sigma$ parameter errors, and fit statistics such as $\chi^2$ and the Bayesian Information Criterion (BIC). The user can then view the normalized residuals, $[(Fit-Data)/(Data$ $Error)]^2$, plotted against wavelength. This serves as a useful diagnostic for locating data points with unusually small error bars than undesirably drive the best fit, and the user has the ability to remove these points through the use of a residual cutoff value. The model is then run again to generate a new best fit. If STIS G140M or E140M data is loaded, the user can compare the recovered profile to the STIS data. An additional window becomes available to scale the STIS data to the COS data for a visual inspection of the two profiles. The integrated attenuated fluxes are also reported as another metric of comparison.
\par
When the `Remove Airglow' button is pressed, the airglow component of the current best fit is subtracted from the COS data. A result plot and a diagnostic plot (see Figures \ref{fig:recover3} and \ref{fig:diagnostic3}, respectively) are then produced and can be saved by the user. If the user desires, a bootstrap of this best fit can then be run following the method described in Section \ref{sss:rrcbb}. This process is then repeated for all of the available airglow lines, and the final recovered spectrum can be saved as a CSV file. A second CSV file is automatically generated which contains all of the user inputs that led to the saved recovered spectrum, along with best fit parameters and their errors for the selected model.

\begin{figure*}[!ht] %the * means it spans both columns
\epsscale{1.15}
\plotone{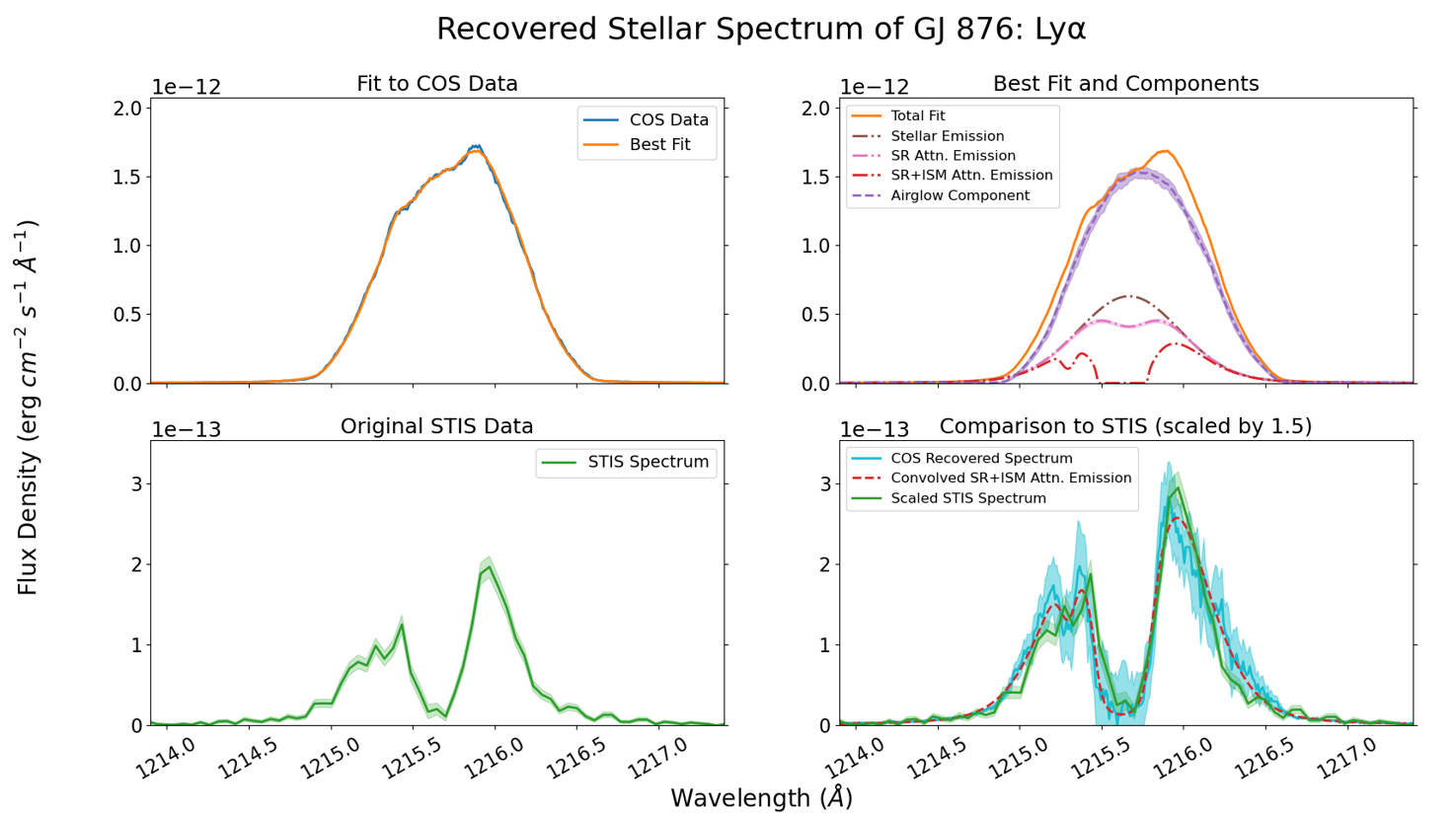}
\caption{The results of the Ly$\alpha$ airglow subtraction for GJ 876  produced by the airglow removal tool (see Section \ref{ss:GUI}). The upper left plot compares the overall fit to the COS data, the upper right plot breaks the fit down to its constituents (SR = self reversal), the lower left plot displays the STIS spectrum when available, and the lower right plot compares the recovered COS spectrum to the model's recovered spectrum, and to a scaled STIS spectrum when available. \vspace{0.25cm} \label{fig:recover1}} %6
\plotone{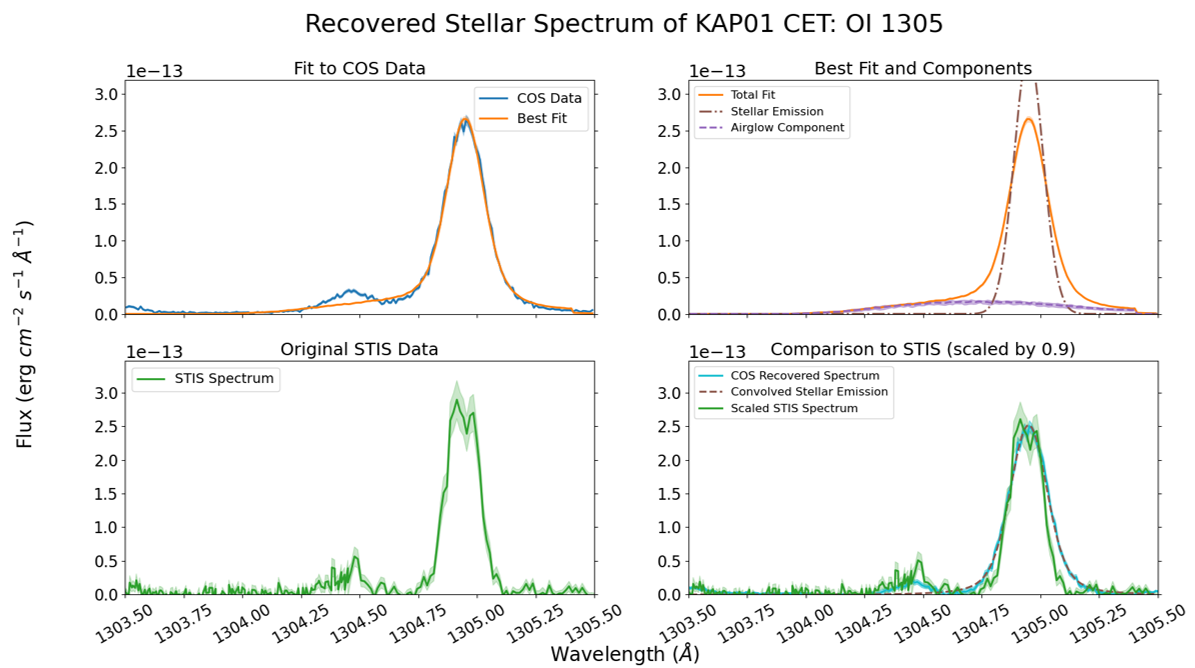}
\caption{The results of the O I 1305 airglow subtraction for $\kappa^1$ Ceti produced by the airglow removal GUI. The descriptions of the subplots are given in Figure \ref{fig:recover1}. The Si II 1304 line is present in both COS and STIS data, however it does not significantly affect fitting the O I 1305 line. \label{fig:recover2}} %7
\end{figure*} 
%putting both under the same figure{} groups them together on a page, and still gives each plot its own figure number

\subsubsection{Residual Resampling Circular Block Bootstrap}\label{sss:rrcbb} %3.3.1

The \texttt{LMFIT} fitting routine does not always report 1$\sigma$ errors for the fitted parameters. This can happen for a number of reasons, such as a parameter remaining stuck at an upper or lower bound, or if varying the parameter does not influence the fit, remaining stuck at its initial guess. Even when parameter errors are reported, these may not accurately represent the underlying parameter distributions. The authors of \texttt{LMFIT} recommend using the reported errors as initial estimates, and to conduct a more thorough exploration of the parameter space.\footnote{\url{https://lmfit.github.io/lmfit-py/index.html}} We perform this exploration through residual resampling circular block bootstrapping (RRCBB). The details of this technique are described in Appendix \ref{s:appendixA}. The user has the option of running the bootstrap in order to generate a 1$\sigma$ confidence interval of the \texttt{LMFIT} best fit parameters and the integrated stellar flux. We utilized this bootstrapping technique for all available emission lines for every star in our sample.

\subsection{Validation of the Method}\label{ss:validity} %3.4

Using the methodology described in Section \ref{ss:GUI}, we remove the airglow contribution from Ly$\alpha$ and the O I triplet, recovering the stellar profiles for the subset of stars where both COS and STIS data were available. The recovered COS profiles were then compared to the STIS spectra, where airglow is already removed by the CALSTIS pipeline. Examples of these recovered spectra are shown in Figures \ref{fig:recover1} and \ref{fig:recover2}. 

\subsubsection{STIS Comparison}\label{sss:STISCompare} %3.4.1
A version of our airglow subtraction model was created to fit the stellar, self reversal, and ISM components to STIS/G140M and STIS/E140M Ly$\alpha$ profiles. This allowed for a better comparison between the best fit parameters of both instrument datasets. Additional STIS LSFs were obtained from STScI for the G140M and E140M gratings for use with the modified model. 
\par
The modified Ly$\alpha$ reconstruction model was run on 21 G140M spectra and 7 E140M spectra, totaling to 28 stars. The results of these reconstructions were then compared to those of the recovered COS profiles. A STIS/E140M comparison is shown as an example in Table \ref{tab:E140M}. Of these 28 stars, there were 18 successful COS stellar recoveries. Of these 10 failed recoveries, four failed due to poorly fitted ISM parameters, while the remaining six failed due to an overall poor fit to the data (See Section \ref{ss:limits} for further explanation on what can cause a failed recovery). 
\par
We note that there can occasionally be large discrepancies between COS and STIS integrated fluxes. This can partially be attributed to variation over a stellar activity cycle. COS and STIS observations were not always taken at similar times; observations taken at different points in the stellar cycle can result in different observed fluxes. This is observed in the solar cycle, where the Ly$\alpha$ flux at solar maximum is $\sim$1.5 - 2.0$\times$ the flux at solar minimum \citep{Tom_Woods2020-pj,2003Natur.422..143V}. 
\par
We also observe discrepancies in literature values for the Ly$\alpha$ flux for a given star. One such example is Kapteyn's Star, with a reported literature value of $5.32 \times 10^{-13}$ erg cm$^{-2}$ s$^{-1}$ using G130M data in \cite{2016ApJ...821...81G}, while this is reported as $(2.88^{+0.16}_{-0.08}) \times 10^{-13}$ erg cm$^{-2}$ s$^{-1}$ in \cite{2022ApJ...926..129Y} using G140M data. We also note that our G130M value of $(5.67^{+0.46}_{-0.12}) \times 10^{-13}$ erg cm$^{-2}$ s$^{-1}$ is similar to that of \cite{2016ApJ...821...81G}, and our G140M value of $(2.83^{+0.36}_{-0.12}) \times 10^{-13}$ erg cm$^{-2}$ s$^{-1}$ agrees with that of \cite{2022ApJ...926..129Y}. This discrepancy may partially be due to the known slit loss of STIS observations due to sub-optimal target centering of narrow STIS slits (See \cite{2016ApJ...824..101Y,2016ApJ...824..102L}). Based primarily on the likely range of intrinsic stellar variability, and a smaller but non-negligible contribution from flux calibration variations, we assume in this work that Ly$\alpha$ fluxes within a factor of two are considered in sufficient agreement.

\vspace{0.4cm}

\begin{table}[ht]
\centering
\begin{adjustwidth}{-1.5cm}{}
\resizebox{4in}{!}{
\begin{tabular}{|c|c|c|c|}
\hline
\textbf{Parameter} & \textbf{COS} & \textbf{STIS} & \textbf{Literature} \\ \hhline{|=|=|=|=|}
Radial Velocity & \multirow{2}*{$21.933^{+3.55}_{-3.534}$} & \multirow{2}*{$12.836^{+2.571}_{-1.871}$} & \multirow{2}*{$13.164$} \\
(km/s) & & & \\ \hline
Gaussian FWHM & \multirow{2}*{$113.926^{+3.036}_{-3.301}$} & \multirow{2}*{$117.651^{+3.518}_{-3.624}$} & \multirow{2}*{$N/A$} \\ 
(km/s) & & & \\ \hline
Lorentz FWHM & \multirow{2}*{$14.489^{+1.504}_{-1.437}$} & \multirow{2}*{$18.12^{+2.163}_{-2.149}$} & \multirow{2}*{$N/A$} \\ 
(km/s) & & & \\ \hline
Flux Amplitude & \multirow{2}*{$-10.844^{+0.086}_{-0.094}$} & \multirow{2}*{$-11.067^{+0.099}_{-0.087}$} & \multirow{2}*{$N/A$} \\ 
(erg cm$^{-2}$ s$^{-1}$ \AA$^{-1}$) & & & \\ \hline
Column Density & \multirow{2}*{$18.056^{+0.065}_{-0.083}$} & \multirow{2}*{$18.103^{+0.057}_{-0.059}$} & \multirow{2}*{$18.2^{+0.03}_{-0.03}$} \\ 
($10^x$ cm$^{-2}$) & & & \\ \hline
ISM Velocity & \multirow{2}*{$-12.526^{+1.8}_{-1.196}$} & \multirow{2}*{$-15.531^{+0.782}_{-1.078}$} & \multirow{2}*{$-17.1^{+0.8}_{-0.7}$} \\ 
(km/s) & & & \\ \hline
Integrated Flux & \multirow{2}*{$1.13^{+0.12}_{-0.11} \times 10^{-12}$} & \multirow{2}*{$8.45^{+0.98}_{-0.88} \times 10^{-13}$} & \multirow{2}*{$9.5^{+0.6}_{-0.6} \times 10^{-13}$} \\ 
(erg cm$^{-2}$ s$^{-1}$) & & & \\ \hline
\end{tabular}}
\end{adjustwidth}
\caption{The best fit results of the COS/G130M and STIS/E140M Ly$\alpha$ profiles of GJ 832. Reported parameter errors were obtained using our bootstrapping method. See Table \ref{tab:literature} for literature references. \vspace{-0.2cm}}
\label{tab:E140M} %3
\end{table}

\subsubsection{Literature Comparison}\label{sss:litCompare} %3.4.2
In addition to comparisons between COS and STIS Ly$\alpha$ reconstructions using the same methods, we also compare these results with values reported in the literature. An example is shown in Table \ref{tab:E140M} which reports literature values for the stellar radial velocity, the ISM column density, the ISM velocity, and the Ly$\alpha$ integrated flux as observed from Earth. We primarily focus on literature agreement for the column density and the integrated flux, however all parameter values are reported for completeness. 
Again, we adopt that fluxes within a factor of 2 were comparable.
\par
\begin{figure*}[!ht]
\centering
\subfigure[Column Density]{\epsscale{0.5} \plotone{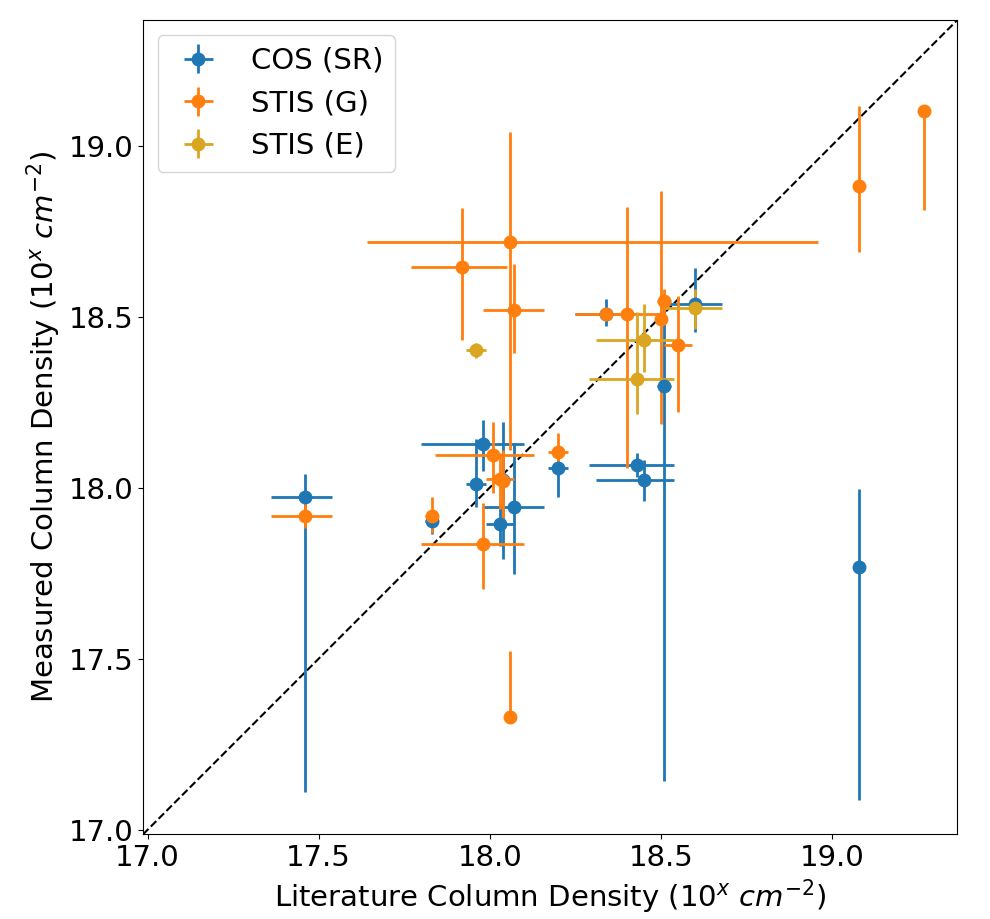} \label{sfig:litCompareN}} %8a
\subfigure[Integrated Flux]{\epsscale{0.5} \plotone{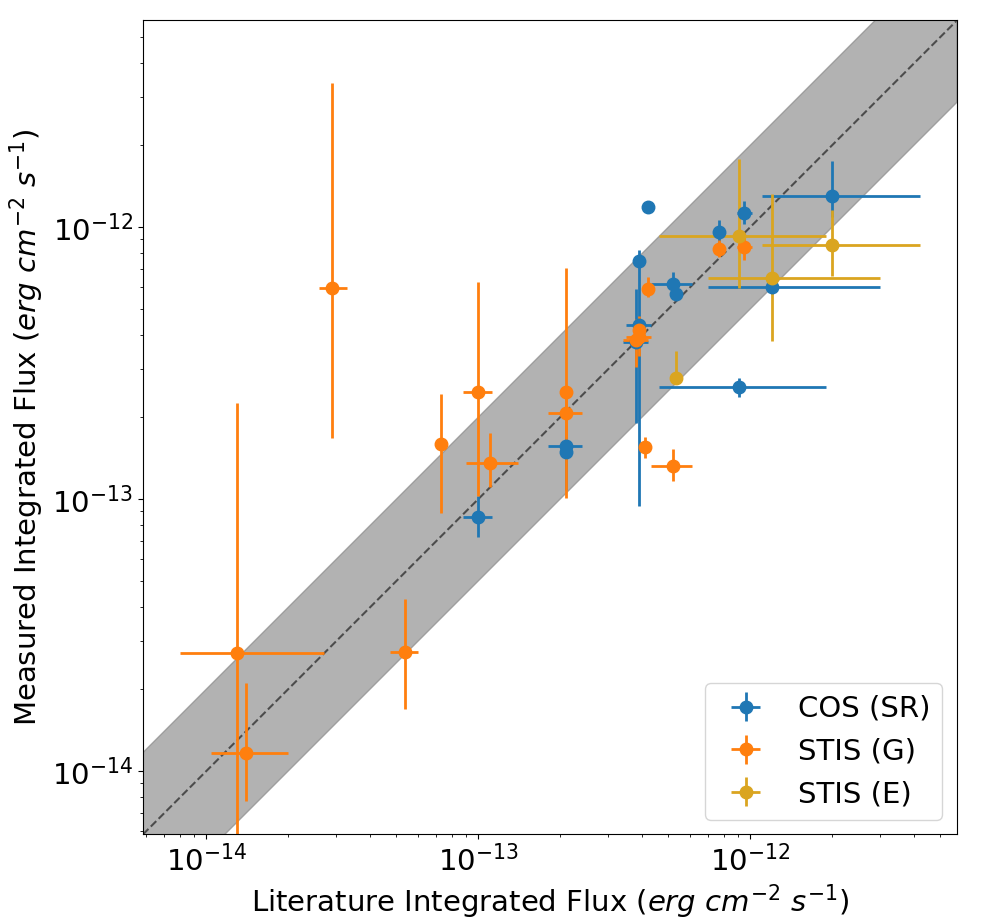} \label{sfig:litCompareF}} %8b
\caption{Comparisons between 18 recovered COS best fit parameters (blue points) and 28 reconstructed STIS best fit parameters (orange points for G140M spectra, and golden points for E140M spectra) to the literature values for (a) ISM column density and (b) integrated Ly$\alpha$ flux. Literature sources are found in Table \ref{tab:literature}. The dashed lines in both subplots is the 1:1 line. The shaded region in (b) represents a factor of 2 region about the 1:1 line.  \vspace{0.2cm} \label{fig:litCompare}} %8
\end{figure*}

Of the subset of 28 stars with COS and STIS Ly$\alpha$ observations, we were able to successfully recover the COS Ly$\alpha$ profiles of 18 stars. We find a generally acceptable agreement between the COS and STIS integrated fluxes of these stars when compared to values reported in the literature. We find less agreement for the column density; this is a consequence of fixing the Doppler b parameter to compensate for the degeneracy with the column density. Comparisons for the column density and integrated fluxes are shown in Figure \ref{fig:litCompare}. In \ref{sfig:litCompareF}, the integrated flux of GJ 1214 (leftmost G140M point) was consistent with a non-detection, represented by the error bar extending below the range of the log-log plot. The Ly$\alpha$ flux of GJ 1214 was not detected in \cite{2013ApJ...763..149F} and marginally detected in \cite{2016ApJ...820...89F}, however the reconstructed emission was poorly constrained.

\subsection{Limitations of the Method}\label{ss:limits} %3.5

\subsubsection{Si III Diagnostics}\label{sss:SNR} %3.5.1

\begin{figure}[!ht]
\epsscale{1.15}
\plotone{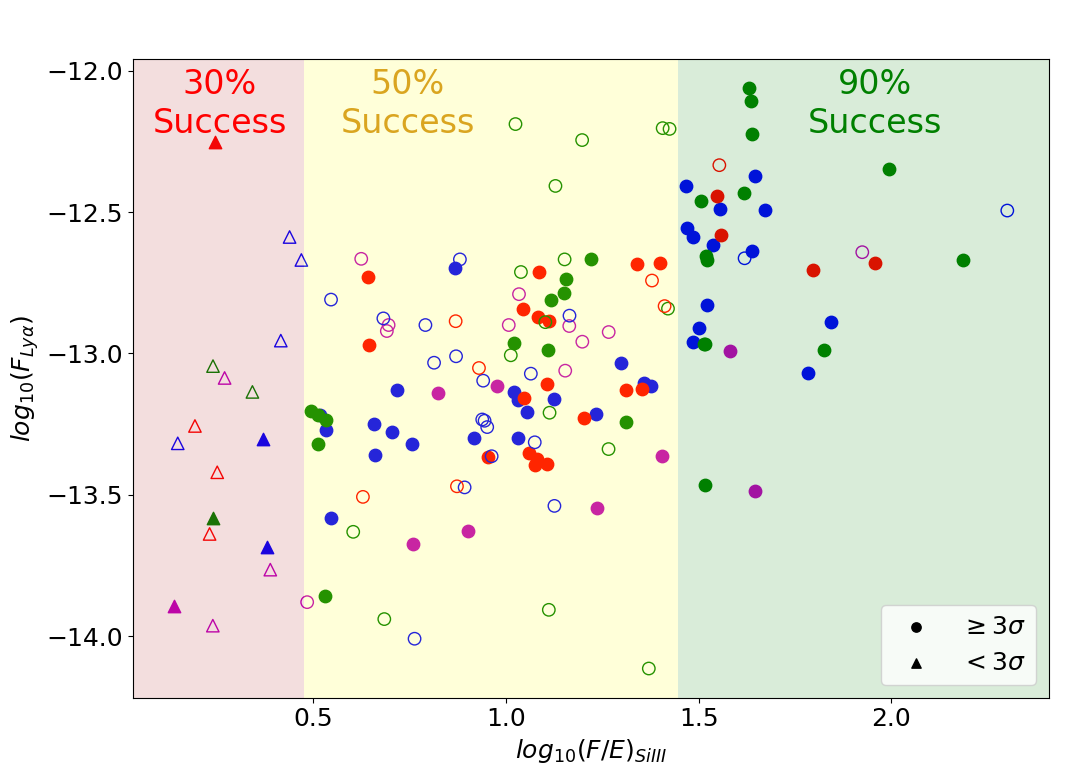}
\caption{Ly$\alpha$ recovered flux vs Si III SNR. Points on the plot are colored and filled in the same way as plots in Section \ref{s:correlations}. We identify three regimes of Si III SNR in which the stellar recovery tool has a 30\% success rate (SNR $<$ 3), a 50\% success rate (3 $\leq$ SNR $<$ 28), and a 90\% success rate (SNR $\geq$ 28) of recovering stellar Ly$\alpha$. \label{fig:Attn Cutoff}} %9
\end{figure}

While the stellar recovery tool is capable of extracting the underlying stellar spectrum from COS data, it does not always do so successfully. The Si III emission line's SNR (flux divided by flux error) is used as a metric for determining the likelihood of the stellar recovery tool producing a reliable recovered Ly$\alpha$ spectrum. Si III fluxes below a 3$\sigma$ detection, but with a measured flux statistically above zero, are plotted in Figure \ref{fig:Attn Cutoff} as triangles. We define three regions of varying success. The first region (Si III SNR $<$ 3) sees a 30\% success rate, the second (3 $\leq$ SNR $<$ 28) a 50\% success rate, and the third (SNR $\geq$ 28) a 90\% success rate of recovering stellar Ly$\alpha$ emission. 

We compare the recoverable Ly$\alpha$ stellar flux to the Si III SNR (Figure \ref{fig:Attn Cutoff}). We find the lowest recoverable flux in the 50\% success region to be $F_{Ly\alpha}=1.39\times10^{-14}$ erg cm$^{-2}$ s$^{-1}$. While the lowest successfully recovered flux observed in the entire sample was $F_{Ly\alpha}=1.28\times10^{-14}$ erg cm$^{-2}$ s$^{-1}$, this star lies within the 30\% success regime. For future observations with the COS instrument, we recommend a combination of an estimated Si III SNR $\geq$ 3 and  an estimated attenuated Ly$\alpha$ flux $\geq$ $1.39\times10^{-14}$ erg cm$^{-2}$ s$^{-1}$ for a 50\% minimum likelihood of recovering and reconstructing the underlying stellar emission from the contaminating airglow.

\subsubsection{Gain Sag and Future Observations}\label{sss:gainsag} %3.5.2

Gain sag will affect the majority of COS stellar spectra in the future, making it increasingly difficult to retrieve faint stellar spectra. Currently there are no airglow spectra available beyond LP3. With the implementation of LP4 in Cycle 25, the COS2025 policy was introduced with the aim of extend the science capabilities of the COS detector to 2025 and beyond by minimizing the ever increasing gain sag caused by geocoronal Ly$\alpha$ \citep{2018AAS...23115038R}. 
\par
Cenwave 1222, implemented in Cycle 19 \citep{2014cos..rept....1R}, places geocoronal (and stellar) Ly$\alpha$ within the gap between detector plates. This Cenwave prevents further gain sag while still capturing other important nearby FUV emission lines such as Si III and N V. Due to the COS2025 policy discouraging unnecessary airglow exposure on the detector face, pure airglow spectra taken at LPs 4, 5, or 6 appear to be unlikely, as these, along with LPs 2 and 3, are currently being utilized \citep{COS_IHB_2022}. 
\par
It may be beneficial to implement a new Cenwave purely for Ly$\alpha$ studies with COS. Such a Cenwave should place Ly$\alpha$ in a location on the detector which is generally free from other major FUV features amongst the various COS settings. The Cenwave could also be used exclusively at LP1, which is otherwise unused. The use of a dedicated Cenwave could extend the Ly$\alpha$ science output of COS without compromising its various other science capabilities.

\section{Recovered Emission Profiles and Stellar Activity Indicators}\label{s:correlations} %4

Of the initial 165 Ly$\alpha$ profiles and 152 O I triplet profiles where a recovery was possible, the stellar recovery tool successfully recovered 93 Ly$\alpha$ profiles, 91 O I 1302 profiles, 112 O I 1305 profiles, and 118 O I 1306 profiles. Section \ref{ss:failure} provides a discussion on why a stellar spectrum may not be successfully recovered/reconstructed. Figure \ref{fig:bars} breaks down the number of successful and failed recoveries for each emission line by spectral type.

The best fit self reversed Ly$\alpha$ and stellar O I profiles were used to calculate integrated line fluxes and line luminosities. The integrated fluxes and luminosities were compared to various stellar activity indicators ($R'_{HK}$, stellar rotation period, stellar effective temperature, Si III ($\lambda = 1206.51$ \AA) luminosity, and N V ($\lambda = 1238.82$ and $1242.80$ \AA) luminosity) to produce the largest direct study of Ly$\alpha$ emission on nearby stars. 

\subsection{Failed Recoveries}\label{ss:failure} %4.1

We identify 72 unsuccessful Ly$\alpha$ profile recoveries. These failures can be divided into two categories. Category A failures are those in which the observed profile fits in either the Definite Stellar, Strong or Weak morphological categories (see Section \ref{ss:contamination}), and the GUI failed to recover the bright stellar emission. Nearly every Category A failure was due to the effects of gain sag being too severe even for bright targets to be properly recovered. Other reasons for Category A failures were that the throughput decreasing effect of detector gridwire shadows were too strong, or that the model was unable to find a solution within the allowed parameter space. Category B failures are those whose observed profile falls under the Questionable Stellar, Strong or Weak categories, and the GUI was unable to properly detect, recover, or reconstruct the dim stellar signal from the dominating airglow contamination. The effects of gain sag are also more severe for fainter targets, resulting in nearly every failed recovery in this category. There were 18 Category A failures and 54 Category B failures. 
\par
A failed recovery is identified if it fails one or more criteria. For the Ly$\alpha$ profiles, the first pass of flagging failures is by their ISM parameters. If a star's best fit column density or ISM velocity hits an upper or lower limit listed in Table \ref{tab:bounds}, it is flagged as a failed recovery. By this criterion, we flag 61 recoveries as failures. Most often, the upper limit of the column density ($10^{19.1}$ cm$^{-2}$) was hit. There are column densities reported in the literature slightly larger than our upper limit, so it is possible that an increase in our upper limit would flag these stars as successes, although many of these instances are also associated with gain sag which causes additional fit complications (see Section \ref{sss:gainsag}). 
\par
Our second pass was based on whether the model was able to visually fit the data or not, and that the parameter errors were not unreasonably large. The definition of a poor visual fit is admittedly subjective, but generally these cases are flagged when the model clearly does not track with the shape of the data. Unreasonably large parameter errors are those such that the reported errors are several orders of magnitude larger than the best fit parameter. We identify 20 failures through this criterion, however we do note that only two recoveries were flagged as a failure for this reason alone, and their inclusion would not significantly change our results. 
\par
Our final pass was based on the ratio between the Lorentzian and Gaussian FWHMs of the Voigt profile. We found it unfeasible to customize fit components and parameter bounds for each star due to the size of our sample, and decided to apply the same model to every star. As a consequence, a small number of stellar Ly$\alpha$ profiles were not well fit by a Voigt profile. If the $(FWHM_L/FWHM_G)$ ratio is very small, the flux amplitude is driven to be larger, which is compensated by an increase in the column density. This effect can lead to reaching the upper limit of the column density, which we observe for $\sim$66\% of failed recoveries with small Lorentzian FWHMs. It is an indicator that the underlying Voigt profile is not performing well in modeling the data, and is flagged as a failed recovery. We characterize the ratio as $Y=\sqrt{ln(2)} \times (FWHM_L/FWHM_G)$, and for $Y \leq 0.01$ ($FWHM_L \lesssim 2$ km s$^{-1}$) we flag the recovery as a failure. We identified 38 fits where $Y \leq 0.01$, flagging them as failed recoveries. Of these, only 7 were flagged as a failed recovery for this reason alone.  
\par
Of the 72 failed Ly$\alpha$ recoveries, one was observed with LP1, one was observed with LP2, 17 were observed with LP3, and 53 were observed with LP4. The LP1 star, HD 103095, was flagged as a Category A failure due to hitting the upper ISM velocity limit. The bootstrap samples for this star also fluctuate between two possible solutions, one of which visibly did not fit the data. The LP2 star, KIC 11560431 is a spectroscopically resolved binary star system, which the GUI was not designed to handle. It is marked as a Category B failure, as both Ly$\alpha$ profiles were dim compared to the airglow. The reasons for the LP3 and LP4 failures are primarily driven by detector gridwire shadows and gain sag, respectively.
\par
Of the 17 failed LP3 Ly$\alpha$ recoveries, 14 of these were part of the SNAP 14633 observing program, in which we ignored the detector gridwire shadow DQ flag to prevent the \texttt{COADD\_X1D} code from deweighting FP3 exposures and significantly altering the shape of the coadded spectrum. We observe both Category A and B failures in the SNAP stars, indicating that the gridwire shadow is affecting the GUI's ability to recover even bright stellar emission. This throughput diminishing effect was more prevalent than we anticipated, and these flags should be ignored with caution in the future. Detector gridwire shadow flags were also found and ignored for two non-SNAP stars, GJ 581 and 2MASSJ02543316-5108313, which both resulted in Category B failures. The final LP3 failure was GJ 1214, another Category B failure. The stellar spectrum of this star was dim enough that \texttt{stistools} was unable to locate the spectrum. A manual extraction was also impossible, as the stellar spectrum location in the 2D detector image was unclear. The Si III fluxes of GJ 1214 also indicates that the Ly$\alpha$ flux is either very dim or nondetectable, consistent with previous work \citep[e.g.,][]{2013ApJ...763..149F,2016ApJ...820...89F}.
\par
Of the 53 failed LP4 Ly$\alpha$ recoveries, 49 were targets flagged with gain sag. We determined that the effects of gain sag were most significant in FP3 data, and our coadditions do not include these spectra. The remaining data can still display significant gain sag effects, and this is the primary cause for failure of LP4 data. This includes both Category A and Category B failures, however we note that stars with bright stellar emission can result in successful recoveries despite the present gain sag. The secondary reason for failure is due to the stellar fluxes likely being too dim or nondetectable, based on their observed Si III fluxes. Often, both effects are seen together in the available LP4 data. 
\par
For a handful of targets, the stellar signal is low due to the large distances ($d_* \gtrsim 50$ pc) to the targets. There is a subset of failed LP4 recoveries which did not suffer from gain sag, and are $\gtrsim 80$ pc away. These were all F type stars, with no reported $R'_{HK}$ or $P_{rot,*}$ in the literature. We note that it may be possible for the H I column density to exceed our upper limit of $10^{19.1}$ cm$^{-2}$ along certain lines of sight at these distances, however in general we do not expect a column density larger than this limit. While raising this upper limit could result in a successful recovery, the dimmer stellar emission at these distances combined with stronger ISM attenuation works against a successful recovery. Because it is unclear whether distance or an unusually large column density results in these failures, the upper limit is kept at $10^{19.1}$ cm$^{-2}$ based on typical values reported in the literature.
\par
There were 61 failed O I 1302 recoveries, 40 failed O I 1305 recoveries, and 34 failed O I 1306 recoveries. The primary reason for these failures was that the stellar emission is too dim to be recovered, and is lost in the noise after the airglow subtraction. Occasionally, the GUI also fails by reporting a very small Gaussian FWHM, resulting in a narrow profile which, when convolved with the COS LSF, can appear to fit the COS data. For O I 1302, ISM attenuation is possible. We did not include this in our models as it is generally not the case that the ISM velocity and the stellar velocity coincide. It is possible that the larger number of failed O I 1302 profiles due to dim stellar emission is related to the ISM obscuring what little stellar signal is present.

\subsection{\texorpdfstring{$R'_{HK}$}{R'\_{HK}} and \texorpdfstring{Ly$\alpha$}{Lyalpha} Luminosity}\label{ss:R'HK} %4.2

\begin{figure}[!ht]
\epsscale{1.09}
\plotone{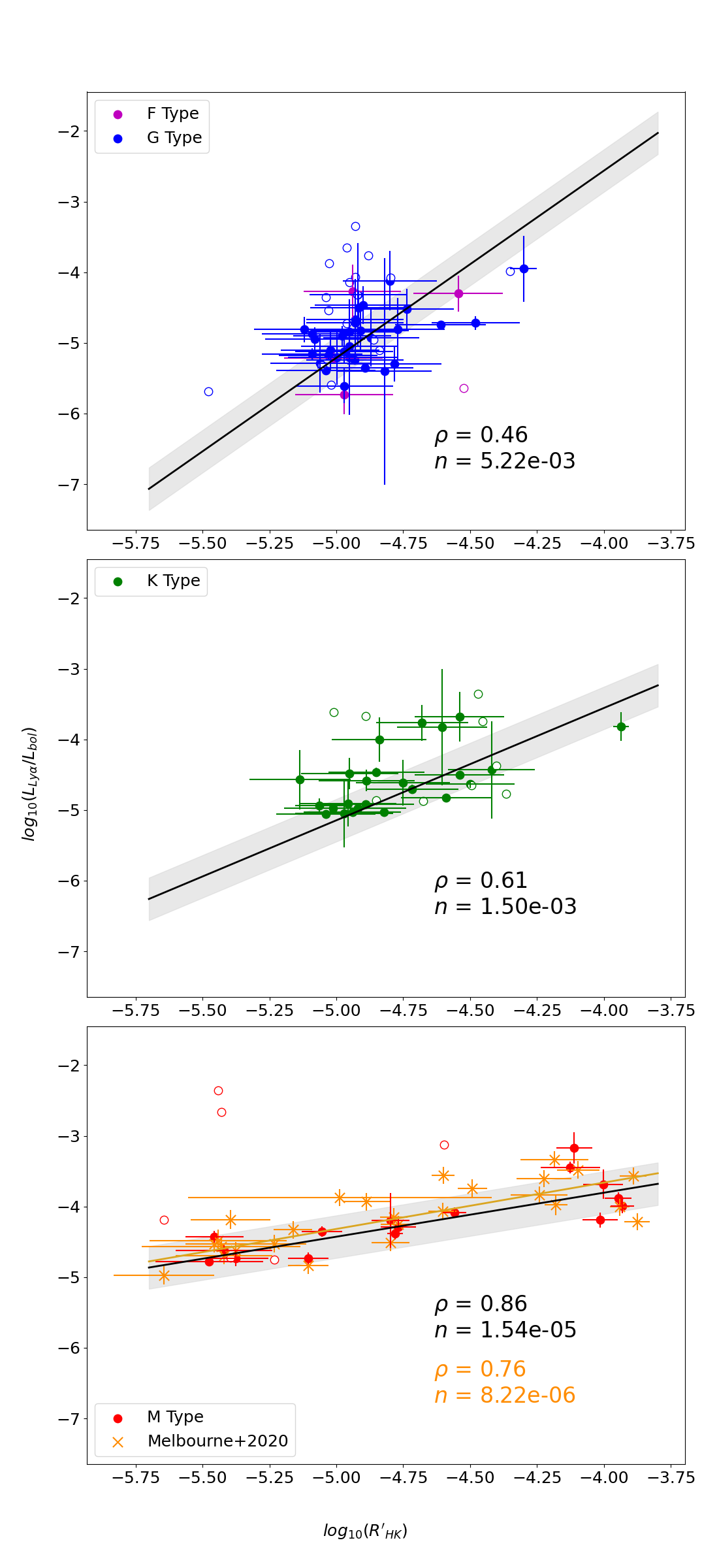}
\caption{Reconstructed Ly$\alpha$ luminosity normalized by stellar bolometric luminosity vs $R'_{HK}$ index. Stars are colored by spectral type. A filled marker represents a successful recovery, while an empty marker represents a failed recovery. Subplots are divided by spectral type. F-type stars are grouped with G-type stars. Best fit lines are shown as a solid line, only considering successful recoveries. The grey region about the best fit line is the factor of 2 region (See Section \ref{sss:STISCompare}). This convention remain consistent for the remainder of this work. A comparison is made to M dwarfs studied in \cite{2020arXiv200907869M}. A best fit line and Spearman statistics are calculated for comparison. \label{fig:RPHK}} %10
\end{figure}

The $R'_{HK}$ index is a chromospheric activity index which measures the strength of the narrow Ca II H ($\lambda = 3968.47$ \AA) and K ($\lambda = 3933.66$  \AA) resonance emission lines without contamination from the surrounding broad photospheric absorption lines, allowing for comparison between spectral types \citep{1984ApJ...279..763N}. These chromospheric emission lines have a formation temperature of $\sim$6,000K \citep{1981ApJS...45..635V}. $R'_{HK}$ is known to correlate well with various chromospheric and transition region lines such as Ly$\alpha$ \citep{1987AJ.....93..920S,2017ApJ...843...31Y,2020arXiv200907869M}. We compared literature $R'_{HK}$ values to the reconstructed Ly$\alpha$ luminosity normalized by the bolometric luminosity of the star (Figure \ref{fig:RPHK}). The bolomeric luminosity is calculated as $L_{bol}=4\pi R_*^2\sigma_{SB}T_{eff,*}^4$, where literature values were obtained for the stellar radius ($R_*$) and the stellar effective temperature ($T_{eff,*}$). For the remainder of this work, we primarily make comparisons with normalized luminosity ratios, as they allow for a better comparison of stellar activity between spectral types \citep{2018ApJS..239...16F,2020arXiv200907869M}.

We fit power laws to the data using \texttt{SciPy}'s Orthogonal Distance Regression (ODR) method \citep{2020SciPy-NMeth}, which considers the data uncertainties in both variables when determining the best fit slope and intercept in log-log space. We calculate separate power law fits to each spectral type, as their best fit parameters differed significantly. F-type stars are grouped with G-type stars, as there were not enough successfully recovered F-types to justify a separate fit.
\par
For each line of best fit in Figure \ref{fig:RPHK}, we report the Spearman rank coefficient ($\rho$), a measure of the monotonic relation between the X data and the Y data. A $\rho$ that approaches $\pm1$ means that the quantities approach a perfect positive or negative monotonic correlation. Below the correlation coefficient is the value $n$, the probability that the X and Y data do not have a monotonic correlation and the data happens to exhibit a correlation that is at least as strong as the reported $\rho$.
\par
Throughout the remainder of this work, several Spearman correlation coefficients will be shown. We assign correlation strengths as follows: very weak correlations ($0.00 < |\rho| \leq 0.19$), weak correlations ($0.20 \leq |\rho| \leq 0.39$), moderate correlations ($0.40 \leq |\rho| \leq 0.59$), strong correlations ($0.60 \leq |\rho| \leq 0.79$), and very strong correlations ($0.80 \leq |\rho| < 1.00$).

We compare the M dwarfs between our sample and that of \cite{2020arXiv200907869M} to check for consistency with previous studies. This is shown in the bottom subplot of Figure \ref{fig:RPHK}. We omit the outlier LP247-13 from the literature dataset, and recalculate the fit using the ODR method. We find good agreement between the two M dwarf datasets. A summary of best fit parameters and Spearman statistics are given in Table \ref{tab:RPHK}. 
\par
We find a very strong correlation between the normalized Ly$\alpha$ luminosity ratio and the $R'_{HK}$ Index for M dwarfs, a strong correlation is observed for the K dwarfs, and a moderate correlation is observed for FG dwarfs. While we observe a strong correlation for M dwarfs, we also note that the normalized Ly$\alpha$ luminosity ratio increases slowly over a wide range of $R'_{HK}$ values. While $R'_{HK}$ is a very strong predictor of Ly$\alpha$ emission of M dwarfs, the shallow slope suggests that the formation location of these emission features within the stellar atmospheres of M dwarfs are not strongly coupled.
\par
While we observe a weaker correlation for K dwarfs, we also find a steeper slope, indicating that Ca II and H I emission in the atmospheres of K dwarfs are more strongly coupled. The FG dwarfs in our stellar sample do not span as wide a range of $R'_{HK}$ values as K and M dwarfs, while also spanning a similar range of normalized Ly$\alpha$ luminosity. This grouping of similar stars in this parameter space makes physical interpretations difficult. The observed correlation of the FG dwarfs is set by the four stars with the largest $R'_{HK}$ values in the upper subplot of Figure \ref{fig:RPHK} (from right to left, HD 62850, HD 25825, HD 36767, and HD 150706), and removing these results in a weak Spearman correlation coefficient of 0.29, with a relatively large 1.04$\times10^{-1}$ probability of no correlation. The clustering of F, G, and K dwarfs is discussed in more detail in Section \ref{ss:FGKtypes}.

\vspace{0.3cm}

\begin{table}[!ht]
 \begin{adjustwidth}{-2.0cm}{}
\centering
\resizebox{4in}{!}{
\begin{tabular}{|c|c|c|c|c|}
\hline
\textbf{Fit Type} & \boldmath{$\alpha$} & \boldmath{$\beta$} & \boldmath{$\rho$} & \textbf{n}\\ 
\hline
FG-Types & 2.65$\pm$0.46 & 8.04$\pm$2.27 & 0.46 & 5.22e-3 \\ \hline
K-Types & 1.59$\pm$0.27 & 2.81$\pm$1.28 & 0.61 & 1.50e-3 \\ \hline
M-types & 0.62$\pm$0.11 & -1.31$\pm$0.50 & 0.86 & 1.54e-5 \\ \hline
Melbourne et al. 2020 & 0.66$\pm$0.11 & -1.04$\pm$0.50 & 0.76 & 8.22e-6 \\ \hline
\end{tabular}}
\end{adjustwidth}
\caption{Best fit parameters for the equation $log_{10} (L_{Ly\alpha}/L_{bol})=\alpha$ $\times$ $ log_{10}(R'_{HK})$ + $\beta$, along with Spearman rank coefficients and probabilities of no correlation. The \cite{2020arXiv200907869M} parameters apply only to M dwarfs.  \vspace{-0.75cm}}
\label{tab:RPHK} %4
\end{table}

\subsection{Stellar Rotation Period and \texorpdfstring{Ly$\alpha$}{Lyalpha} Luminosity}\label{ss:PRot} %4.3

\begin{table*}[!ht]
 \begin{adjustwidth}{-2.4cm}{}
\centering
\resizebox{5in}{!}{
\begin{tabular}{|c|c|c|c|c|c|}
\hline
\textbf{Fit Type} & \boldmath{$\alpha$} & \boldmath{$\beta$} & \boldmath{$log_{10}(P_{rot,sat})$} & \boldmath{$\rho$} & \textbf{n}\\ 
\hline
FGK-types (L) & -0.66$\pm$0.20 & -4.01$\pm$0.29 & - & -0.21 & 1.56e-1 \\ \hline
M-types (LS) & -0.67$\pm$0.04 & -3.37$\pm$0.07 & 0.63$\pm$0.08 & - & - \\ \hline
\end{tabular}}
\end{adjustwidth}
\caption{Best Fit parameters for the equation $log_{10} (L_{Ly\alpha}/L_{bol})=\alpha$ $\times$ $ log_{10}(P_{rot})$ + $\beta$ for (L). A piecewise function is used for (LS), where the two regions are separated by the saturation period $P_{rot,sat}$. For $log_{10}(P_{rot}) > log_{10}(P_{rot,sat})$ the same equation as (L) is used, and for $log_{10}(P_{rot}) \leq log_{10}(P_{rot,sat})$ the equation becomes $log_{10} (L_{Ly\alpha}/L_{bol})=\alpha$ $\times$ $ log_{10}(P_{rot,sat})$ + $\beta$. The Spearman rank coefficient and probability of no correlation is reported for the linear fit. \vspace{-0.3cm}}
\label{tab:Prot} %5
\end{table*}

\begin{figure}[!ht]
\epsscale{1.15}
\plotone{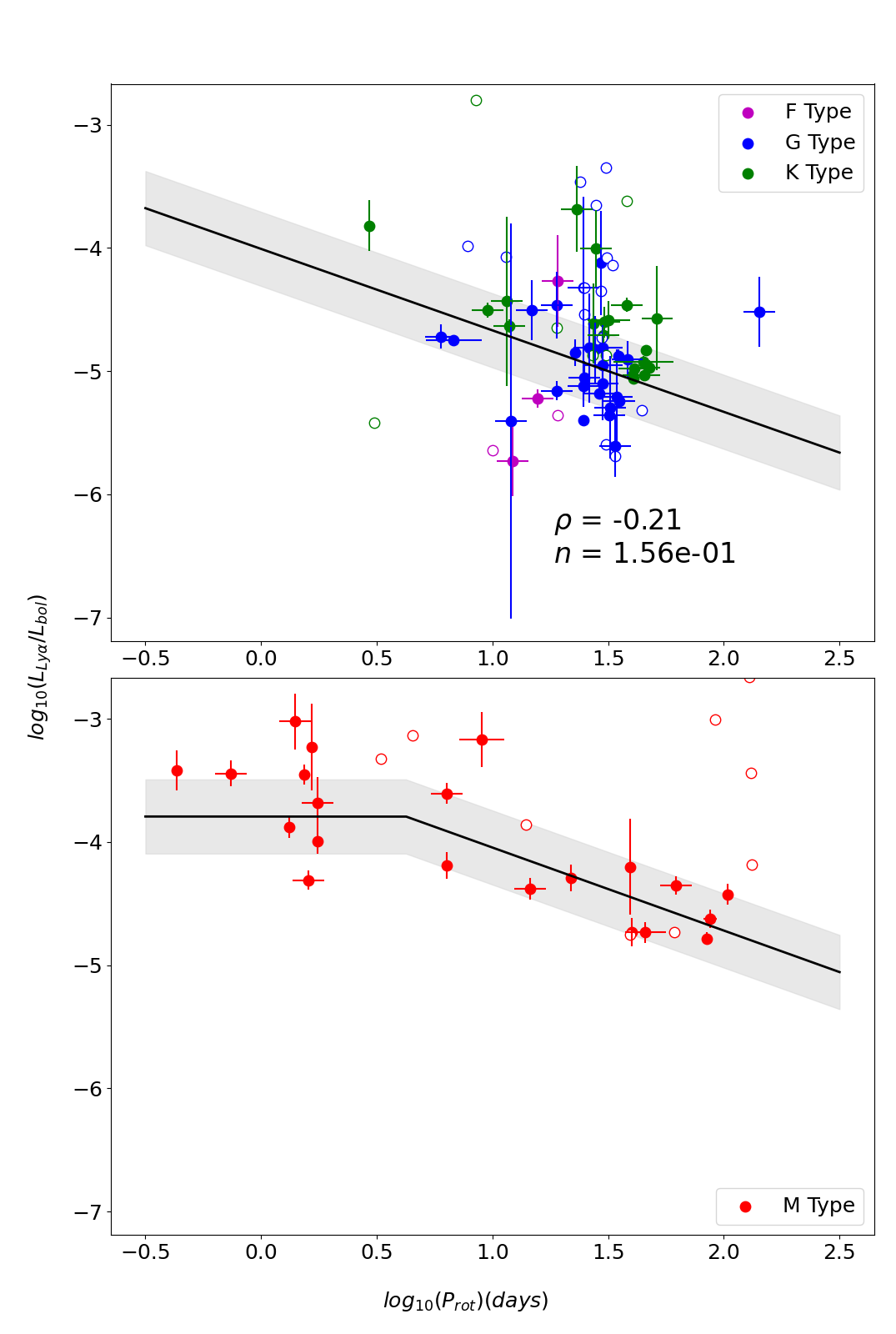}
\caption{Reconstructed Ly$\alpha$ luminosity normalized by stellar bolometric luminosity vs stellar rotation period. Saturation was only seen for M dwarfs in our sample. The Spearman statistics are relevant only to FGK dwarfs. \label{fig:Prot}} %11
\end{figure}

The stellar rotation period ($P_{rot,*}$) is known to strongly correlate with chromospheric activity \citep{1984ApJ...279..763N}. $P_{rot,*}$ is related to differential rotation within the star, which drives magnetic heating from the stellar dynamo, ultimately leading to stellar emission in the chromosphere, the transition region, and the corona \citep{1984Natur.310...22D,2011ApJ...743...48W,2013ApJ...766...69L}. Figure \ref{fig:Prot} shows the relations between normalized Ly$\alpha$ luminosity and $P_{rot,*}$ for M dwarfs and FGK dwarfs separately. We do not separate the K dwarfs as in Section \ref{ss:R'HK} as the separate fits do not significantly differ from each other.
\par
As stellar rotation period decreases, chromospheric activity indicators (e.g. Ly$\alpha$, X-ray emission) of FGKM dwarf stars reach a saturation point \citep{1993ApJ...410..387F,2003AnA...397..147P}. The normalized luminosity in the saturated regime becomes independent of spectral type \citep[e.g.,][]{2003AnA...397..147P,2011ApJ...743...48W,2017ApJ...834...85N,2018ApJS..239...16F}. While the luminosity ratio in the saturated regime is the same for different spectral types, departure from the saturated regime occurs at different $P_{rot,*}$ \citep{2003AnA...397..147P,2020JASTP.21105456K}. As stellar mass increases, the critical rotation period which separates the saturated and unsaturated regimes ($P_{rot,sat}$) decreases \citep{2017AstL...43..202N,2020JASTP.21105456K}. 
\par
There is not enough stellar diversity of FGK dwarfs in our sample (that had reported $P_{rot,*}$ in the literature) to properly determine when these spectral types reach the saturation point, so a power law fit was used. For the M dwarfs, a saturated fit was possible and is shown below the FGK dwarfs. The saturated fit was calculated using the Differential Evolution (DE) algorithm available through \texttt{SciPy} \citep{2020SciPy-NMeth}. 

We summarize the saturated and linear fits in Table \ref{tab:Prot}. We find a weak correlation (-0.21) between the normalized Ly$\alpha$ luminosity ratio and stellar rotation period for our FGK dwarfs. This is due to most of these stars occupying a narrow range of rotation periods in log space (1.25 $\lesssim log_{10}(P_{rot,*}) \lesssim$ 1.75), unlike the M dwarfs which span a larger range. Further discussion on this is provided in Section \ref{ss:FGKtypes}.
\par
The slopes ($\alpha$) in the unsaturated regime of the FGK and M dwarf fits are found to be -0.66$\pm$0.20 and -0.67, respectively, indicating that the rate at which activity from the stellar dynamo decreases is similar for FGKM dwarfs. The offset between the two populations in the unsaturated regime (the difference in $\beta$) indicates that the higher mass FGK stars depart from the saturated regime at shorter rotation periods than M dwarfs, as we expect. We see this departure for the M dwarfs at $P_{rot,sat}$=4.27$\pm$0.79 days. We made additional DE saturated fits to M dwarfs against Si III and N V emission, and see similar departures at 5.89$\pm$0.27 and 8.32$\pm$0.57 days, respectively. With additional observations of M dwarfs with rotation periods between $\sim$2 - 20 days, a more precise $P_{rot,sat}$ could be measured for Ly$\alpha$, Si III, and N V saturation.
\par
In \cite{2021ApJ...907...91L}, the value of $P_{rot,sat}$ is simultaneously fit for several FUV surface fluxes of M dwarfs, and was found to be $\sim$7.01 days. We convert the reconstructed Ly$\alpha$, Si III, and N V fluxes of the M dwarfs in our sample to surface fluxes, and find saturation periods of 3.98$\pm$0.73, 6.92$\pm$0.27, and 8.91$\pm$0.62 days, respectively. For the FGK dwarfs, we also convert the reconstructed Ly$\alpha$ fluxes to surface fluxes and compare these with the study of Ly$\alpha$ fluxes recovered from echelle-mode STIS data from \cite{2005ApJS..159..118W}. For the region of overlap in rotation period between the two data sets, we find good agreement between our best fit results.

\subsection{Effective Temperature and \texorpdfstring{Ly$\alpha$}{Lyalpha} Luminosity}\label{ss:Teff} %4.4

\begin{figure}[!ht]
\epsscale{1.15}
\plotone{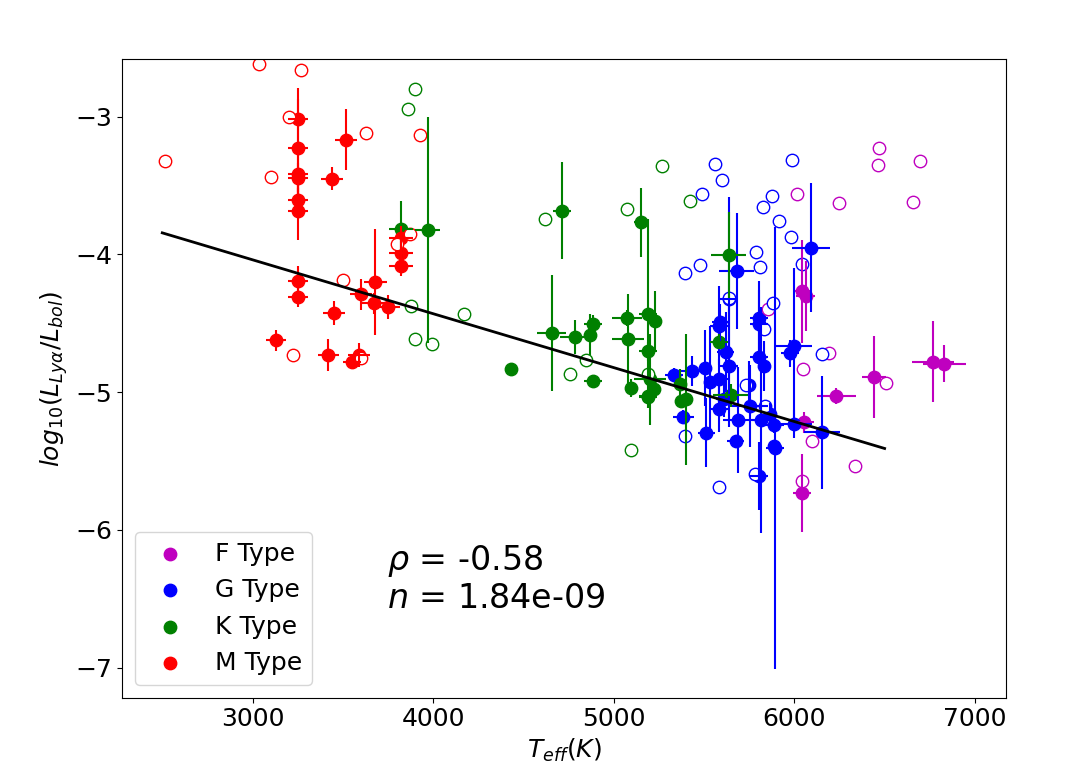}
\caption{Reconstructed Ly$\alpha$ luminosity normalized by stellar bolometric luminosity vs stellar effective temperature. \vspace{-0.5cm} \label{fig:Teff}} %12
\end{figure}

It has been shown that stellar effective temperature ($T_{eff,*}$) can be used to roughly predict the Ly$\alpha$ flux of a star \citep{2013ApJ...766...69L}. We expect that normalized Ly$\alpha$ luminosity will also display a correlation with $T_{eff,*}$, and is shown in Figure \ref{fig:Teff}. All spectral types are considered when calculating this power law fit. With a Spearman rank coefficient of -0.58 and a probability $1.85\times10^{-9}$ of no correlation, the two data sets exhibit a moderate correlation. The large spread of normalized Ly$\alpha$ luminosity at a given $T_{eff,*}$ observed in Figure \ref{fig:Teff} is due to variations in stellar activity at a given $T_{eff,*}$ \citep{2013ApJ...766...69L}. For the equation $log_{10} (L_{Ly\alpha}/L_{bol})=\alpha$ $\times$ $ T_{eff,*}$ + $\beta$, we find that $\alpha$=(-3.91$\pm$0.38)$\times10^{-4}$ and $\beta$=-2.87$\pm$0.19 are the best fit parameters. 

\subsection{Line Luminosities} \label{ss:fluxes} %4.5

Chromospheric and transition region emission lines such as Ly$\alpha$, Si III, N V, and the O I triplet, are expected to strongly correlate with each other \citep[e.g.,][]{1993ApJ...408..305L,2013ApJ...766...69L,2017ApJ...843...31Y,2020arXiv200907869M}. These emission features all arise from dynamo driven magnetic heating (See Section \ref{ss:PRot}). Si III and N V are transition region emission lines with formation temperatures of $\sim$50,000 K and $\sim$160,000 K, respectively \citep{1997AnAS..125..149D,2012ApJ...744...99L}. Integrated fluxes of Si III and N V were calculated from the same COS coadded spectra developed in Section \ref{ss:stellar obs}. The fluxes of the individual N V emission lines were summed together, and all N V fluxes and luminosities hereafter represent combined emission unless otherwise stated. For this work, we only consider 3$\sigma$ detections of Si III and N V fluxes for comparison with Ly$\alpha$ and O I, unless otherwise stated. The nearby Si II ($\lambda=$ 1304 \AA) did not interfere when calculating O I 1305 integrated fluxes (See Figure \ref{fig:recover2} for their spectral locations).

\subsubsection{Si III and \texorpdfstring{Ly$\alpha$}{Lyalpha} Luminosities}\label{sss:silicon} %4.5.1

\begin{figure}[!ht]
\epsscale{1.15}
\plotone{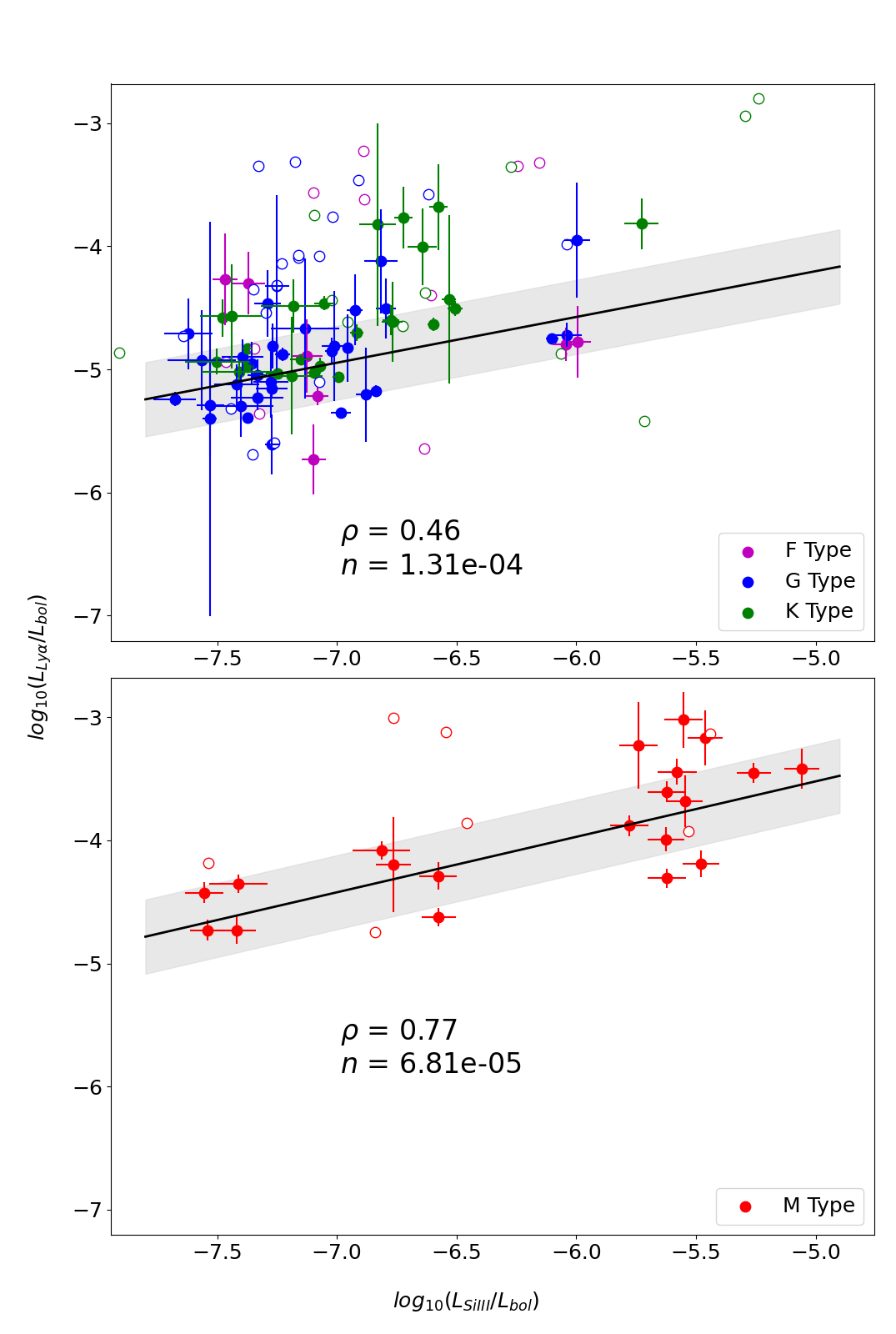}
\caption{Reconstructed Ly$\alpha$ luminosity vs Si III luminosity, both normalized by stellar bolometric luminosity. M-types are separated from FGK-types, and a comparison is made to M-types studied in \cite{2017ApJ...843...31Y}. \label{fig:SiIII_L}} %13
\end{figure}

Figure \ref{fig:SiIII_L} compares Si III normalized luminosity and the reconstructed Ly$\alpha$ normalized luminosity. We separate the M dwarfs from FGK dwarfs as the y-intercept differs. This correlation is found to be very strong in \cite{2017ApJ...843...31Y} for M dwarf surface fluxes of Si III and reconstructed Ly$\alpha$, and we expect a similar correlation for our FGK and M dwarfs. The power law parameters and Spearman statistics are reported in Table \ref{tab:SiIII}. We find a strong correlation for M dwarfs, and a moderate correlation for FGK dwarfs.
\par
There is a noticeable gap in our sample ranging from $-6.5 \lesssim log_{10}(L_{SiIII}/L_{bol}) \lesssim -5.8$, with few stars within it. This gap in parameter space is attributed to an observational bias, discussed in further detail in Section \ref{ss:FGKtypes}. The stars above the gap ($\gtrsim -5.8$) are the same stars that appear in the saturated regime of Figure \ref{fig:Prot}, while stars below the gap ($\lesssim -6.5$) appear in the unsaturated regime. Five stars are located within this gap, three G dwarfs (HD 62850, HD 150706, and HD 25825) and two F dwarfs (HD 24636 and HD 39755). The two F dwarfs and HD 62850 did not have reported rotation periods in the literature. The remaining two G dwarfs are the two fastest rotating G dwarfs in Figure \ref{fig:Prot}, with $\sim$6 day rotation periods. While we do not have enough stellar diversity of rapidly rotating F and G dwarfs to definitively say that these stars are in the saturated regime or not, Figure \ref{fig:SiIII_L} suggests that these F and G dwarfs are near the saturation limit for their spectral types.
\par
The slopes of the FGK and M fits are consistent with each other, however their y intercepts differ at the 1$\sigma$ level, with the M dwarfs located slightly above the FGK dwarfs. This is similar to the observed difference in intercepts in Figure \ref{fig:Prot}. It is possible that the M dwarfs are exhibiting Ly$\alpha$ saturation with Si III, similar to saturation with $P_{rot,*}$, and the difference in intercept indicates that the spectral types depart saturation at different $log_{10}(L_{SiIII}/L_{bol})$. Additional observations of stars within $-6.5 \lesssim log_{10}(L_{SiIII}/L_{bol}) \lesssim -5.8$ are needed to better interpret this result.

\vspace{0.25cm}

\begin{table}[!ht]
 \begin{adjustwidth}{-1.7cm}{}
\centering
\resizebox{4in}{!}{
\begin{tabular}{|c|c|c|c|c|}
\hline
\textbf{Fit Type} & \boldmath{$\alpha$} & \boldmath{$\beta$} & \boldmath{$\rho$} & \textbf{n}\\ 
\hline
FGK-Types & 0.37$\pm$0.08 & -2.34$\pm$0.56 & 0.46 & 1.31e-4 \\ \hline
M-types & 0.45$\pm$0.08 & -1.27$\pm$0.53 & 0.77 & 6.81e-5 \\ \hline
\end{tabular}}
\end{adjustwidth}
\caption{Best Fit parameters for the equation $log_{10}(L_{Ly\alpha}/L_{bol})=\alpha$ $\times$ $ log_{10}(L_{SiIII}/L_{bol})$ + $\beta$. The Spearman rank coefficients and probabilities of no correlation are reported.  \vspace{-0.5cm}}
\label{tab:SiIII} %6
\end{table}

\subsubsection{N V and \texorpdfstring{Ly$\alpha$}{Lyalpha} Luminosities}\label{sss:nitrogen} %4.5.2

\begin{figure}[!ht]
\epsscale{1.15}
\plotone{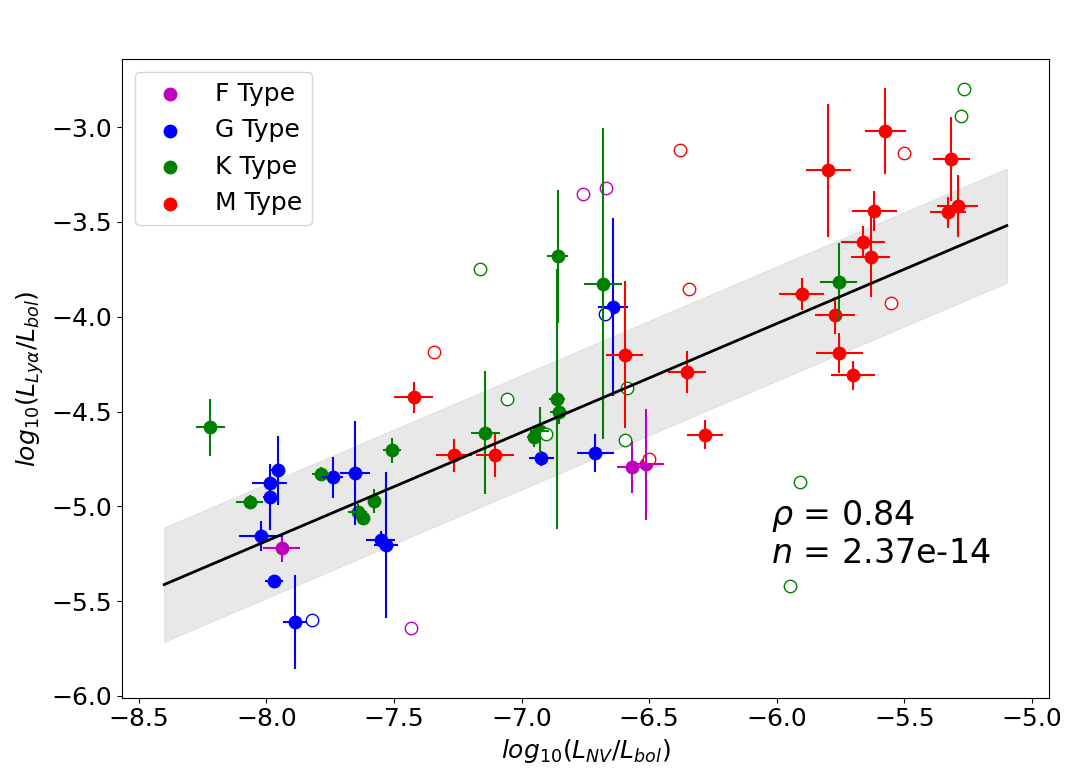}
\caption{Reconstructed Ly$\alpha$ luminosity vs N V luminosity, both normalized by stellar bolometric luminosity. \label{fig:NV_L}} %14
\end{figure}

We compare Ly$\alpha$ normalized luminosity to the combined N V normalized luminosity, show in Figure \ref{fig:NV_L}. For M dwarfs, \cite{2017ApJ...843...31Y} finds a very strong correlation between surface fluxes of combined N V and reconstructed Ly$\alpha$, and we expect to see a similar correlation for our sample. Similar to Figure \ref{fig:SiIII_L}, a gap is present between $-6.25 \lesssim log_{10}(L_{NV}/L_{bol}) \lesssim -5.95$ for the same reasons as with Si III. We do not separate the M dwarfs as in Section \ref{sss:silicon} due to the two fits were consistent with each other, indicating that the conditions in which these emission features form is similar across the four spectral types. For the equation $log_{10}(L_{Ly\alpha}/L_{bol})=\alpha$ $\times$ $ log_{10}(L_{NV}/L_{bol})$ + $\beta$, we find the best fit parameters to be $\alpha$=0.57$\pm$0.04 and $\beta$=-0.59$\pm$0.33. We observe a very strong correlation with a Spearman correlation coefficient of 0.84 and a 2.37$\times10^{-14}$ probability of no correlation. 

\subsubsection{OI Luminosities}\label{sss:oxygen} %4.5.3

\begin{figure*}[!ht]
\epsscale{0.97}
\hspace*{-0.5cm} 
\plotone{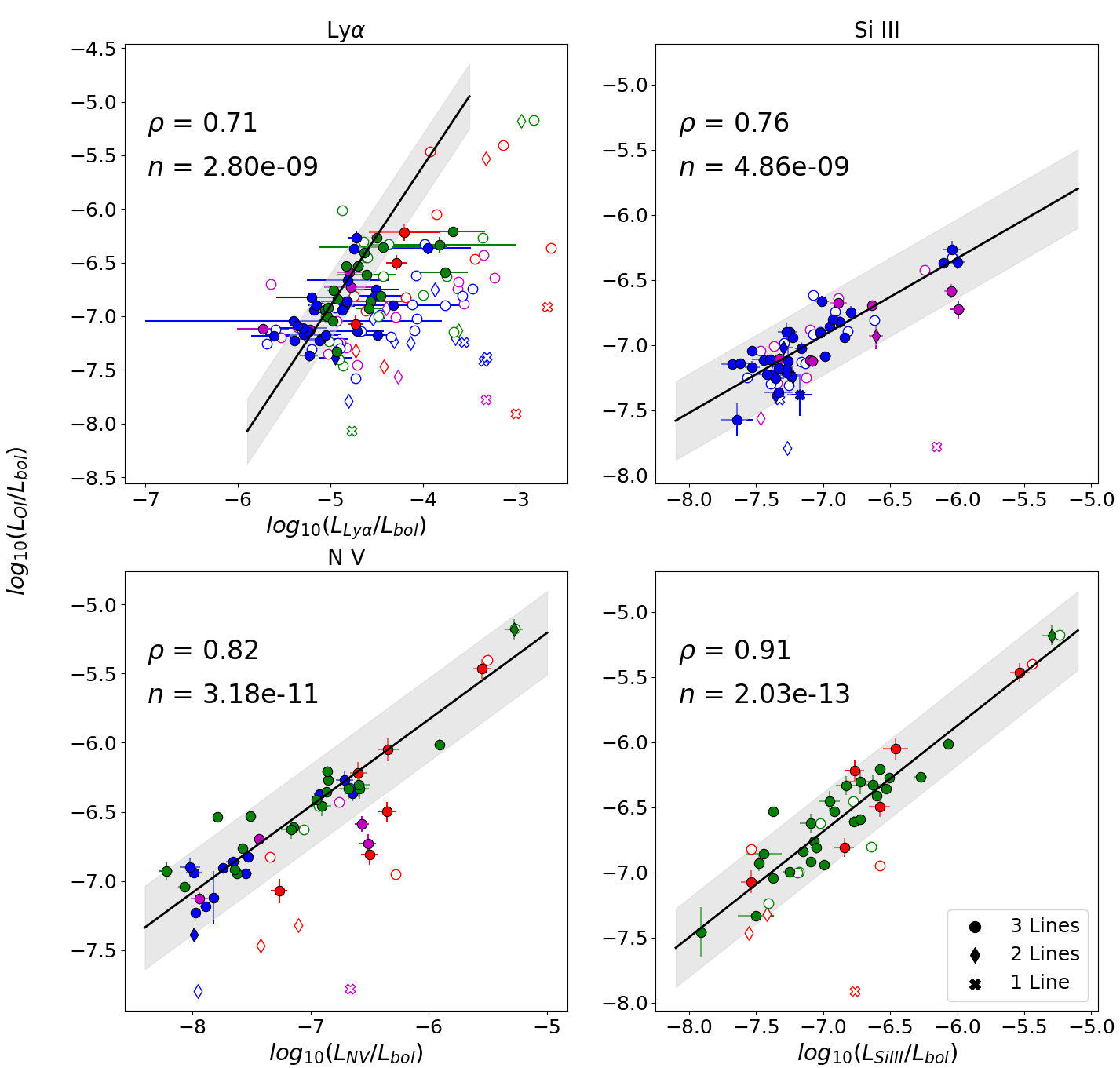}
\caption{Recovered O I triplet luminosity vs reconstructed Ly$\alpha$ luminosity (top left), Si III luminosity (right), and N V luminosity (bottom left), all normalized by stellar bolometric luminosity. Sums of all three O I lines are represented as circular markers, sums of two are represented as diamond markers, and a single line is represented as an X marker. Si III is split between FG and KM dwarfs and share X and Y axes. \vspace{0.35cm} \label{fig:OI_All}} %15
\end{figure*}

Figure \ref{fig:OI_All} displays the correlation between the normalized luminosities of the recovered O I triplet and normalized luminosities of reconstructed Ly$\alpha$, Si III, and N V, respectively. The recovered flux of individual oxygen lines are summed together. The number of summed O I lines determines the marker shape in Figure \ref{fig:OI_All}.
\par
O I triplet emission originates in the chromosphere, the same region where the Ly$\alpha$ emission originates. We expect to see these two to strongly correlate. For the comparison between O I and Ly$\alpha$, only stars that saw successful recoveries for both Ly$\alpha$ and O I were considered when determining the correlation and the line of best fit. Si III and N V are transition region lines, however as mentioned in Section \ref{ss:PRot}, dynamo driven magnetic heating gives rise to the flux in both regions of the stellar atmosphere. We observe strong or very strong correlations between these various chromospheric and transition region emission lines, and the best fit lines and Spearman statistics for each luminosity comparison is summarized in Table \ref{tab:OI}. 

\vspace{0.25cm}

\begin{table}[!ht]
 \begin{adjustwidth}{-2.0cm}{}
\centering
\resizebox{4in}{!}{
\begin{tabular}{|c|c|c|c|c|}
\hline
\textbf{Fit Type} & \boldmath{$\alpha$} & \boldmath{$\beta$} & \boldmath{$\rho$} & \textbf{n}\\ 
\hline
O I - Ly$\alpha$ & 1.30$\pm$0.14 & -0.40$\pm$0.72 & 0.71 & 2.80e-9 \\ \hline
O I - Si III (FG) & 0.59$\pm$0.05 & -2.77$\pm$0.34 & 0.76 & 4.86e-9 \\ \hline
O I - Si III (KM) & 0.81$\pm$0.08 & -1.00$\pm$0.59 & 0.91 & 2.03e-13 \\ \hline
O I - N V & 0.63$\pm$0.05 & -2.08$\pm$0.40 & 0.82 & 3.18e-11 \\ \hline
\end{tabular}}
\end{adjustwidth}
\caption{Best Fit parameters for the equation $log_{10}(L_{OI}/L_{bol})=\alpha$ $\times$ $log_{10} (L_{UV}/L_{bol})$ + $\beta$, where UV = Ly$\alpha$, Si III, or N V. Spearman rank coefficients and probabilities of no correlation are reported. \vspace{-0.25cm}}
\label{tab:OI} %7
\end{table}

We successfully recover a similar number of FGK O I profiles as we do Ly$\alpha$ profiles, however we do not successfully recover as many M dwarf O I profiles as with Ly$\alpha$. Due to the few recovered M dwarfs, we group them with K dwarfs. We find no significant difference in separating FG and KM dwarfs for the Ly$\alpha$ and N V power law fits, however we do find a difference for Si III in both slope and intercept. As with the Ly$\alpha$ comparison between Si III and N V, we see a gap structure when comparing O I to Si III and N V. With more observations of dwarf stars in the range $-6.5 \lesssim log_{10}(L_{SiIII}/L_{bol}) \lesssim -5.8$ would allow for determining if FG and KM dwarfs truly follow separate trend lines. 

\subsection{The Narrow Range of F, G, and K Dwarfs  in~the~COS~Data~Archive}\label{ss:FGKtypes} %4.6

Of all four spectral types, only F dwarfs resulted in more failed Ly$\alpha$ recoveries than successful ones. The causes for failure are a mixture of the effects of gain sag, detector gridwire shadow, and weak stellar signals due to the large distances to these stars (See Section \ref{ss:failure}). Even if we were able to recover more F dwarf profiles, very few had reported $R'_{HK}$ and $P_{rot,*}$ in the literature. It would be impossible to properly sample both the saturated and unsaturated regimes for F dwarfs with the currently available COS data alone. 
\par
The G-type stars in our sample are mostly limited to low activity stars, with narrow ranges of $P_{rot,*}$ ($\sim$ 15 to 38 days, i.e,. the unsaturated regime) and $log_{10}(R'_{HK})$ ($\sim$ -5.1 to -4.7) as seen in Figures \ref{fig:Prot} and \ref{fig:RPHK}, respectively. Of the G dwarfs in Figure \ref{fig:Prot}, 76\% were obtained from the SNAP 14633 observing program, which observed G and K-type exoplanet host stars in the solar neighborhood \citep{2018ApJS..239...16F}. Exoplanet detection techniques have an observational bias for low-activity, and often slowly rotating, stars \citep{2016MNRAS.459.3565V, 2018ApJS..239...16F}. 
\par
There are four G dwarfs in our sample with more rapid ($\lesssim$ 10 day) rotation periods, however these were observed only using Segment A to protect the COS detector from these bright targets, and Ly$\alpha$ could not be observed. A sample of nearby, active G dwarfs observed with COS does not currently exist because this region of parameter space is best explored with STIS, which can readily observe these bright targets. Outside of the solar neighborhood, it would be possible to observe faint, active G dwarfs with COS to begin to sample the saturated regime, however the effects of gain sag will strongly work against successfully recovering these stellar profiles. 
\par
The K dwarfs in our sample come from a variety of observing programs. While they do cover larger stellar activity spaces than G dwarfs, we still do not see enough stellar diversity to properly sample the saturated regime. The K Dwarf Advantage (GO 15955) surveys several K dwarfs that may begin to sample the saturated regime. Literature values of $R'_{HK}$ and $P_{rot,*}$ were unavailable for many of these stars. Severe gain sag is also prevalent in the Ly$\alpha$ profiles, largely the result of these stars being observed recently, when the COS detector sensitivity at airglow wavelengths have been completely depleted. 

\vspace{-0.5cm}

\section{Conclusions}\label{s:conclude} %5

Ly$\alpha$ and O I triplet fluxes serve as tracers of the stellar activity of F, G, K, and M dwarfs. These fluxes drive photochemistry in the atmospheres of orbiting exoplanets, altering the chemistry and abundances of key constituents in exoplanet atmospheres. Both Ly$\alpha$ and the O I triplet can additionally probe the atmospheric escape of exoplanets. The heavily attenuated EUV flux, critical for atmospheric escape models, can also be estimated from Ly$\alpha$ flux. 
\par
We have shown that the airglow which contaminates COS spectra can often be removed, revealing the underlying stellar emission. We created templates for the Ly$\alpha$ and O I triplet geocoronal emission lines based on observations taken with the COS instrument at LPs 1, 2, and 3. These templates were then used to recover the spectra on a sample of 171 archival F, G, K, and M dwarf stars. Moderate to very strong correlations were found with stellar properties and activity indicators.
\par
We developed an airglow subtraction and stellar recovery tool which simultaneously fits underlying stellar emission, attenuation from the ISM, and airglow contamination. The tool is made available through this \dataset[DOI]{https://doi.org/10.5281/zenodo.7435191} \citep{GUI}. The comparisons to STIS data and the available literature in Section \ref{ss:validity} and the subsequent analysis of the results presented in Section \ref{s:correlations} demonstrate the capability of the stellar recovery tool to recover the underlying stellar spectrum. 
\par
We discuss the limitations of this method and develop criteria to assist future observers aiming to study stellar Ly$\alpha$ with the COS instrument, recommending that estimated attenuated Ly$\alpha$ flux be greater than $1.39\times10^{-14}$ erg cm$^{-2}$ s$^{-1}$. We also recommend a Si III SNR of at least 3 for a 50\% successful recovery rate, and an SNR of at least 28 for a 90\% successful recovery rate. In the future, stellar Ly$\alpha$ studies may benefit from a dedicated Cenwave at the currently unused LP1, extending the science capabilities of COS without compromising currently used detector regions.
\par
The stellar recovery tool has demonstrated its ability to separate stellar and geocoronal signals for a majority of archival dwarf stars. This tool provides the opportunity to make the fluxes of critical photochemical drivers available for future investigations of exoplanet host stars with the COS instrument. 

\vspace{1cm}

\section*{Acknowledgements}

The authors thank the anonymous referees for their thorough review of and constructive comments on this manuscript and data products.
\par
Support for Program number AR 15635 was provided by NASA through a grant from the Space Telescope Science Institute, which is operated by the Association of Universities for Research in Astronomy, Incorporated, under NASA contract NAS5-26555.
\par
This project has received funding from the European Research Council (ERC) under the European Union's Horizon 2020 research and innovation programme (project {\sc Spice Dune}).
\par
This research has made use of the NASA Exoplanet Archive, which is operated by the California Institute of Technology, under contract with the National Aeronautics and Space Administration under the Exoplanet Exploration Program.
\par
The authors thank the members of the Colorado Ultraviolet Spectroscopy Program (CUSP) for helpful feedback and discussion on the stellar recovery tool developed for this work. Author Cruz Aguirre additionally thanks P. C. Hinton, Daniel Sega, Adriana Valverde, and Jim Green for useful discussion.

\appendix
\vspace{-0.5cm}
\setcounter{figure}{0} %reset figure count to 1
\renewcommand{\thefigure}{A\arabic{figure}} %start figures with A
\renewcommand*{\theHfigure}{\thefigure} %get hyperlinks to go to correct figure
\setcounter{table}{0} %reset table count to 1
\renewcommand{\thetable}{A\arabic{table}} %start tables with A
\renewcommand*{\theHtable}{\thetable} %get hyperlinks to go to correct table

\section{Residual Resampling Circular Block Bootstrap}\label{s:appendixA}

Bootstrapping is a technique that can be used to estimate the underlying distribution from which a data set is sampled. This is accomplished by generating a number of samples from the original data set. Each individual sample is generated by resampling with replacement. For a data set consisting of M data points, a random value is drawn from the original data set (resampling). Each point has equal probability of being drawn, and can be drawn multiple times when generating a bootstrap sample (replacement). Values are drawn until the bootstrap sample has the same length as the original data set, and the process is repeated N times, generating N bootstrap samples of length M. Each of these bootstrap samples will have their own statistics, such as the sample mean. By taking the 15.865th and 85.135th percentiles of the distribution of all bootstrap sample means, a 1$\sigma$ confidence interval is constructed.
\par
This idea can be extended to creating a distribution of best fit parameters from N bootstrap samples, however the simple bootstrap is not applicable in our case. An assumption of the bootstrap is that each point in the original data set is independent and identically distributed. This assumption is not true for spectral data, as there is an inherent order and structure to the data points in a spectrum. A compromise to this is the block bootstrap, where instead of resampling individual data points, blocks of data are resampled. Within each block, the inherent order and structure is preserved. Blocks are constructed with a fixed length B data points, and these blocks are resampled with replacement in the same way, each having equal probability of being drawn and can be drawn multiple times. There are a number of techniques to assure that each bootstrap sample has the correct length M. 
\par
We implement a circular block bootstrap technique, in which the ordered data in the final block will loop back to the data in the beginning of the spectrum if M/B does not result in an integer. For example, if a data set has length M=15 and a block length B=4 is chosen, the generated blocks would consist of the points \{1,2,3,4\}, \{5,6,7,8\}, \{9,10,11,12\}, and \{13,14,15,1\}, where it circles back to the first data point when constructing the final block. These blocks could be used to generate a bootstrap sample, say of data points \{5,6,7,8,13,14,15,1,5,6,7,8,1,2,3\}, where the final block \{1,2,3,4\} had its final point cut off in order to keep this sample at length M.
\par
In order to generate best fit parameter distributions, we generate circular block bootstrap samples using the residuals of the best fit subtracted from the stellar flux. These residuals are grouped into blocks of length B, and are then resampled with replacement to generate N bootstrap samples. These residual bootstrap samples are then added to the best fit profile to the spectral data, and the \texttt{LMFIT} model is run on this bootstrap profile. The best fit parameters of this bootstrap sample are then saved, and this process is repeated for all N bootstrap samples. Parameter distributions can then be generated, and a 1$\sigma$ confidence interval is calculated.
\par
The accuracy of the block bootstrap depends critically on the choice of block length B \citep{blockBootstrap,Buhlmann1999}. We utilize the \texttt{recombinator} python package to estimate the optimal block length for a circular block bootstrap.\footnote{\url{https://github.com/InvestmentSystems/recombinator}} The optimal block length is determined using the methodology described in \cite{Politis2004,Patton2009}. In the case where this method returns a block length that results in less than 5 blocks, 1000 unique samples cannot be generated, and the block length is decreased such that 5 blocks are created. 
\par
In practice, N=1000 bootstrap samples are generated using an estimate of the optimal block length B as described above. We originally chose 1000 samples as a balance between having enough bootstrap samples to properly generate the underlying parameter distributions and to minimize the computation time required to generate the parameter distributions. We demonstrate that 1000 samples is sufficient by running several bootstraps on a given data set, which all produced parameter distributions which were not significantly different from one another.

\clearpage

\section{Additional Figures and Tables}\label{s:appendixB}

\begin{table}[ht!]
    \centering
    \begin{adjustwidth}{1.95cm}{}
    \begin{tabular}{|c|c|c|}
        \hline
         \textbf{LP} & \textbf{Start Date or Start Cycle} & \textbf{Notes}  \\ \hline
         LP1 & May 2009 & All Cenwaves, All Gratings \\ \hline
         LP2 & 07-23-2012 & All Cenwaves, All Gratings \\ \hline
         LP3 & 02-09-2015 & All Cenwaves, All Gratings \\ \hline
         LP4 & 10-02-2017 & All Cenwaves, All Gratings \\ \hline
         LP5 & Cycle 29 (2021) & G130M Cenwaves 1291 and Longer \\ \hline
         LP6 & Cycle 30 (2022) & Long ($\gtrsim$1/2 Orbit) G160M  \\ \hline
    \end{tabular}
    \end{adjustwidth}
    \caption{The Lifetime Positions of the COS FUV detector. The LP is changed every few years to combat the effects of gain sag \citep{COS_IHB_2022}. No LP5 or LP6 data was available for this work.}
    \label{tab:LP} %A1
\end{table}

\begin{figure}[!ht]
\epsscale{1.1}
\plotone{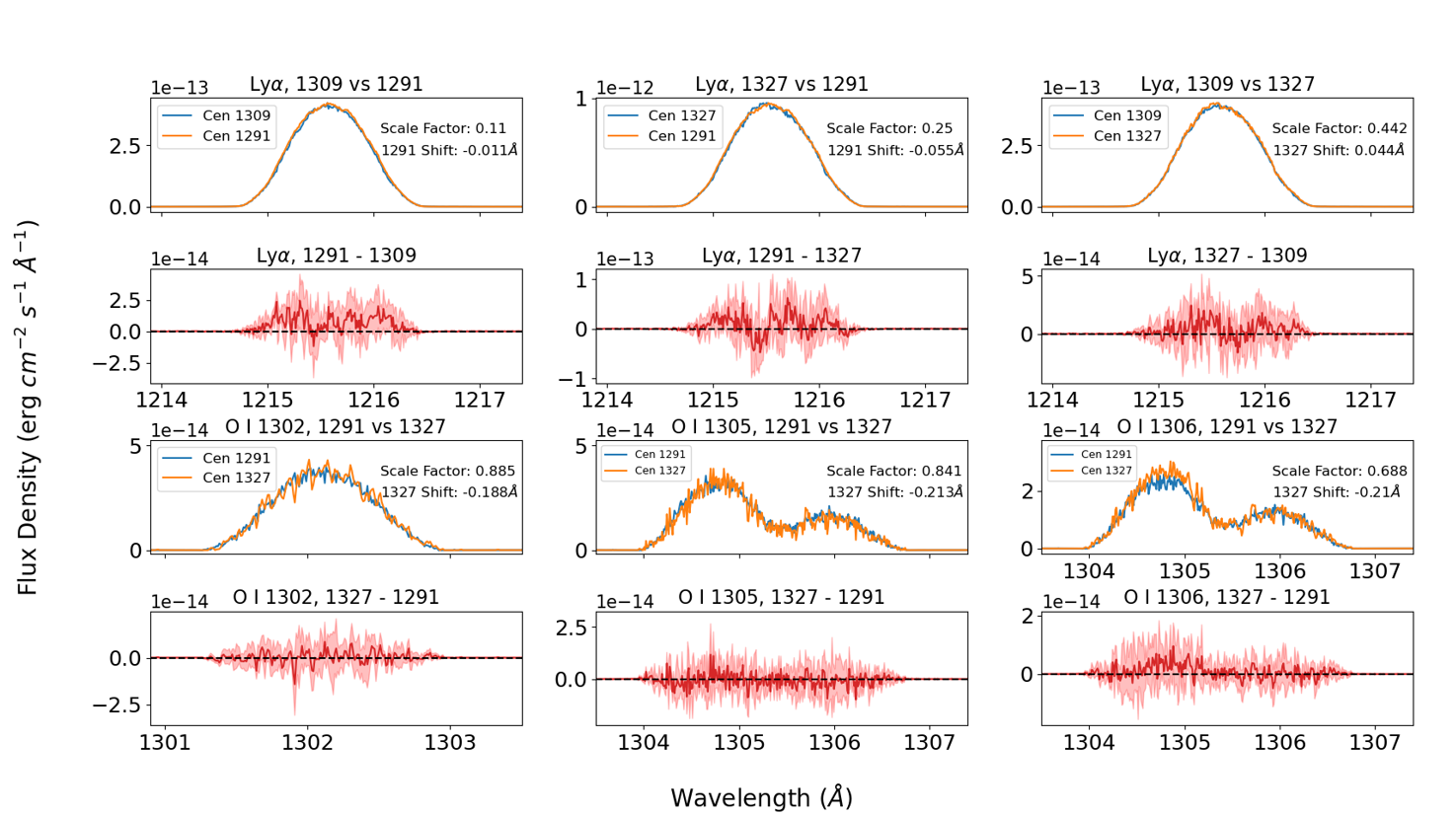}
\caption{All Cenwave comparisons made between the available airglow templates for LP1. Shaded regions in the difference plots show the propagated error uncertainties. \label{fig:All_LP1}} %A1
\end{figure}

\begin{figure}[!ht]
\epsscale{1.1}
\plotone{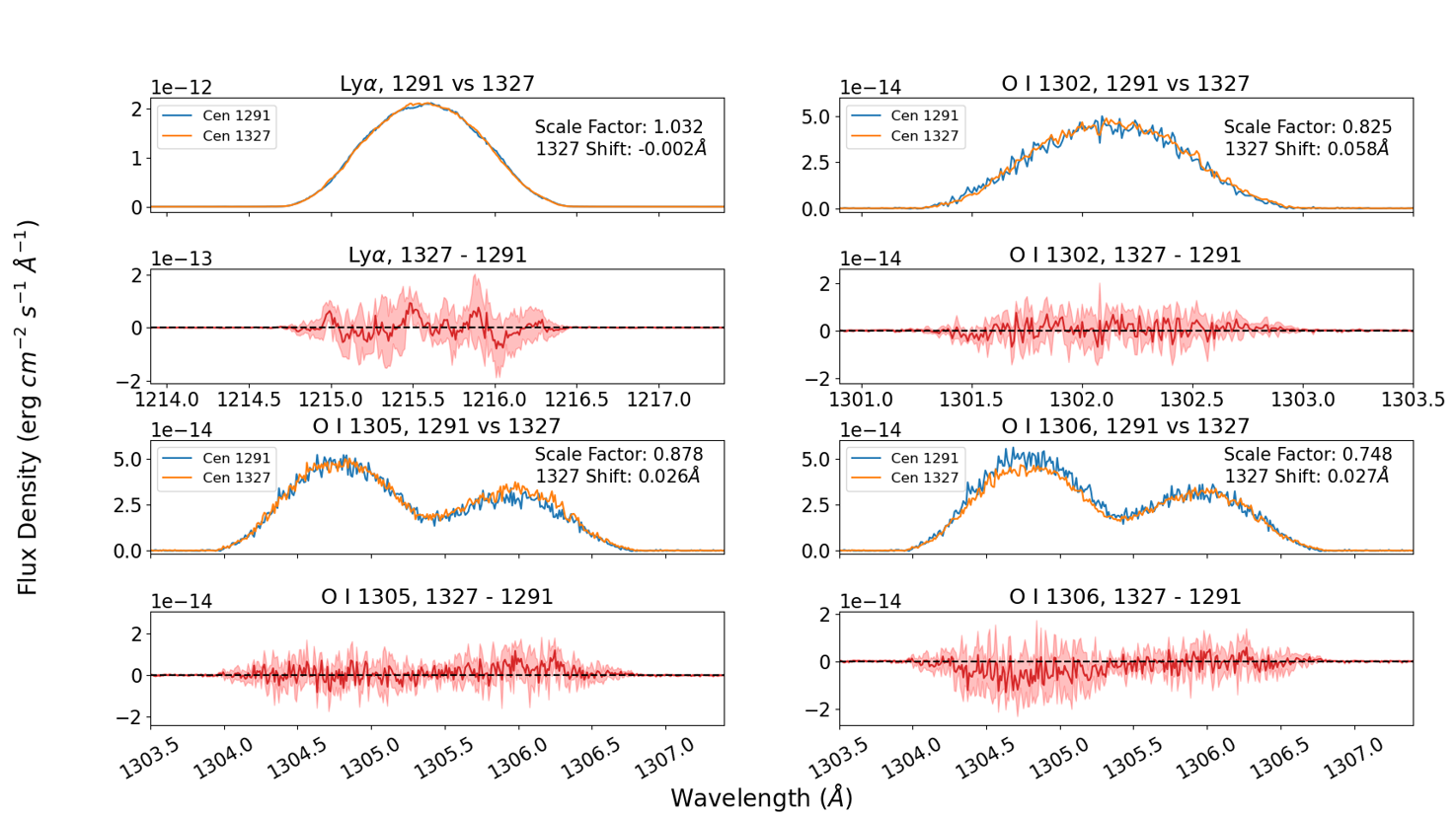}
\caption{Same as Figure \ref{fig:All_LP1} but for LP2. \label{fig:All_LP2}} %A2
\end{figure}

\begin{figure}[!ht]
\epsscale{1.1}
\plotone{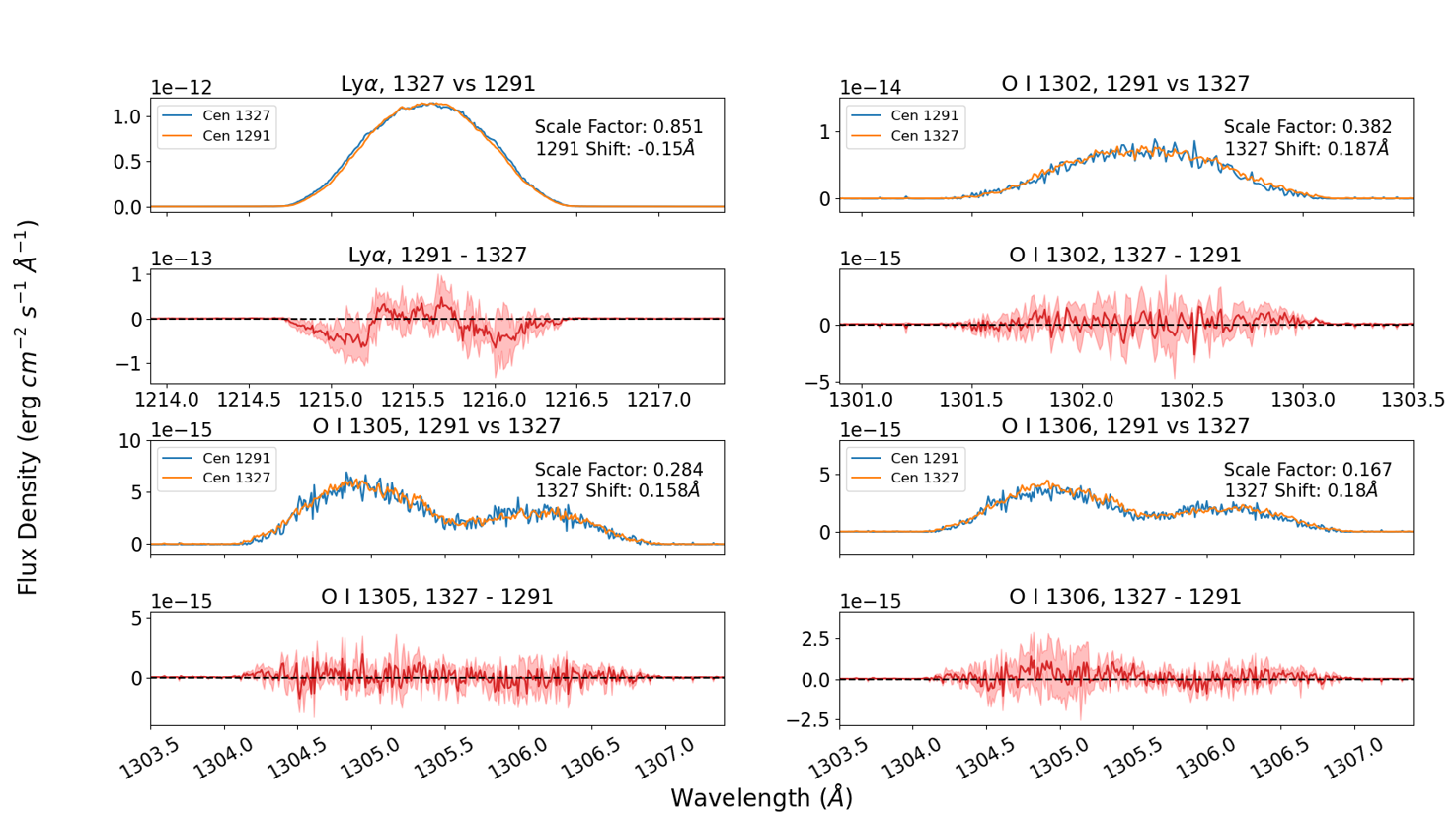}
\caption{Same as Figure \ref{fig:All_LP1} but for LP3. \label{fig:All_LP3}} %A3
\end{figure}

\begin{figure}[!ht]
\epsscale{1.1}
\plotone{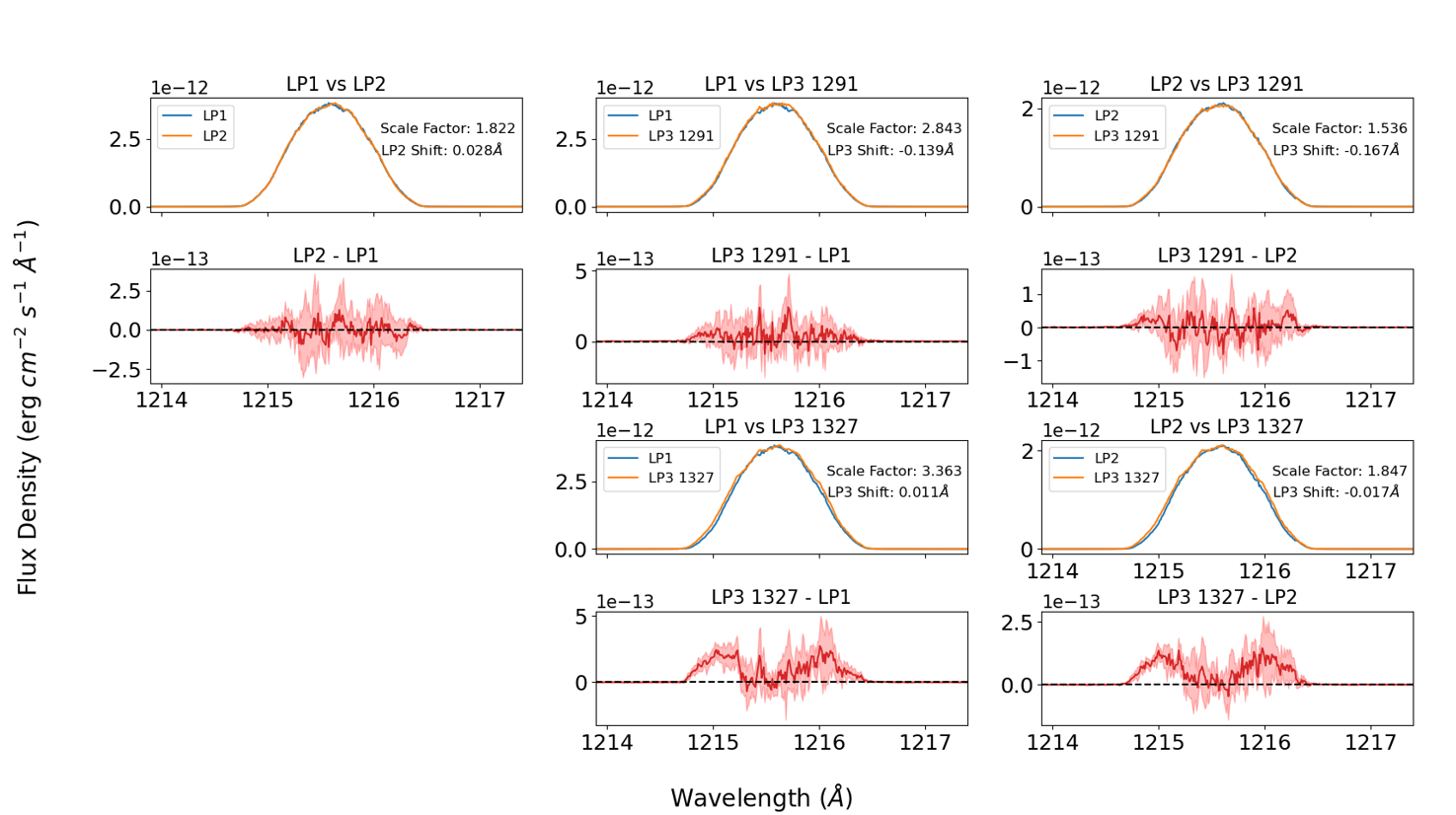}
\caption{All LP comparisons made between the available airglow templates for Ly$\alpha$. Unless otherwise noted, all templates used in these comparisons are the Cenwave 1291 templates. Shaded regions in the difference plots show the propagated error uncertainties. \label{fig:All_LyA}} %A4
\end{figure}

\begin{figure}[!ht]
\epsscale{1.1}
\plotone{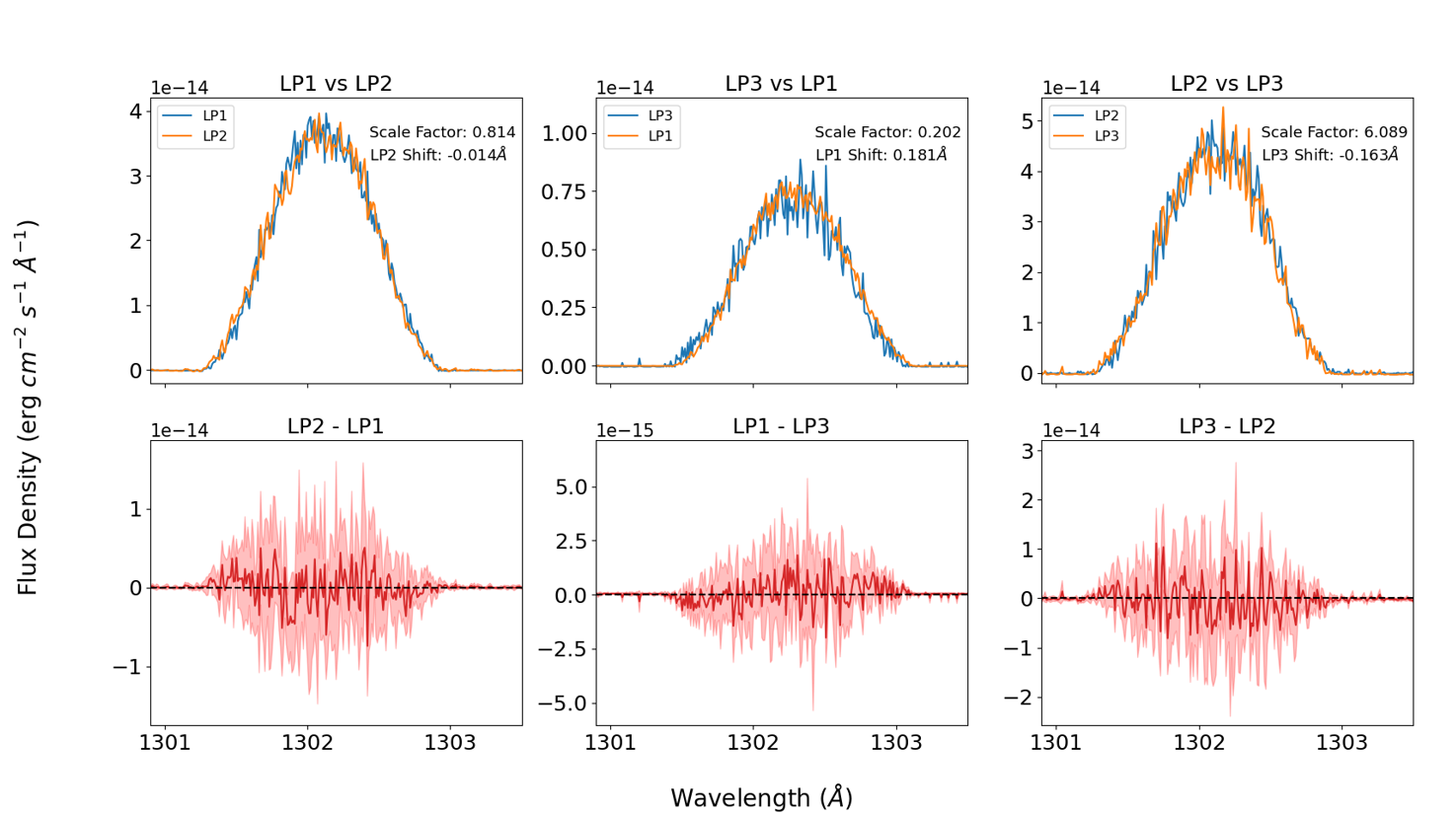}
\caption{Same as Figure \ref{fig:All_LyA} but for O I 1302. \label{fig:All_OI2}} %A5
\end{figure}

\begin{figure}[!ht]
\epsscale{1.1}
\plotone{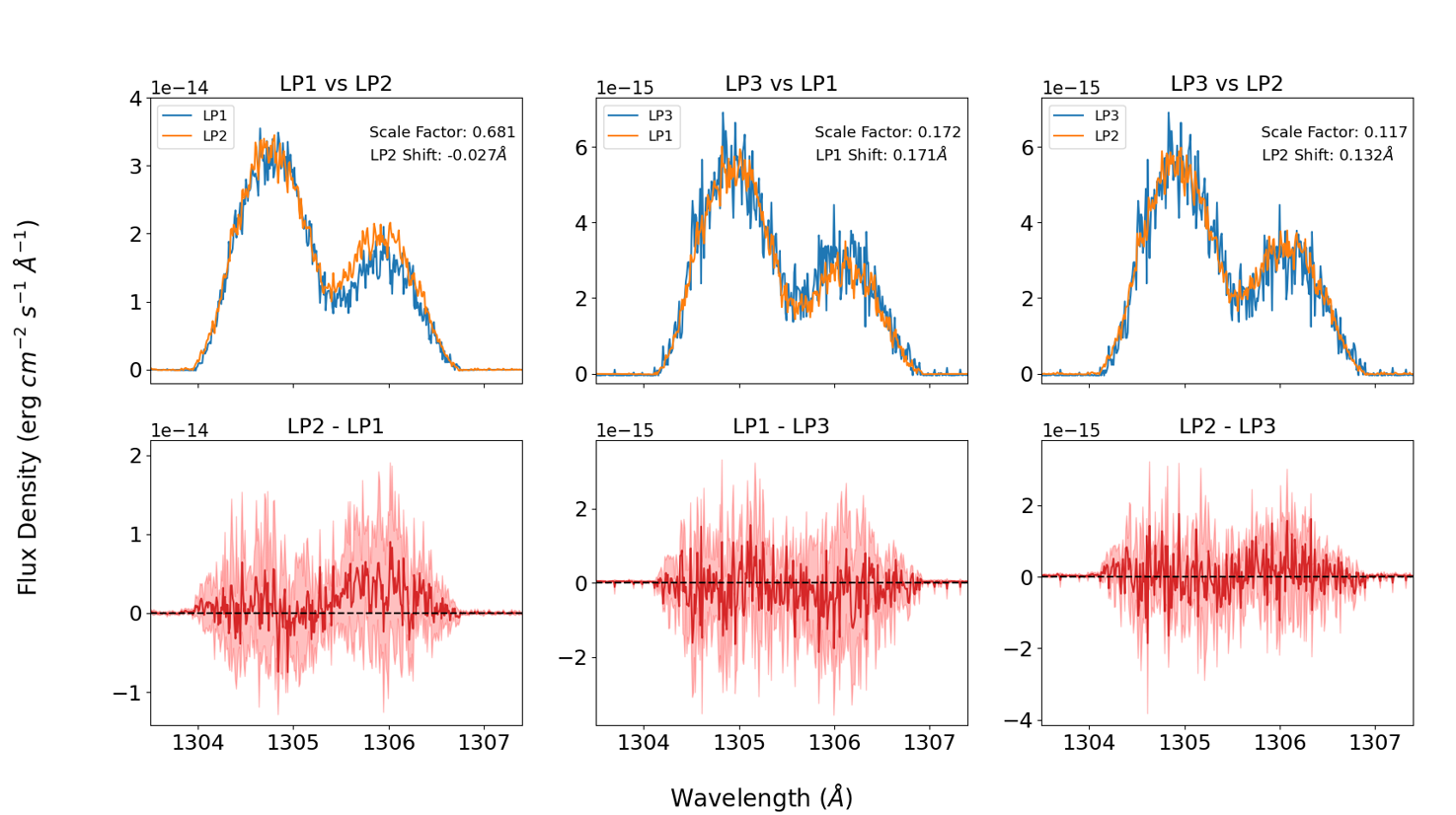}
\caption{Same as Figure \ref{fig:All_LyA} but for O I 1305. \label{fig:All_OI5}} %A6
\end{figure}

\begin{figure}[!ht]
\epsscale{1.1}
\plotone{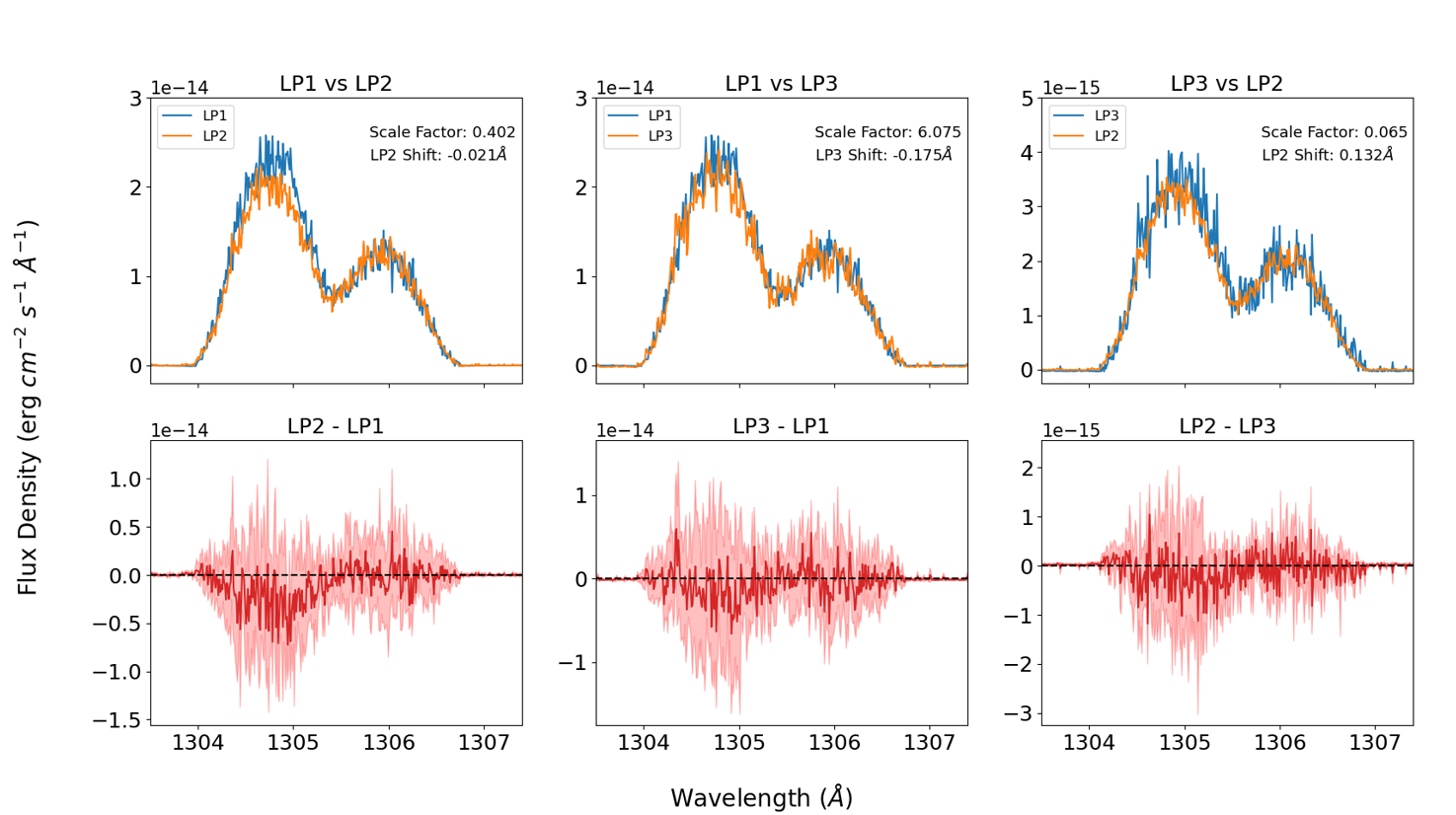}
\caption{Same as Figure \ref{fig:All_LyA} but for O I 1306. \label{fig:All_OI6}} %A7
\end{figure}

\begin{figure*}[!ht]
\centering
\subfigure{\epsscale{0.45} \plotone{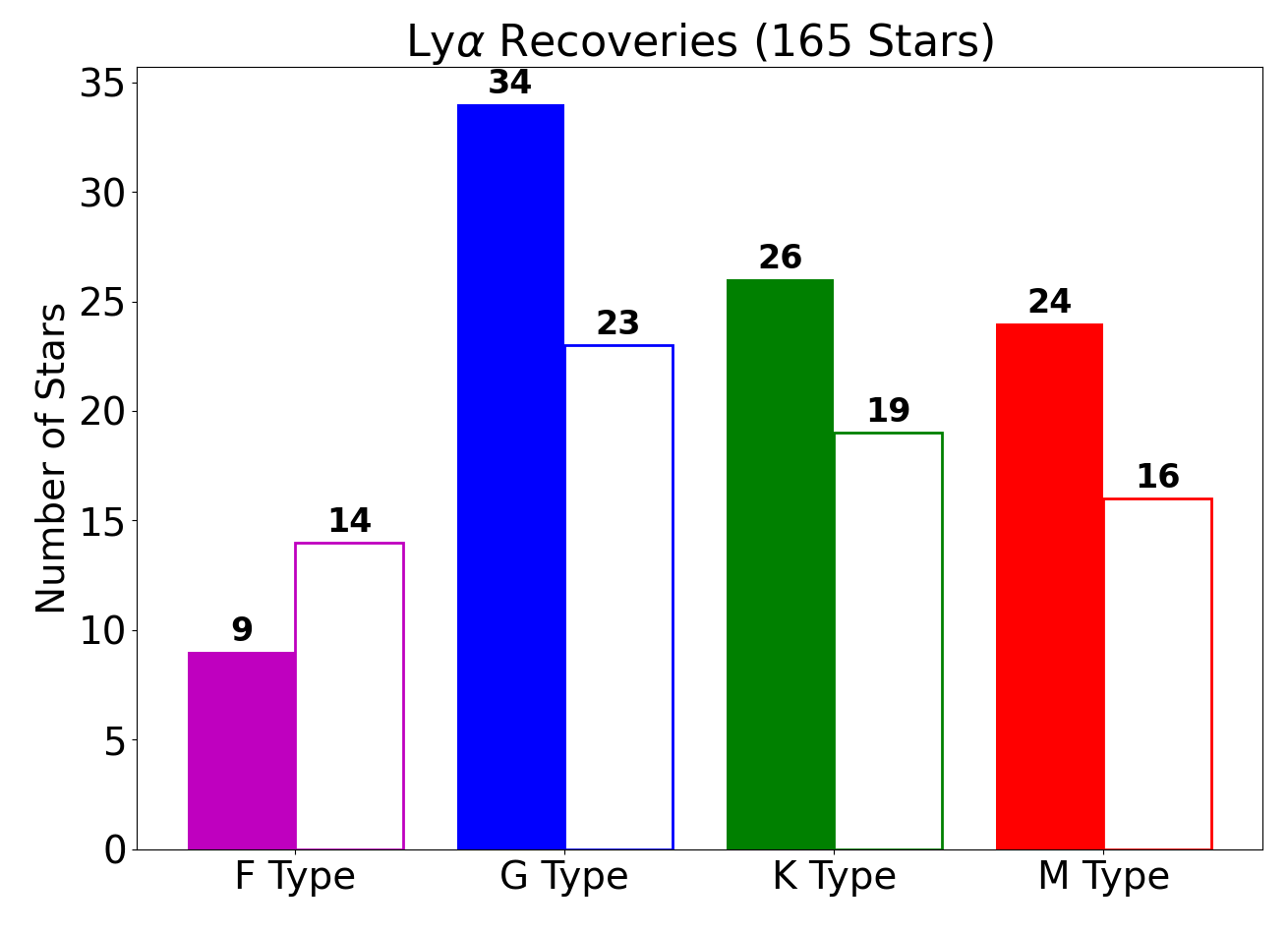} \label{sfig:barL}} %A8a
\subfigure{\epsscale{0.45} \plotone{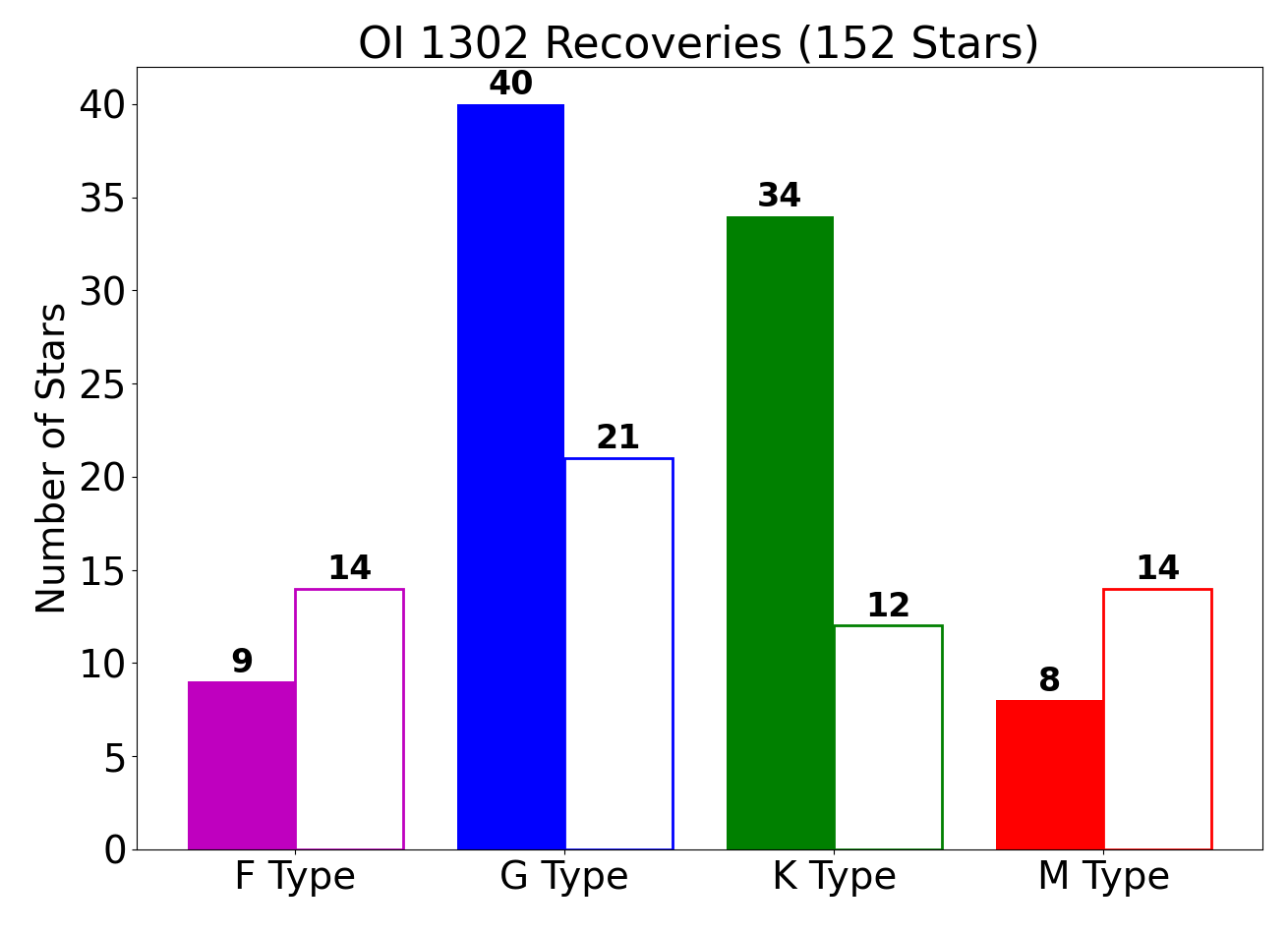} \label{sfig:bar2}} %A8b
\subfigure{\epsscale{0.45} \plotone{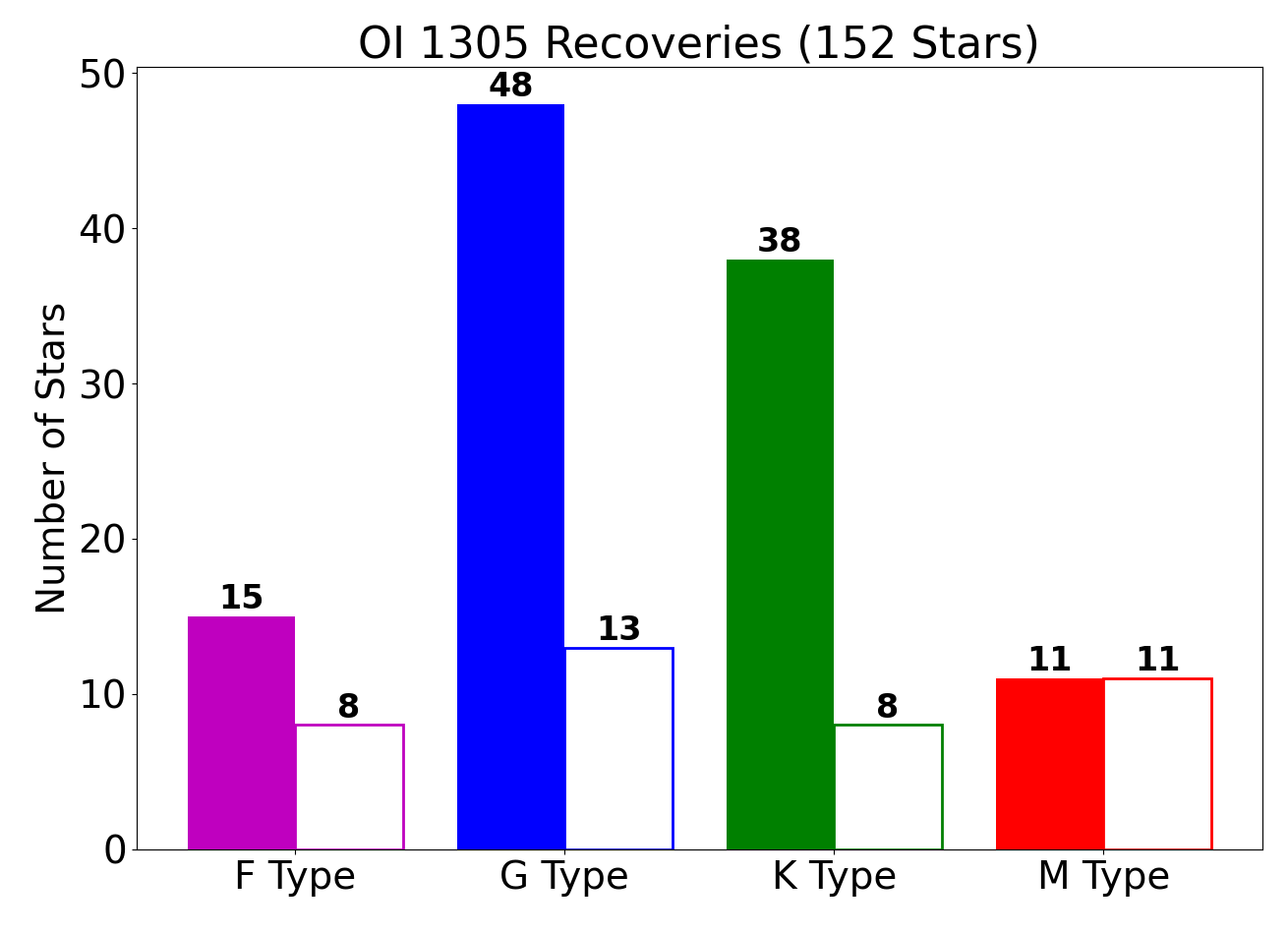} \label{sfig:bar5}} %A8c
\subfigure{\epsscale{0.45} \plotone{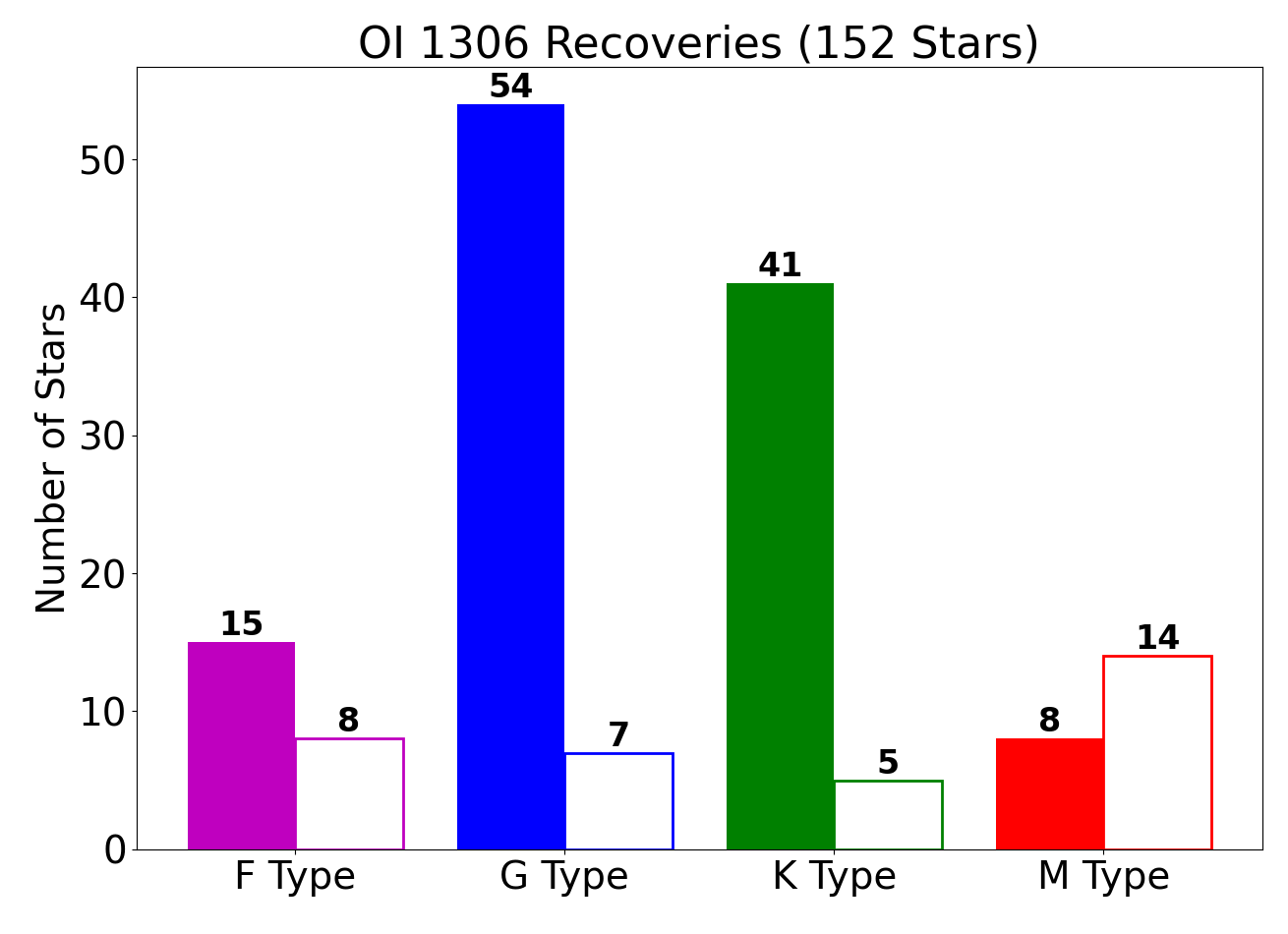} \label{sfig:bar6}} %A8d
\caption{Bar charts representing the number of successful and failed recoveries of each contaminated emission line. Filled bars represent successful recoveries, while empty bars represent failed recoveries. \label{fig:bars}} %A8
\end{figure*}

\begin{sidewaystable}[ht!]
\begin{adjustwidth}{-4.2cm}{}
\centering
\resizebox{10.5in}{!}{
\begin{tabular}{|c|c|c|c|c|c|c|c|c|c|c|c|c|c|}
\hline
\textbf{Star} & \textbf{Spectral} & \textbf{Observing} & \textbf{Date of} & \textbf{Exoplanet} & \textbf{Radial Vel.} & \textbf{ISM Vel.} & \multirow{2}*{\boldmath{$log_{10}(N_{col,HI})$}} & \textbf{Ly$\alpha$ Flux} & \multirow{2}*{\boldmath{$log_{10}(R'_{HK})$}} & \boldmath{$T_{eff}$} & \boldmath{$R_*$} & \textbf{Dist.} & \boldmath{$P_{rot,*}$} \\ 
\textbf{Name} & \textbf{Type} & \textbf{Program} & \textbf{First Obs.} & \textbf{Host?} & \textbf{(km s$^{-1}$)} & \textbf{(km s$^{-1}$)} &  & \textbf{(erg cm$^{-2}$ s$^{-1}$)} &  & \boldmath{(K)} & \boldmath{(R$_{\odot}$)} & \textbf{(pc)} & \boldmath{(days)} \\ \hhline{|=|=|=|=|=|=|=|=|=|=|=|=|=|=|}
\multirow{2}*{16 Cyg A} & \multirow{2}*{G1.5 Vb$^{(1)}$} & \multirow{2}*{GO 13861} & 2015-10-23 &  & \multirow{2}*{-27.61$^{(2)}$} & -6.65 &  &  & \multirow{2}*{$-5.12\pm0.188157$$^{(83,}$$^{*)}$} & \multirow{2}*{$5830.0\pm7.0$$^{(95)}$} & \multirow{2}*{$1.294\pm0.024$$^{(102)}$} & \multirow{2}*{$21.1519\pm0.0146$$^{(3)}$} & \multirow{2}*{$30.0\pm6.0$$^{(123)}$}\\ 
 &  &  & 01:58:56 &  &  &  &  &  &  &  &  &  & \\ \hline
\vdots & \vdots & \vdots & \vdots & \vdots & \vdots & \vdots & \vdots & \vdots & \vdots & \vdots & \vdots & \vdots & \vdots\\ \hline
\multirow{2}*{WASP-80} & \multirow{2}*{K7 V-M0 V$^{(70)}$} & \multirow{2}*{GO 14767} & 2018-07-22 & \multirow{2}*{Yes} & \multirow{2}*{9.82$^{(3)}$} & -17.47 & \multirow{2}*{$19.27^{+nan}_{-nan}$$^{(76)}$} & \multirow{2}*{$7.3^{+nan}_{-nan} \times 10^{-14}$$^{(76)}$} &  &  &  & \multirow{2}*{$49.8591\pm0.1155$$^{(3)}$} & \\ 
 &  &  & 09:53:35 &  &  &  &  &  &  &  &  &  & \\ \hline
\end{tabular}}
\end{adjustwidth}
\caption{For each star, the spectral type, COS observing program ID, and the observation                      start date are listed. If the star is a known exoplanet host, a ``Yes'' will be listed in                      the fourth column. Columns 6 and onward list literature values of various quantities. All                      ISM velocities were calculated using the Local ISM (LISM) Kinematic Calculator \citep{2008ApJ...673..283R}.                      The full version of this table is available online.}
\footnotesize{$^{1}$ \cite{1989ApJS...71..245K}, $^{2}$ \cite{2002ApJS..141..503N}, $^{3}$ \cite{2018yCat.1345....0G}, $^{4}$ \cite{2018AnA...616A...7S}, $^{5}$ \cite{2006AnA...460..695T}, $^{6}$ \cite{2014AJ....147..146K}, $^{7}$ \cite{2017ApJ...840...87R}, $^{8}$ \cite{2006AJ....132..866R}, $^{9}$ \cite{2006ApJ...649..436E}, $^{10}$ \cite{2003AJ....126.2048G}, $^{11}$ \cite{1986AJ.....92..139S}, $^{12}$ \cite{2011AAS...21743412C}, $^{13}$ \cite{2005AnA...442..211S}, $^{14}$ \cite{1985ApJS...59..197B}, $^{15}$ \cite{2013AJ....146..134K}, $^{16}$ \cite{2014AJ....148...91L}, $^{17}$ \cite{2014AJ....147...20N}, $^{18}$ \cite{2010MNRAS.403.1949K}, $^{19}$ \cite{1996AJ....112.2799H}, $^{20}$ \cite{2014MNRAS.438.2413V}, $^{21}$ \cite{2018AJ....155..122S}, $^{22}$ \cite{2013AJ....145..102L}, $^{23}$ \cite{2018MNRAS.475.1960F}, $^{24}$ \cite{1991ApJS...77..417K}, $^{25}$ \cite{2002ApJ...564..466G}, $^{26}$ \cite{1978MSS...C02....0H}, $^{27}$ \cite{1974ApJS...28....1J}, $^{28}$ \cite{1983SAAOC...7..106W}, $^{29}$ \cite{2014ApJ...788...81M}, $^{30}$ \cite{2013MNRAS.433.2097F}, $^{31}$ \cite{2006AJ....132..161G}, $^{32}$ \cite{1975MSS...C01....0H}, $^{33}$ \cite{2001AJ....121.2148G}, $^{34}$ \cite{1988MSS...C04....0H}, $^{35}$ \cite{1999MSS...C05....0H}, $^{36}$ \cite{2001MNRAS.328...45M}, $^{37}$ \cite{2021ApJ...910...71A}, $^{38}$ \cite{1993yCat.3135....0C}, $^{39}$ \cite{1950ApJ...111..221W}, $^{40}$ \cite{2018MNRAS.479.1332M}, $^{41}$ \cite{2007MNRAS.374..664C}, $^{42}$ \cite{2006AstL...32..759G}, $^{43}$ \cite{1955ApJS....2..195R}, $^{44}$ \cite{2007AJ....133.2524W}, $^{45}$ \cite{1950ApJ...112...48M}, $^{46}$ \cite{2018ApJS..239...16F}, $^{47}$ \cite{1982MSS...C03....0H}, $^{48}$ \cite{2017AJ....154..184W}, $^{49}$ \cite{2018AnA...619A..81H}, $^{50}$ \cite{2009MNRAS.396.1895C}, $^{51}$ \cite{2005ApJS..159..141V}, $^{52}$ \cite{2000ApJ...536..902V}, $^{53}$ \cite{2011AnA...526A..71D}, $^{54}$ \cite{2018AnA...615A..31D}, $^{55}$ \cite{2014AnA...563A..22M}, $^{56}$ \cite{2019AJ....158...87D}, $^{57}$ \cite{2018AnA...612A..96F}, $^{58}$ \cite{2019ApJ...871L..24V}, $^{59}$ \cite{2014MNRAS.443.2561G}, $^{60}$ \cite{2017Natur.544..333D}, $^{61}$ \cite{2015AnA...577A.128A}, $^{62}$ \cite{2015ApJS..220...16T}, $^{63}$ \cite{2017AJ....153...75K}, $^{64}$ \cite{1968ApJ...151..605P}, $^{65}$ \cite{2006AJ....132.1234C}, $^{66}$ \cite{2015ApJS..220...18B}, $^{67}$ \cite{2022ApJ...929..169R}, $^{68}$ \cite{2011AnA...529A.136E}, $^{69}$ \cite{2014MNRAS.445.1114A}, $^{70}$ \cite{2013AnA...551A..80T}, $^{71}$ \cite{2016ApJ...824..101Y}, $^{72}$ \cite{2019AJ....158...50W}, $^{73}$ \cite{2018AnA...620A.147B}, $^{74}$ \cite{2021MNRAS.501.4383V}, $^{75}$ \cite{2020AJ....160..269M}, $^{76}$ \cite{2018MNRAS.478.1193K}, $^{77}$ \cite{2020MNRAS.493..559B}, $^{78}$ \cite{2019ApJ...880..117E}, $^{79}$ \cite{2020ApJ...888L..21G}, $^{80}$ \cite{2016ApJ...821...81G}, $^{81}$ \cite{2021ApJ...911...18W}, $^{82}$ \cite{2021AnA...649A..40D}, $^{83}$ \cite{2011ARep...55.1123K}, $^{84}$ \cite{2004ApJS..152..261W}, $^{85}$ \cite{2012AnA...537A.147H}, $^{86}$ \cite{2018AnA...616A.108B}, $^{87}$ \cite{2017AnA...600A..13A}, $^{88}$ \cite{2009AnA...493.1099S}, $^{89}$ \cite{2018AnA...612A..89S}, $^{90}$ \cite{1996AJ....111..439H}, $^{91}$ \cite{2013AnA...552A..27M}, $^{92}$ \cite{2010ApJ...725..875I}, $^{93}$ \cite{2011AnA...531A...8J}, $^{94}$ \cite{2015ApJ...812L..35F}, $^{95}$ \cite{2014ApJ...790L..25T}, $^{96}$ \cite{2012ApJ...746..101B}, $^{97}$ \cite{2003AnA...411..559K}, $^{98}$ \cite{2008AnA...478..507M}, $^{99}$ \cite{2011MNRAS.411..435B}, $^{100}$ \cite{2017ApJ...848...34H}, $^{101}$ \cite{2009ApJ...694.1085V}, $^{102}$ \cite{2006AnA...450..735M}, $^{103}$ \cite{2013AnA...553A..95M}, $^{104}$ \cite{2012MNRAS.427..343M}, $^{105}$ \cite{2003AnA...398..721T}, $^{106}$ \cite{2012ApJ...749...39R}, $^{107}$ \cite{2014MNRAS.443L..89A}, $^{108}$ \cite{2015PASJ...67...32N}, $^{109}$ \cite{2019ApJ...886..142M}, $^{110}$ \cite{2020AnA...642A.115C}, $^{111}$ \cite{2021AnA...656A.142K}, $^{112}$ \cite{2011AnA...526L...4B}, $^{113}$ \cite{2001AnA...367..521P}, $^{114}$ \cite{2017ApJ...834...85N}, $^{115}$ \cite{2019AnA...623A..72K}, $^{116}$ \cite{2017AJ....153..136S}, $^{117}$ \cite{2016AnA...585A...5B}, $^{118}$ \cite{2018AnA...615A..76S}, $^{119}$ \cite{2010ApJ...720.1290G}, $^{120}$ \cite{2010MNRAS.407.1657H}, $^{121}$ \cite{2013AnA...557L..10N}, $^{122}$ \cite{2019ApJS..241...12S}, $^{123}$ \cite{2014ApJ...790L..23D}, $^{124}$ \cite{2008MNRAS.388...80P}, $^{125}$ \cite{2010AnA...520A..15M}, $^{126}$ \cite{2021ApJ...907...91L}, $^{127}$ \cite{2016AnA...586A..14H}, $^{128}$ \cite{1996ApJ...466..384D}, $^{129}$ \cite{2016ApJ...821...93N}, $^{130}$ \cite{2018AJ....156..217N}, $^{131}$ \cite{2018AnA...614A..35M}, $^{132}$ \cite{2015MNRAS.452.2745S}, $^{133}$ \cite{2020MNRAS.491.5216G}, $^{134}$ \cite{2010ApJ...722..343D}, $^{135}$ \cite{2018AnA...612L...2K}, $^{136}$ \cite{2016AnA...595A..12S}, $^{137}$ \cite{2005AnA...437.1121L}, $^{138}$ \cite{2013PASP..125.1436O}, $^{139}$ \cite{2013AnA...551A..90M}, $^{140}$ \cite{2001ApJ...561.1095B}, $^{141}$ \cite{2011ApJ...743...48W}, $^{142}$ \cite{2002PASP..114..529F}, $^{143}$ \cite{2010MNRAS.408.1666S}, $^{144}$ \cite{2011ApJ...726...73H}, $^{145}$ \cite{2011AnA...534A..58P}, $^{146}$ \cite{2001AnA...379..999S}, $^{147}$ \cite{2002AnA...392..215S}, $^{148}$ \cite{2009AnA...496..527B}, $^{149}$ \cite{2009ApJ...696...75H}, $^{150}$ \cite{2011ApJ...727..117M}, $^{151}$ \cite{2016MNRAS.463.1844S}, $^{152}$ \cite{2017ApJ...841..124V}, $^{153}$ \cite{2019ApJ...879...39J}\\}
\footnotesize{$^*$ Error estimated based on other reported errors.}\\ 
\footnotesize{$^{\dagger}$ Calculated using Ly$\alpha$ reconstruction method developed in \cite{2016ApJ...824..101Y}.} \\ 
\footnotesize{$^{\ddagger}$ Upper limit based on $v$ sin $i$}
\label{tab:literature} %A2
\end{sidewaystable}

\begin{sidewaystable}[ht!]
\begin{adjustwidth}{-3.75cm}{}
\centering
\resizebox{10.5in}{!}{
\begin{tabular}{|c|c|c|c|c|c|c|c|c|c|c|c|}
\hline
\textbf{Star} & \textbf{Emission} & \boldmath{Radial Vel.} & \boldmath{FWHM$_G$} & \boldmath{FWHM$_L$} & \multirow{2}*{\boldmath{$log_{10}(F_{amp})$}} & \multirow{2}*{\boldmath{$log_{10}(N_{col,HI})$}} & \boldmath{ISM Vel.} & \textbf{Airglow Shift} & \textbf{Airglow} & \boldmath{Rec. Flux} & \textbf{Morph.} \\ 
\textbf{Name} & \textbf{Line} & \boldmath{(km s$^{-1}$)} & \boldmath{(km s$^{-1}$)} & \boldmath{(km s$^{-1}$)} &  &  & \boldmath{(km s$^{-1}$)} & \textbf{(\AA)} & \textbf{Scale Factor} & \boldmath{(erg cm$^{-2}$ s$^{-1}$)} & \textbf{Cat.} \\ \hhline{|=|=|=|=|=|=|=|=|=|=|=|=|}
\multirow{2}*{16 Cyg A} & Ly$\alpha$ & $-27.398^{+2.412}_{-4.066}$ & $138.813^{+4.702}_{-4.283}$ & $6.665^{+2.373}_{-1.185}$ & $-10.24^{+0.224}_{-0.334}$ & $18.78^{+0.07}_{-0.103}$ & $-16.858^{+4.365}_{-0.753}$ & $-0.126^{+0.004}_{-0.001}$ & $1.731^{+0.006}_{-0.006}$ & $1.93^{+0.78}_{-0.81} \times 10^{-12}$ & QS,W\\ 
 & O I 1306 & $-36.221^{+5.314}_{-5.191}$ & $38.161^{+6.175}_{-22.467}$ & $-$ & $-13.547^{+0.455}_{-0.079}$ & $-$ & $-$ & $0.002^{+0.007}_{-0.007}$ & $6.43^{+0.205}_{-0.152}$ & $5.01^{+0.99}_{-0.88} \times 10^{-15}$ & ME\\ \hline
\vdots & \vdots & \vdots & \vdots & \vdots & \vdots & \vdots & \vdots & \vdots & \vdots & \vdots & \vdots\\ \hline 
\multirow{3}*{WASP-80} & O I 1302 & $6.267^{+0.745}_{-0.655}$ & $28.825^{+2.391}_{-2.558}$ & $-$ & $-14.972^{+0.03}_{-0.029}$ & $-$ & $-$ & $0.017^{+0.012}_{-0.007}$ & $0.007^{+0.0}_{-0.0}$ & $1.42^{+0.06}_{-0.06} \times 10^{-16}$ & ME\\ 
 & O I 1305 & $-1.136^{+0.824}_{-0.747}$ & $19.151^{+2.829}_{-2.836}$ & $-$ & $-14.884^{+0.061}_{-0.053}$ & $-$ & $-$ & $0.034^{+0.01}_{-0.007}$ & $0.006^{+0.0}_{-0.0}$ & $1.16^{+0.05}_{-0.06} \times 10^{-16}$ & ME\\ 
 & O I 1306 & $3.776^{+1.519}_{-1.159}$ & $31.43^{+3.954}_{-4.82}$ & $-$ & $-15.206^{+0.057}_{-0.043}$ & $-$ & $-$ & $-0.003^{+0.013}_{-0.011}$ & $0.004^{+0.001}_{-0.001}$ & $9.07^{+0.61}_{-0.68} \times 10^{-17}$ & ME\\ \hline
\end{tabular}}
\end{adjustwidth}
\caption{Best fit results for all stars where the GUI successfully recovered the stellar                     emission. Fit parameters are described in Section \ref{sss:components}. All parameter errors                     were calculated from the bootstrapping technique described in Section \ref{sss:rrcbb}.                     Reconstructed stellar flux for Ly$\alpha$ and recovered stellar flux for the O I triplet                     are reported after best fit parameters. The final column (Morph. Cat.) displays the                     morphological category of the observed Ly$\alpha$ and O I profiles (DS,S = Definite Stellar,                     Strong, DS,W = Definite Stellar, Weak, QS,S = Questionable Stellar, Strong, QS,W = Questionable                     Stellar, Weak, SD = Stellar Dominant, ME = Mixed Emission, AD = Airglow Dominant). The full                     version of this table is available online.}
\label{tab:GUI params} %A3
\end{sidewaystable}

\begin{table}[ht!]
\centering
\begin{adjustwidth}{1.1cm}{}
\begin{tabular}{|c|c|c|c|}
\hline
    \textbf{Star Name} & \textbf{Observing Program} & \textbf{First Obs. Date}  & \textbf{Grating Used}\\ \hline
    55 Cnc & GO/DD 12681 & 3/7/2012 & G140M \\ \hline
    eps Eri & GO 13650 & 2/1/2015 & G140M \\ \hline
    \multirow{2}*{GJ 1132} & GO 14462 & \multirow{2}*{2/13/2016} & \multirow{2}*{G140M} \\ 
     & GO 14757 & & \\ \hline
    GJ 1214 & GO 13650 & 8/19/2015 & G140M \\ \hline
    GJ 163 & GO 15071 & 7/10/2019 & G140M \\ \hline
    GJ 173 & GO 14084 & 9/22/2016 & E140M \\ \hline
    GJ 176 & GO 13650 & 3/3/2015  & G140M \\ \hline
    GJ 3470 & GO 14767 & 11/28/2017 & G140M \\ \hline
    GJ 436 & GO 13650 & 6/24/2015 & G140M \\ \hline
    GJ 581 & GO 13650 & 8/10/2015 & G140M \\ \hline
    GJ 667C & GO 13650 & 8/10/2015 & G140M \\ \hline
    GJ 832 & GO 13650 & 10/10/2014 & G140M \\ \hline
    GJ 849 & GO 15071 & 6/9/2019 & G140M \\ \hline
    GJ 876 & GO 13650 & 7/6/2015 & G140M \\ \hline
    \multirow{2}*{HAT-P-11} & GO 14625 & \multirow{2}*{10/23/2016} & \multirow{2}*{G140M} \\
     & GO 14767 & & \\ \hline
    HD 103095 & GO 9455 & 7/19/2003 & E140M \\ \hline
    HD 129333 & GO 12566 & 3/27/2012 & E140M \\ \hline
    \multirow{2}*{HD 189733} & GO 11673 & \multirow{2}*{4/6/2010} & \multirow{2}*{G140M} \\
     & GO 12920 & & \\ \hline
    HD 192310 & GO 12475 & 11/28/2011 & E140M \\ \hline
    \multirow{2}*{HD 209458} & GO 7508 & \multirow{2}*{8/28/2001} & \multirow{2}*{G140M} \\
     & GO 9064 & & \\ \hline
    HD 39091 & GO 15699 & 7/24/2019 & G140M \\ \hline
    HD 39587 & GO 8280 & 10/3/2000 & E140M \\ \hline
    HD 40307 & GO 13650 & 3/16/2015 & E140M \\ \hline
    HD 72905 & GO 12596 & 9/25/2012 & E140M \\ \hline
    HD 85512 & GO 13650 & 3/13/2015 & E140M \\ \hline
    HD 97658 & GO 13650 & 2/12/2015 & E140M \\ \hline
    KAP01 CET & GO 8280 & 9/19/2000 & E140M \\ \hline
    Kapteyn's Star & GO 15190 & 4/3/2019 & E140M \\ \hline
    L 980-5 & GO 15071 & 3/17/2019 & G140M \\ \hline
    \multirow{2}*{LHS 1140} & GO 14888 & \multirow{2}*{8/17/2017} & \multirow{2}*{G140M} \\
     & GO 15264 & & \\ \hline
    LP 756-18 & GO 15071 & 8/7/2019 & G140M \\ \hline
    TRAPPIST-1 & GO 15071 & 12/9/2018 & G140M \\ \hline
    WASP-29 & GO 14767 & 8/9/2019 & G140M \\ \hline
    WASP-69 & GO 14767 & 6/5/2018 & G140M \\ \hline
    \multirow{2}*{WASP-80} & SNAP 13487 & \multirow{2}*{7/14/2014} & \multirow{2}*{G140M} \\
     & GO 14767 & & \\ \hline
\end{tabular}
\end{adjustwidth}
\caption{Observing program, date of first observation, and grating used for STIS observations of stars in this study.}
\label{tab:literature_STIS} %A4
\end{table}

\FloatBarrier

\begin{figure}[!ht]
\epsscale{1.19}
\plotone{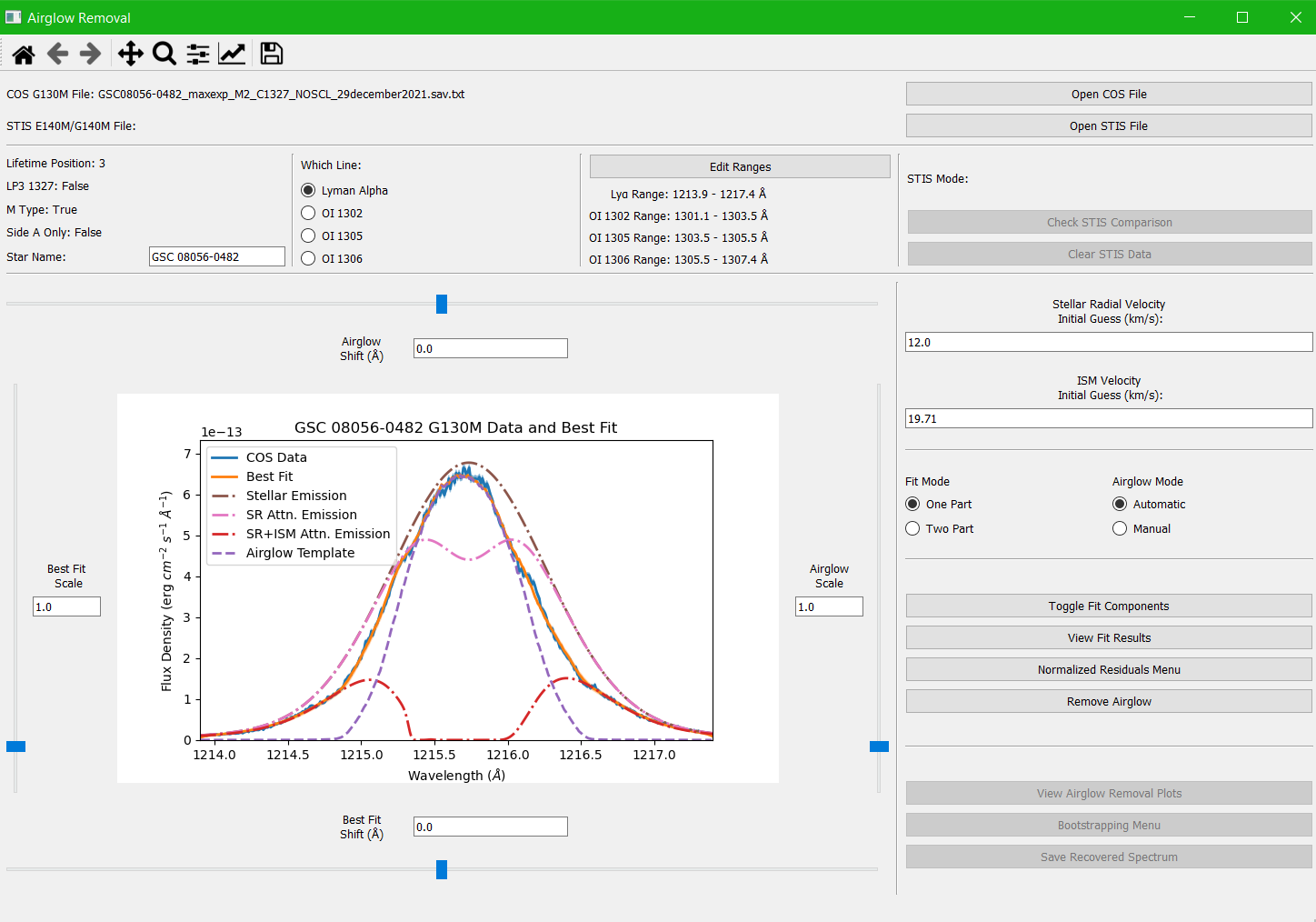}
\caption{The GUI developed in this work, in the process of an airglow subtraction for GSC 08056-0482. See Section \ref{ss:model} for details on the model used by the GUI and Section \ref{ss:GUI} for details on using the GUI. \label{fig:GUI}} %A9
\end{figure}

\begin{figure}[!ht]
\epsscale{1.1}
\plotone{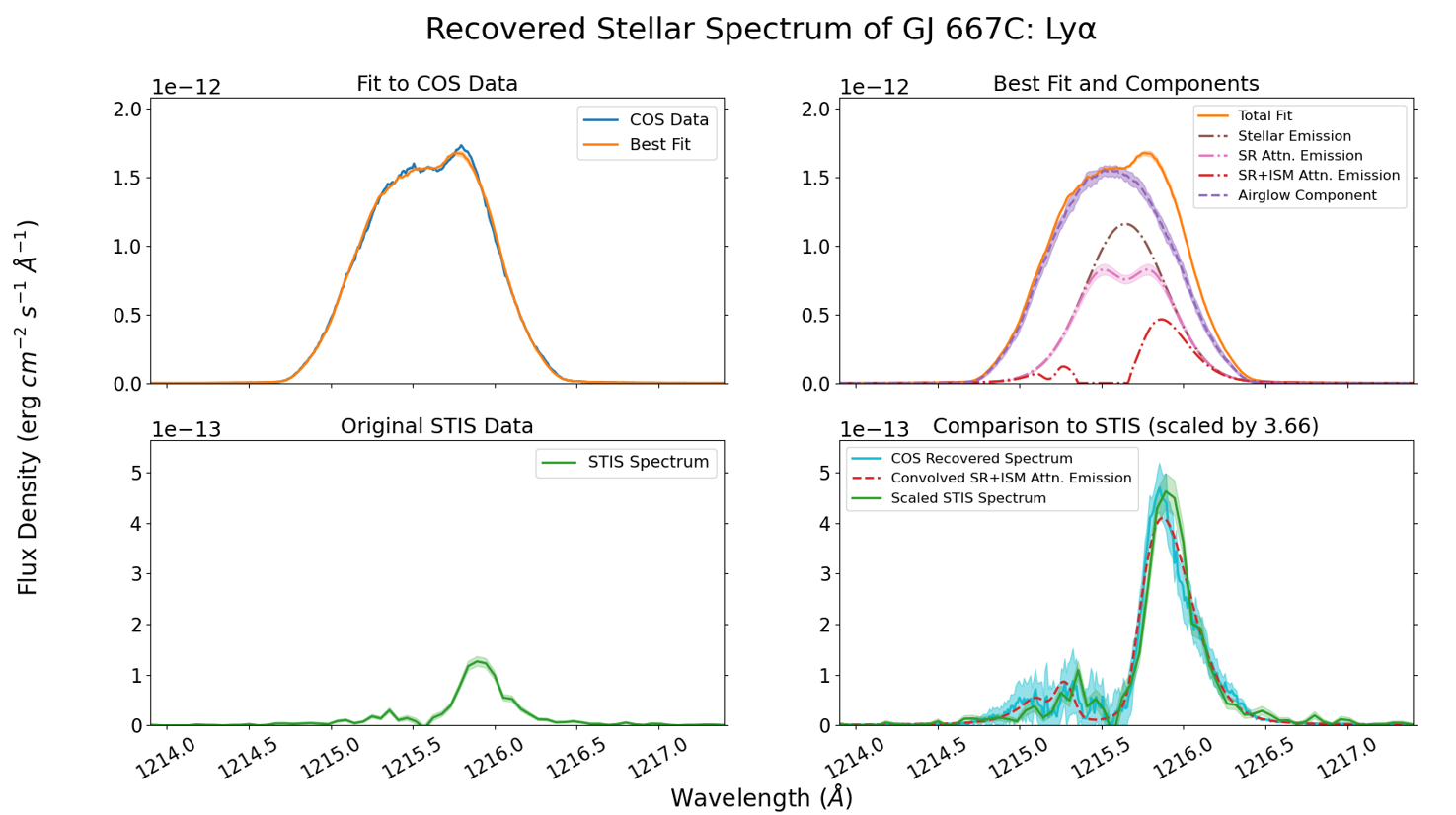}
\caption{The results of the Ly$\alpha$ airglow subtraction for GJ 667C. Plot descriptions are the same as Figure \ref{fig:recover1}. \label{fig:recover3}} %A10
\end{figure}

\begin{figure}[!ht]
\epsscale{1.1}
\plotone{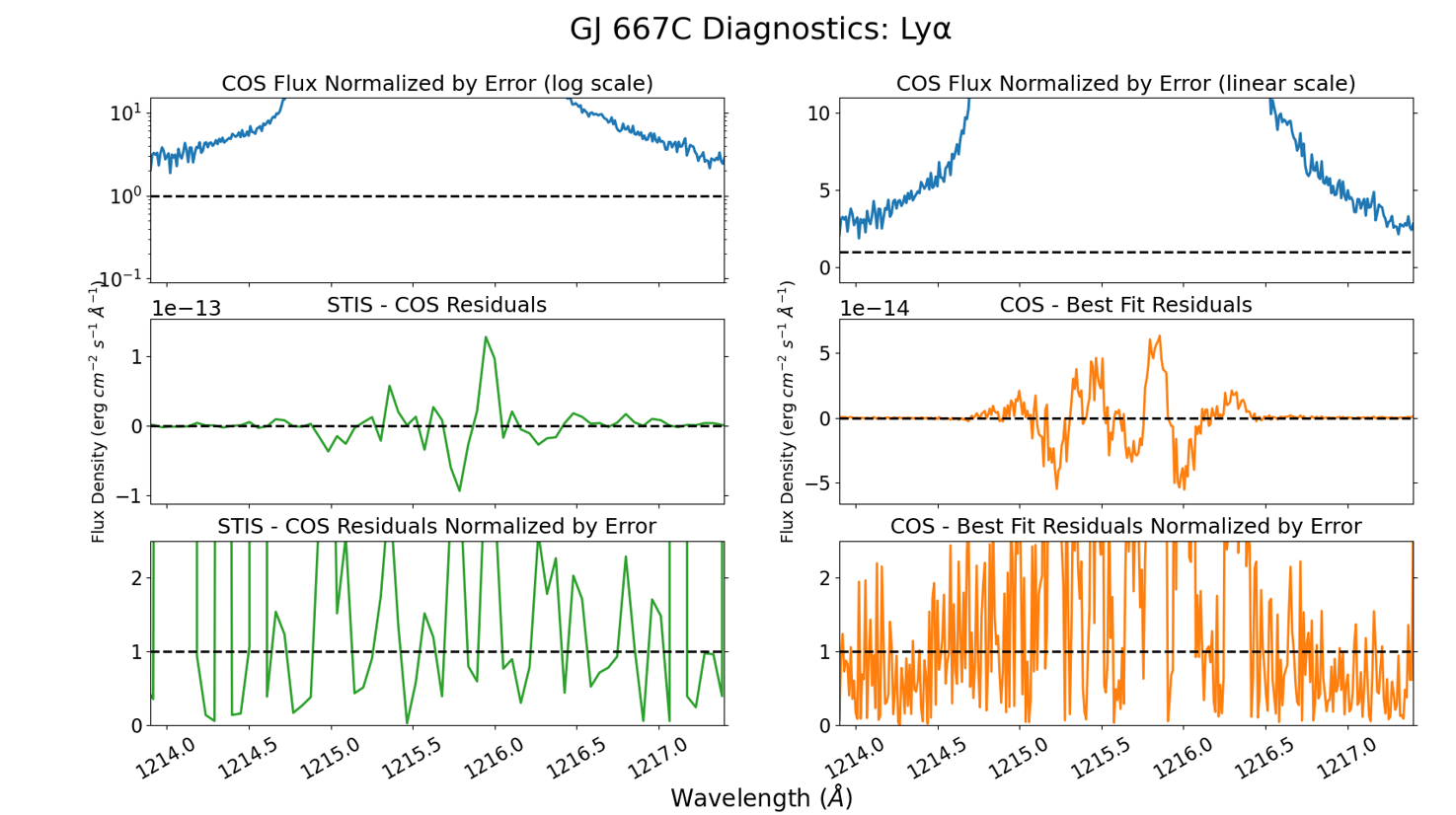}
\caption{The diagnostics of the Ly$\alpha$ airglow subtraction for GJ 667C produced by the airglow removal GUI. The upper left and right plots show the original COS flux normalized by the original COS error, in log space and linear space respectively, for cases where one is more useful than the other. The two lower left plots show the recovered COS - STIS residual, and this residual normalized by the STIS error. The two lower right plots show the observed COS - best fit residual, and this residual normalized by the observed COS error. Both normalized plots on the bottom show the absolute value of the residuals. \label{fig:diagnostic3}} %A11
\end{figure}

\FloatBarrier %keep all Appendix tables and figures out of the refs

\begin{figure*}[!ht]
\centering
\subfigure{\epsscale{0.5} \plotone{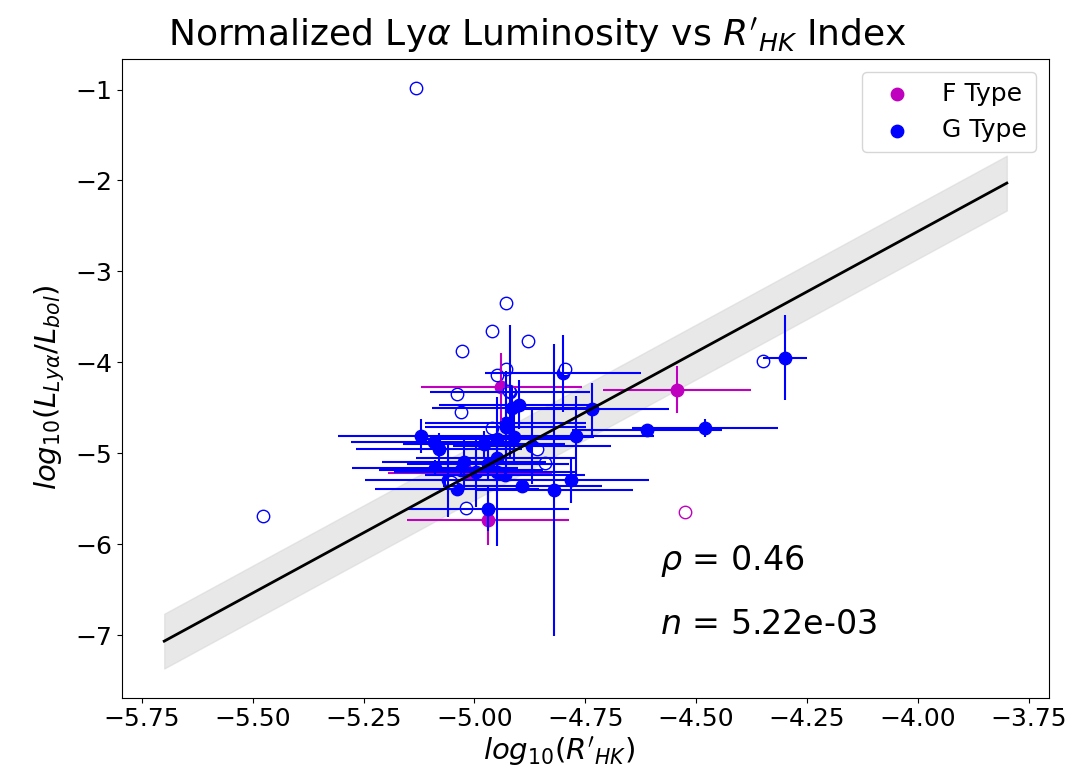} \label{sfig:rphk}} %A12a
\subfigure{\epsscale{0.5} \plotone{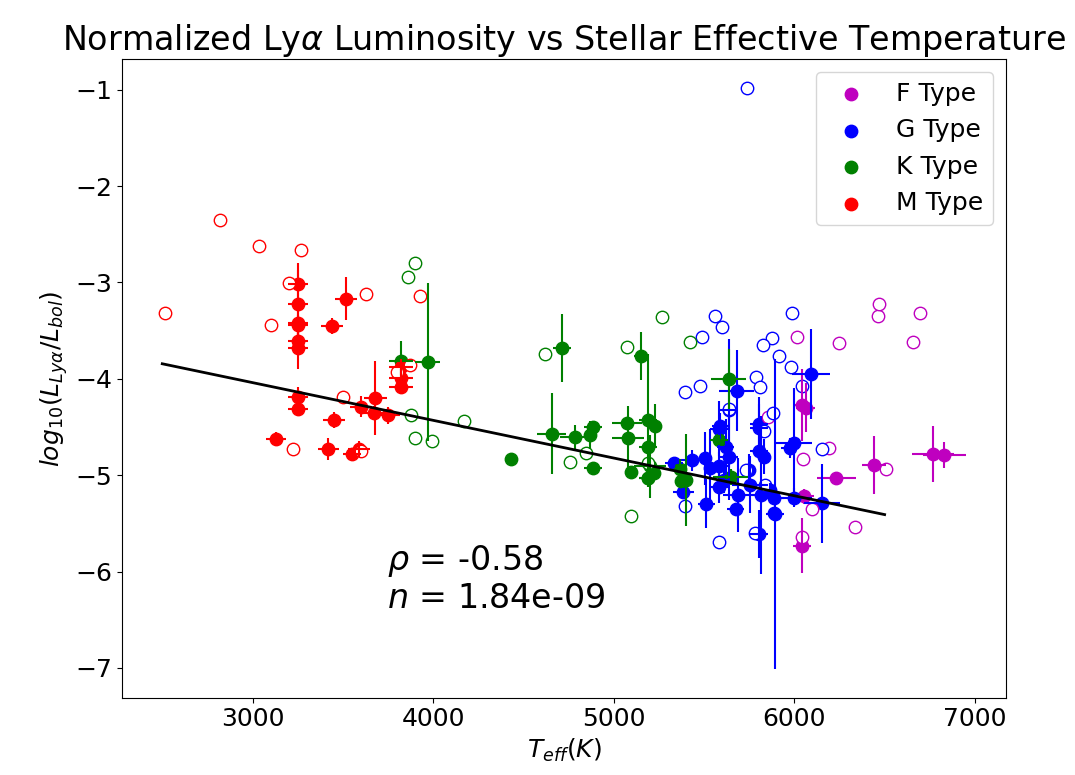} \label{sfig:teff}} %A12b
\subfigure{\epsscale{0.5} \plotone{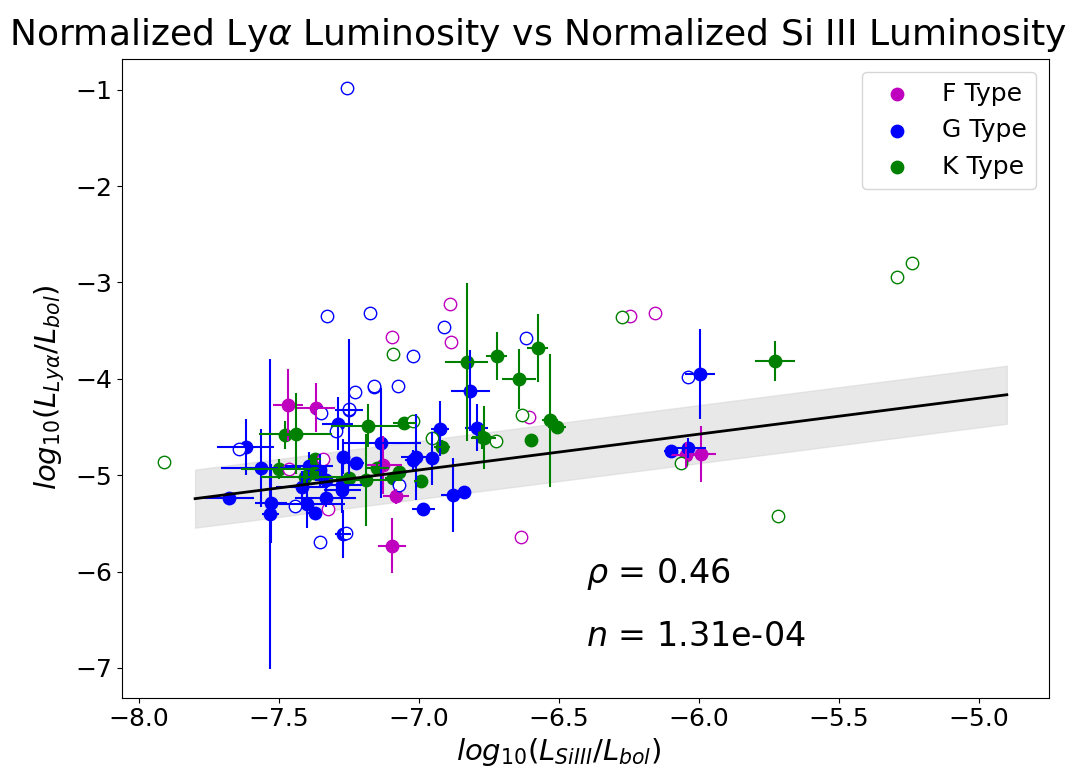} \label{sfig:si3}} %A12c
\subfigure{\epsscale{0.5} \plotone{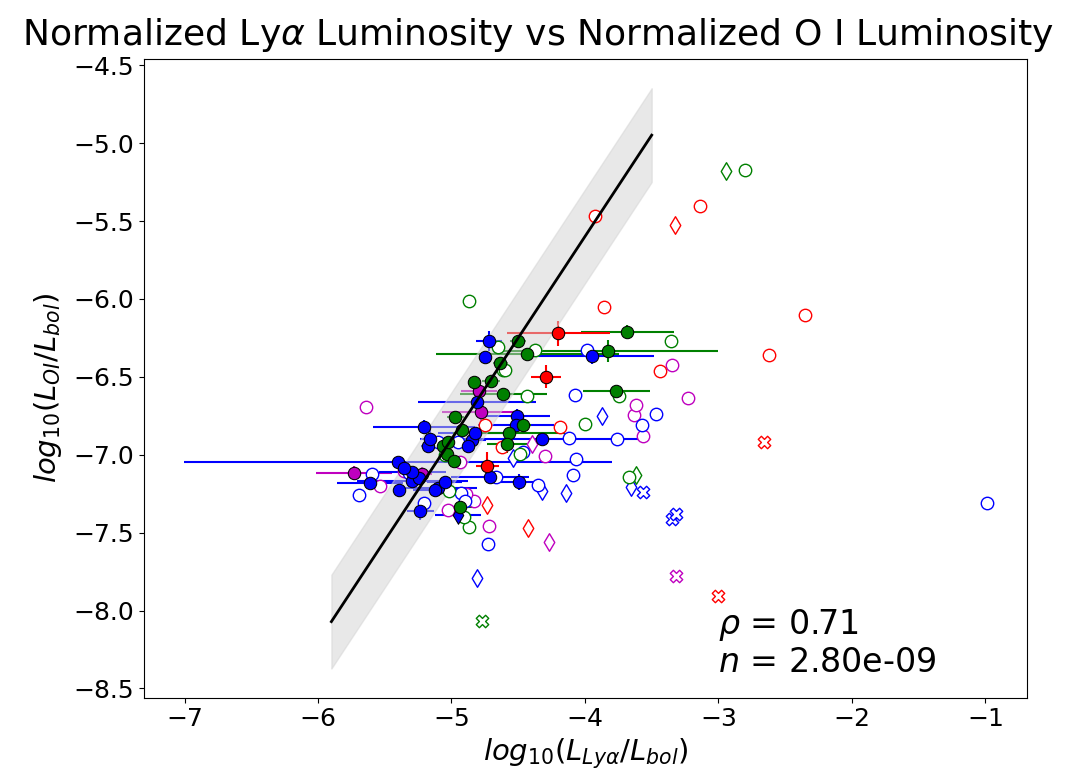} \label{sfig:O1}} %A12d
\subfigure{\epsscale{0.98} \plotone{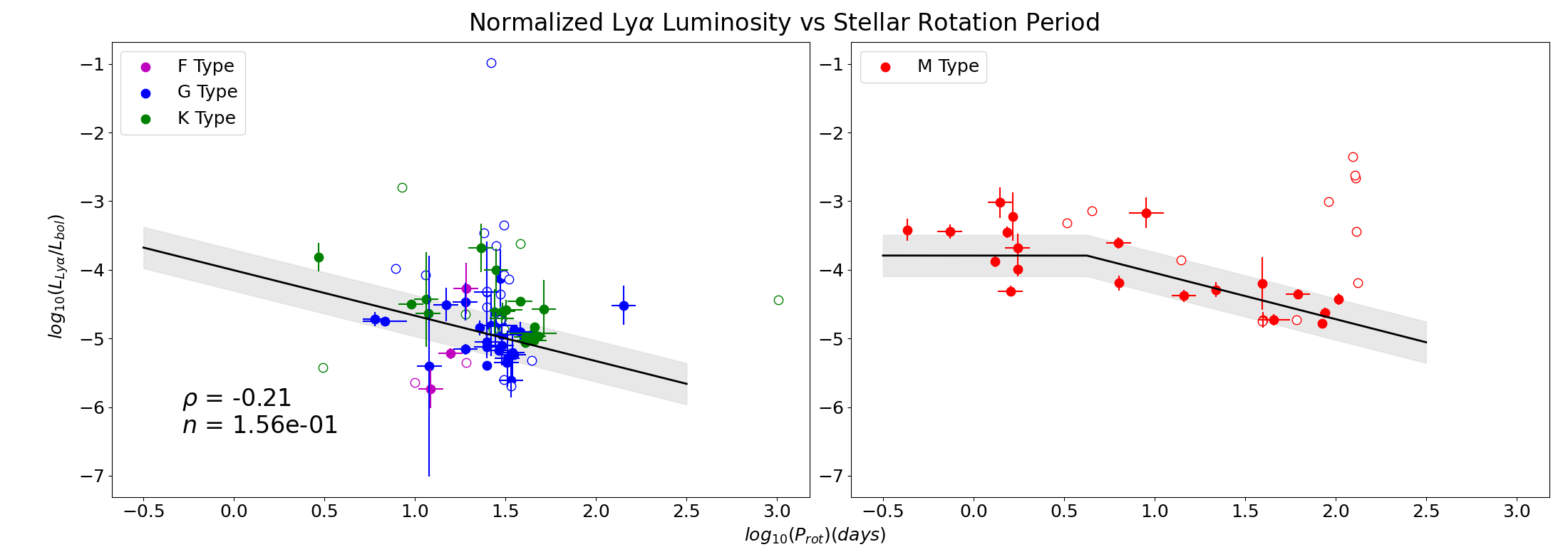} \label{sfig:prot}} %A12e
\caption{Versions of Figures \labelcref{fig:RPHK,fig:Teff,fig:SiIII_L,fig:OI_All,fig:Prot,} that display previously cropped failed recoveries. \label{fig:failed}} %A12
\end{figure*}

%\clearpage %force bibliography to start after the appendix figures
\bibliography{citations}{}
\bibliographystyle{aasjournal}

\end{document}